\documentclass{elsart}
\usepackage{natbib} 
\usepackage{graphicx} 
\usepackage{times}
\newcommand{\myemail}{gmellema@astron.nl}
\newcommand{\be}{\begin{equation}} 
\newcommand{\ba}{\begin{eqnarray}}
\newcommand{\ee}{\end{equation}} 
\newcommand{\ea}{\end{eqnarray}}
\def\lesssim{\mathrel{\hbox{\rlap{\hbox{\lower4pt\hbox{$\sim$}}}\hbox{$<$}}}}
\def\gtrsim{\mathrel{\hbox{\rlap{\hbox{\lower4pt\hbox{$\sim$}}}\hbox{$>$}}}}

\def\gtsima{$\;\buildrel > \over \sim \;$} 
\def\ltsima{$\; \buildrel < \over \sim \;$} \def\gsim{\lower.5ex\hbox{\gtsima}}
\def\lsim{\lower.5ex\hbox{\ltsima}}
\def\simgt{\lower.5ex\hbox{\gtsima}}
\def\simlt{\lower.5ex\hbox{\ltsima}}
\def\simpr{\lower.5ex\hbox{\prosima}}

 \newcommand{\apjl}{ApJ}
\def\simless{\mathbin{\lower 3pt\hbox {$\rlap{\raise
5pt\hbox{$\char'074$}}\mathchar''7218$}}} 
\def\simgreat{\mathbin{\lower 3pt\hbox {$\rlap{\raise
5pt\hbox{$\char'076$}}\mathchar''7218$}}} 
\newcommand{\apj}{ApJ} \newcommand{\aj}{AJ} \newcommand{\apjs}{ApJS}
\newcommand{\mnras}{MNRAS} \newcommand{\aap}{A\&A}
\newcommand{\beq}{\begin{equation}} \newcommand{\eeq}{\end{equation}}

\begin{document}
\runauthor{Mellema, Iliev, Alvarez \& Shapiro}
\begin{frontmatter}
  \title{C$^2$-Ray: A new method for photon-conserving transport of 
ionizing radiation}
  \author[ASTRON,Leiden]{Garrelt~Mellema} \author[CITA]{Ilian~T.~Iliev}
  \author[Austin]{Marcelo~A.~Alvarez} \author[Austin]{Paul~R.~Shapiro}
  \thanks[]{E-mail addresses:\tt ~\myemail, iliev@cita.utoronto.ca,
    marcelo@astro.as.utexas.edu, shapiro@astro.as.utexas.edu}
  \address[ASTRON]{ASTRON, P.O. Box 1, NL-7990 AA Dwingeloo, The Netherlands}
  \address[Leiden]{Sterrewacht Leiden, P.O. Box 9513, NL-2300 RA Leiden, The
    Netherlands} \address[CITA]{Canadian Institute for Theoretical
    Astrophysics, University of Toronto, 60 St. George Street, Toronto, ON M5S
    3H8, Canada} \address[Austin]{Department of Astronomy, University of
    Texas, Austin, TX 78712-1083}
\begin{abstract}
  We present a new numerical method for calculating the transfer of ionizing
  radiation, called C$^2$-Ray (Conservative, Causal Ray-tracing method). The
  transfer of ionizing radiation in diffuse gas presents a special challenge
  to most numerical methods which involve time- and spatial-differencing. 
  Standard approaches to radiative transport require that grid cells must be 
  small enough to be optically-thin while time steps are small enough that 
  ionization fronts do not cross a cell in a single time step. This quickly
  becomes prohibitively expensive. We have developed an algorithm which
  overcomes these limitations and is, therefore, orders of magnitude more
  efficient. The method is explicitly photon-conserving, so the depletion of
  ionizing photons by bound-free opacity is guaranteed to equal the
  photoionizations these photons caused. As a result, grid cells can be large
  and very optically-thick without loss of accuracy. The method also uses an
  analytical relaxation solution for the ionization rate equations for each
  time step which can accommodate time steps which greatly exceed the
  characteristic ionization and ionization front crossing times. Together,
  these features make it possible to integrate the equation of transfer along
  a ray with many fewer cells and time steps than previous methods. For 
  multi-dimensional calculations, the code utilizes short-characteristics ray 
  tracing. The method scales as the product of the number of grid cells and
  the number of sources. C$^2$-Ray is well-suited for coupling radiative
  transfer to gas and N-body dynamics methods, on both fixed and adaptive
  grids, without imposing additional limitations on the time step and grid
  spacing. We present several tests of the code involving propagation of
  ionization fronts in one and three dimensions, in both homogeneous and 
  inhomogeneous 
  density fields. We compare to analytical solutions for the ionization front
  position and velocity, some of which we derive here for the first time.
  As an illustration, we apply  C$^2$-Ray to simulate cosmic
  reionization in three dimensional inhomogeneous cosmological density field. 
  We also apply it to the problem of I-front trapping 
  in a dense clump, using both a fixed and an adaptive grid.   
\end{abstract}

\end{frontmatter}

\section{Introduction}
The interplay between matter and radiation plays a crucial role in many
astrophysical processes. The radiative feedback effects on hydrodynamic flows
are particularly strong for the case of hydrogen- and helium-ionizing
radiation. The most dramatic manifestation of such radiative feedback, with
far-reaching consequences for the present-day universe, was the reionization of
hydrogen between redshifts $z\sim30$ and $z\sim6$ and helium by $z\sim3-4$.
The formation of the first ionizing sources in the universe created expanding
intergalactic H~II regions. These eventually overlapped, leaving the universe 
largely ionized, as demonstrated by the absence of a
Gunn-Peterson trough in the spectra of high-redshift QSO's and galaxies. On
smaller scales, the change in pressure due to the heating associated with 
photoionization drives powerful flow phenomena, such as 
photoevaporation of minihalos during cosmic reionization 
\citep{2004MNRAS.348..753S,ISR05}, triggering and regulating star formation in 
Damped Lyman-$\alpha$ systems \citep{IHF05}, the formation
of pillars in H~II regions, evaporation of planetary disks in the Orion nebula
\citep{1998AJ....116..322H}, the formation of planetary nebulae
\citep{1994A&A...290..915M}, and other nebulae, such as the famous ring around
SN1987A \citep{1995ApJ...452L..45C}.

Precise modeling of processes of radiative feedback is thus very important in
order to understand all these phenomena. However, the development of effective
and robust numerical implementations of radiative feedback, particularly in
the case of optically-thick media, presents huge challenges.  The full
radiative transfer problem adds to the usual three spatial and three velocity
coordinates also angular and frequency dependencies, leading to a complicated,
multi-dimensional problem. It also adds non-locality to the fluid equations, 
since distant sources can affect local dynamics, and introduces new requirements 
for the maximum size of the numerical time steps and cell sizes.
Any na\"ive, brute-force attempts to solve such
problems are currently beyond the capabilities of even the fastest computers.
The development of more efficient and robust radiative transfer algorithms is 
thus crucial.

In the last few years there has been an intense development of numerical 
radiative transfer methods suitable for cosmology.
However, compared to the much
more sophisticated and mature methods developed for N-body and gas-dynamical
simulations, the modeling of radiative transfer effects is still in its
infancy. Most of the current state-of-the-art cosmological radiative transfer
codes, with very few exceptions, transport the radiation on pre-computed 
density fields, thus ignoring the dynamical back-reaction of the gas. While 
such approaches have their legitimate uses, the lack of gasdynamical feedback 
severely limits the questions that can be answered. Additional important 
limitations on all cosmological radiative transfer codes at present are 
imposed by their low resolution, which currently rarely exceeds $128^3$ 
computational cells in three dimensions.

There are two basic classes of computational radiative transfer methods 
currently in use, moment methods
\citep{2001NewA....6..437G,2002ApJS..141..211C,2003ApJS..147..197H}, and
ray-tracing methods \citep{1998A&A...331..335M,1999MNRAS.309..287R,
1999ApJ...523...66A,2000MNRAS.314..611C,2001MNRAS.321..593N,2001NewA....6..359S,
2002ApJ...572..695R,2003A&A...405..189L,2003MNRAS.345..379M,2004MNRAS.348..753S,
  2004MNRAS.348L..43B,ISR05}. Each class of methods has its own advantages and
disadvantages. Generally, moment methods are fast and largely independent of
the number of ionizing sources, but are also fairly diffusive, which can lead
to incorrect results in some situations like e.g.\ producing incorrect shadows
\citep{2001NewA....6..437G,2002ApJS..141..211C,2003ApJS..147..197H}. On the
other hand, ray-tracing approaches can be very accurate, but care should be
taken to cover properly the space with rays, which often makes them
computationally-expensive. This limits the number of sources that can be
handled, and complicates the coupling of such methods to gas- and N-body
dynamics. A whole new approach which may combine the best of both worlds, 
is being explored by \citet{Ritzi}.

The coupling of radiative transfer and hydrodynamics to model ionization
fronts (I-fronts) is difficult because accurate finite-differencing in time
and space generally requires time steps smaller than the cell-crossing time of
the I-front (which can be highly supersonic). In addition, cell sizes must be
small enough to be optically thin to ionizing radiation prior to the passage
of the I-front. If cell sizes are limited to be optically thin, the Courant
time drives the time step way down and the number of time steps way up.   
To make matters much worse, if the 
usual Courant condition, which limits the time step size to be less than the 
sound crossing time of a cell, has to be replaced by an "I-front Courant 
condition" involving the I-front speed rather than the speed of sound, this 
problem is greatly exacerbated. It is this combination of small cell size and
small time step that makes the coupling of radiative transfer of
ionizing radiation to gas and gravitational dynamics so computationally 
difficult by traditional finite-differencing approaches. In this paper we
present a method that relaxes these requirements on both the spatial- and
time-resolution.  

Previously we have developed an adaptive-grid (AMR), axisymmetric 
radiative-transfer and hydrodynamics code CORAL, which also contains 
non-equilibrium chemistry and cooling due to H, He, and metals (C, O, N, S and Ne)
\citep{1997ApJS..109..517R, 1998A&A...331..335M,2004MNRAS.348..753S}. The
latest versions of this code properly track fast, R-type I-fronts as well as
slow, D-type I-fronts, but can require fairly large number of time steps in
order to assure accuracy. We have applied this code to a variety of
astrophysical and cosmological problems, including photoevaporation of dense
clumps in planetary nebulae \citep{1998A&A...331..335M} and photoevaporation
of cosmological minihalos during reionization
\citep{2004MNRAS.348..753S,ISR05}. The improved, three-dimensional (3D) 
successor of CORAL, different versions of which use either AMR or uniform 
grid, is 
known as Yguaz\'u \citep{2000RMxAA..36...67R}. The uniform grid version of 
Yguaz\'u  was used e.g.\ to calculate pre-ionization from a jet bow shock
\citep{1999RMxAA..35..123R} and stellar jets moving into H~II regions
\citep{2000MNRAS.314..681R}. \citet{2003A&A...405..189L} used the to same
basic methodology as in Yguaz\'u, but in a different implementation, to study
the interaction of two photoevaporating clouds in 3D. With adaptive mesh
refinement these methods can reach effective resolution of $512^3$, or more.

In order to improve upon the current implementation of radiative feedback in 
the CORAL and Yguaz\'u codes, we present here a new approach to the
photoionization calculations. Ensuring a high level of photon conservation
helps relax the spatial resolution requirements of the code. All published 
ray-tracing methods have strong constraints on the time step in order to
conserve photons. This makes combined hydrodynamics and photoionization
calculations expensive. Our method relaxes these constraints on the time step.
The ultimate goal is to combine it with a hydrodynamics method, and hence
speed and efficiency are essential. In the interests of length, in the current
paper we will the describe our photoionization calculation and ray-tracing
method, without discussing its coupling to hydrodynamics, which we will
present in a follow-up paper. Our method is in fact also useful for
`stand-alone' or post-processing photoionization calculations, and that is
how it is presented here.

The structure of this paper is as follows. In \S~\ref{conserv_sect} we present
our photon-conserving method for transferring the ionizing radiation and
calculating the ionization rate. We also present a relaxation scheme to
advance the non-equilibrium ionization rate equations across a finite time
step, which is not limited by the ionization time. In order to use the method 
in a multidimensional setting, we need to cast rays from the sources. We
describe our causal ray tracing scheme in \S~\ref{tracing_sect0} and 
Appendix~\ref{tracing_sect}. The treatment of multiple sources
is discussed in \S~\ref{multiple_sect}. In \S~\ref{tests_sect} we present the
tests we have performed to verify our method, in both 1D and 3D. Finally, in 
\S~\ref{applic_sect} we present the first illustrative applications of our method.

\section{Conserving photons}
\label{conserv_sect}
Consider a continuum radiation field produced by an ionizing source with a
spectral energy distribution of $L_\nu$, traveling through a gas with a
frequency-dependent optical depth $\tau_\nu$. The flux of hydrogen-ionizing
photons arriving at a distance $r$ from the source is given by
\begin{equation}
    F(r)={1 \over 4\pi r^2}\int_{\nu_{\rm th}}^\infty {L_\nu
  e^{-\tau_\nu(r)} \over h\nu}{\rm d}\nu\,,
\label{opt_thin}
\end{equation}
where $h\nu_{\rm th}=13.6$ eV is the ionization threshold of hydrogen.  The
exact expression for the local ionization rate at a distance $r$ from the 
ionizing source for hydrogen atoms with a cross section for ionizing photons 
$\sigma_\nu$ is \citep{1989agna.book.....O}
\begin{equation}
  \Gamma_{\rm local}(r)={1 \over 4\pi r^2}\int_{\nu_{\rm th}}^\infty
  {L_\nu \sigma_\nu e^{-\tau_\nu(r)} \over h\nu}{\rm d}\nu\,.
  \label{stdgamma}
\end{equation}
The optical depth is defined, as usual, as
\begin{equation}
  \tau_\nu = \sigma_\nu N_{\rm HI}\,,
\end{equation}
where $N_{\rm HI}$ is the column density of neutral hydrogen. 
The expression in Eq.~(\ref{stdgamma}) is exact only at a given 
point in space and moment in time. However, in numerical simulations 
both space and time are necessarily discretized into finite-size cells 
and finite time steps. For finite cells the expression in
Eq.~(\ref{stdgamma}) needs to be finite-differenced in a correct manner 
to ensure explicit photon conservation, which we discuss next.

\subsection{Spatial discretization}
\label{spat_discret_sect}
In a spatially discretized volume (a `grid'), each spatial element does not
have a single distance to the source, but spans a certain range $\Delta r$.
Taking one ionization rate to be representative for this range is an
approximation that is valid only if the grid cells are limited in size so 
that each cell is optically thin to the ionizing radiation. Since radiative 
transfer 
is computationally expensive, in general limiting the cell size in this way 
is prohibitive. As a result, the spatial discretization is often coarse, with 
very optically-thick cells. The
effect of the approximation is that the number of photons absorbed by a `grid
cell' is no longer equal to the number of ionizations calculated for that
cell. In other words, photons are not conserved, and ionization fronts will
not travel at the correct speed. This problem was previously noted by
\citet{1999ApJ...523...66A}, who suggested that a better approach would be 
to force the ionization rate inside each halo cell to equal the absorption rate 
per cell used to attenuate the radiation in the transport algorithm. We shall
adopt this approach and develop it further as follows.

Consider a spherical shell of central radius $r$ and width $\Delta r$, filled
with neutral hydrogen of number density $n_{\rm HI}$. Let $\dot{N}(r-\Delta
r/2)$ be the number of ionizing photons arriving at the shell per unit time,
and $\dot{N}(r+\Delta r/2)$ the number of photons leaving the shell. The
difference between these two numbers gives us the number of photons (per unit
time) which were absorbed in the shell. These photons ionized a fraction of the 
$n_{\rm HI}V_{\rm shell}$ hydrogen atoms in the shell, where $V_{\rm shell}$ is the volume
of the shell.  The photoionization rate is then given by
\begin{equation}
  \Gamma = {\dot{N}(r-{\Delta r\over 2}) - \dot{N}(r+{\Delta r \over 2}) \over
    n_{\rm HI}V_{\rm shell}}\,,
\end{equation}
with
\begin{equation}
  V_{\rm shell}={4\pi \over 3}\left[\left(r+{\Delta r \over
  2}\right)^3- \left(r-{\Delta r \over 2}\right)^3\right]\,.
  \label{photorate}
\end{equation}
Defining the optical depth from the source to $r-\Delta r/2$ as $\tau_\nu$,
and the optical depth between $r-\Delta r/2$ and $r+\Delta r/2$ (i.e.\ the
optical depth of the cell) as $\Delta\tau_\nu$, we can re-write
Eq.~(\ref{photorate}) as
\begin{equation}
  \Gamma=\int_{\nu_{\rm th}}^\infty {L_\nu e^{-\tau_\nu} \over h\nu}
  {1-e^{-\Delta\tau_\nu}\over n_{\rm HI} V_{\rm shell}}{\rm d}\nu\,,
  \label{spatialgamma}
\end{equation}
Taking the limit of $\Delta\tau_\nu\ll 1$ and $\Delta r\ll r$ (i.e.\ low optical
depth per cell and also the distance from the source to the cell much larger
than the size of the cell), one retrieves Eq.~(\ref{stdgamma}).
Since transport is radial, this formula is also valid if we are only
considering a small part of the shell. Equation~(\ref{spatialgamma}) is
equivalent to Eq.~(15) from \citet{1999ApJ...523...66A} if we identify
their $V_{\rm cell}$ with $V_{\rm shell}$.

\begin{figure}
  \includegraphics[scale=0.35]{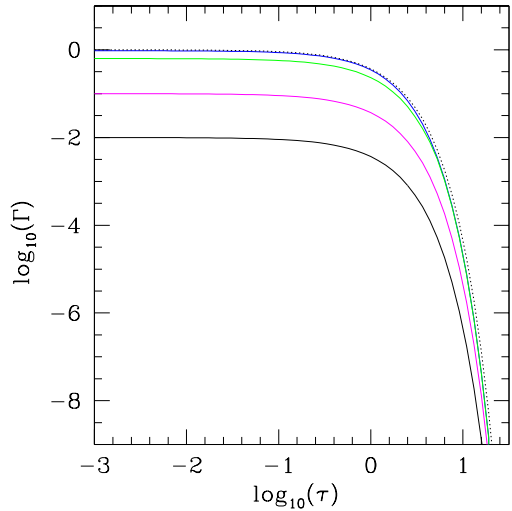}
\caption{Photoionization rate from Eq.~(\ref{stdgamma}) evaluated based on
the optical depth, $\tau$, up to the cell (dotted line) compared to the 
photon-conserving photoionization rates from Eq.~(\ref{spatialgamma}) for 
$\Delta\tau=0.1, 1, 10,$ and 100 (solid lines; top to bottom) vs. optical depth 
$\tau$ from the source to the cell. We normalized the rates to 1 for 
$\tau\ll1$ and $\Delta\tau\ll1$, assuming $r\gg\Delta r$ and gray
opacities (i.e.\ $\sigma_\nu=$ constant).
\label{photorates_fig}}
\end{figure}

In Figure~\ref{photorates_fig} we illustrate the difference between the
local photoionization rate, $\Gamma_{\rm local}(r)$ from 
Eq.~(\ref{stdgamma}) 
evaluated at the entrance point of a cell and the photon-conserving 
photoionization rate, $\Gamma$, from Eq.~(\ref{spatialgamma}) over the 
finite cell. We plot $\Gamma$ for optical depths per cell 
$\Delta\tau=0.1, 1, 10,$ and 100, and $\Gamma_{\rm local}(r)$ (which is 
independent of $\Delta\tau$) vs. the optical
depth up to the cell, $\tau$. All rates are calculated in gray-opacity
approximation (photoionization cross-section independent of
frequency). We clearly see that, as expected, the local photoionization
rate agrees well with the photon-conserving rates only for optically-thin
cells ($\Delta\tau<0.1$), in which case the two curves are barely
distinguishable. However, the two rates start to differ significantly even for
moderately optically-thick cells ($\Delta\tau\sim1$) and are completely
different for highly optically-thick cells ($\Delta\tau>10$). In
practice, the main differences when using photon-conserving finite-differenced 
rates would arise for moderately optically-thick cells for which the optical 
depth 
from the source $\tau$ is low to moderate. For a cell which is either shielded or
self-shielded (i.e.\ either $\tau$ or $\Delta\tau$ is large), the photoionization 
rate would be very low, and hence even incorrect photoionization rates would have 
no appreciable impact on the ionized fraction.

Another way to think of the photon-conserving photoionization rate 
in Eq.~(\ref{spatialgamma}) is to notice that for thin shell,
$\Delta r\ll r$, and gray opacity ($\sigma_\nu=$constant) we have
\be
\Gamma=\frac{\Gamma_{\rm local}(r) - 
\Gamma_{\rm local}(r+\Delta r)}{\Delta \tau}.
\label{spatial_gamma_2}
\ee
This shows that the fractional error introduced by using 
Eq.~(\ref{stdgamma}) evaluated at the shell inner edge radius to 
give the ionization rate for all atoms in the shell, is given by
$[\Gamma_{\rm local}(r) - \Gamma_{\rm local}(r+\Delta r)]/\Gamma=\Delta \tau$
and hence grows as $\Delta\tau$ grows. This again demonstrates that 
$\Gamma_{\rm local}$ evaluated at the inner cell boundary is a good 
approximation to the finite-cell photon-conserving rate only for 
optically-thin cells.

\subsection{Temporal discretization}
\label{time_discr_sect}
When using the photoionization rate $\Gamma$ to integrate the ionization
equations over a time step $\Delta t$, one normally assumes it to be constant
during this time step. However, since the ionizations and recombinations in
the cell will change the density of neutral hydrogen, the optical depths 
$\tau_\nu$ and $\Delta \tau_\nu$ will also change. The usual solution is to 
take $\Delta t$ which is small enough so the
optical depths do not change appreciably over one step. For fast moving I-fronts 
this demands very short time steps. Various authors use different criteria in
determining what the time step should be: a fraction of the photon travel time
over a cell $\Delta r/c$ \citep{1999ApJ...523...66A}, several times the
ionization time $\Gamma^{-1}$ \citep{2004MNRAS.348L..43B}, a small fraction of
the timescale of change of the H~I fraction, $n_{\rm HI}/({\rm d}n_{\rm
  HI}/{\rm d}t)$, \citep[which sets the speed of the ionization
front,][]{2004MNRAS.348..753S}.  Such small time steps make for expensive
calculations if high accuracy is to be achieved.

Since the above rate $\Gamma$ is really a time-averaged rate (we only know
what went in and what came out of the cell, not what happened in detail within
the cell), a time-averaged value for $\Delta \tau_\nu$ can be expected to
relax the constraint on the time step.  We illustrate this with the following
simplified model. Consider an infinitely-thin parcel of hydrogen gas of number
density $n$ illuminated by a time-varying photon flux $F_{\nu}(t)$. Neglecting
for the moment the effect of recombinations and collisional ionizations, the
ionization rate equation can be written as
\begin{equation}
\label{ndionization}
\frac{{\rm d}y_{\rm HI}}{{\rm d}t}=-\Gamma y_{\rm HI}.
\end{equation}
where $y_{\rm HI}$ is the neutral fraction of hydrogen. If we define a
time-averaged photoionization rate, $\langle\Gamma\rangle$, as
\begin{equation}
\langle\Gamma\rangle = \frac{1}{\Delta t}\int_t^{t+\Delta t}\Gamma(t'){\rm d}t',
\end{equation}
then the solution to Eq.~(\ref{ndionization}) is identical to that of
\begin{equation}
\frac{{\rm d}y_{\rm HI}}{{\rm d}t}=-\langle \Gamma\rangle y_{\rm HI}.
\end{equation}
This shows that, if only photoionizations contribute to the rate equation, it
is correct to treat the ionization rate as a constant over the time step, no matter 
how large $\Delta t$ is, as long as we take its value to be the time-average of 
$\Gamma$ over $\Delta t$. If recombinations
and collisional ionizations are taken into account, then this is not
necessarily true since photoionizations that happen early in the time interval
are more likely to be canceled by recombinations than those that happen later.
However, as we shall show, using the time-averaged flux only introduces appreciable
deviations from photon-conservation when the time step becomes comparable to the 
recombination time of the cell, and even in such cases the approximation holds fairly 
well.

Solving for the ionization fractions is in itself not entirely trivial since
we are dealing with stiff partial differential equations. Still considering
only hydrogen, then the evolution of the ionized fraction 
$x\equiv1-y_{\rm HI}$ can be written as
\begin{equation}
  {{\rm d} x \over {\rm d} t} = (1-x)(\Gamma+n_{\rm e}C_H) - xn_{\rm e}\alpha_H\,,
  \label{ionization}
\end{equation}
where $n_{\rm e}$ is the electron density (itself dependent on the ionization
fraction $x$) and $C_H$ and $\alpha_H$ are, respectively, the
(temperature-dependent) collisional ionization and recombination coefficients
for hydrogen. In this paper we assume the On-The-Spot (OTS) approximation (i.e.\ that 
the recombinations to the ground state are locally reabsorbed), in which case
the recombination coefficient $\alpha_H$ is equal to 
$\alpha_B = 2.59 \times 10^{-13}(T/10^4\,K)^{-0.7}$ cm$^3$ s$^{-1}$,
the Case~B recombination coefficient for hydrogen at gas temperature $T$.

If one takes $\Gamma$, $n_{\rm e}$, $C_H$ and $\alpha_H$ to be
constant, this equation has an analytical solution
\begin{equation}
   x(t) = x_{\rm eq} + (x_0 - x_{\rm eq})e^{- t / t_{\rm i}}\,,
\end{equation}
with
\begin{equation}
  t_{\rm i} = 1 / (\Gamma + n_{\rm e}C_H+n_{\rm e}\alpha_H)\,,
\end{equation}
and
\begin{equation}
  x_{\rm eq} = {\Gamma+n_{\rm e}C_H \over \Gamma +n_{\rm e}C_H+n_{\rm e}\alpha_H}\,.
\end{equation}
\citet{1987A&A...174..211S} suggested using this solution, and iterating for
the electron density. The approach can be expanded to include other atomic
species such as helium and metals. This method was successfully used by e.g.\ 
\citet{1994A&A...289..937F, 1997ApJS..109..517R,
  1998A&A...331..335M,2004MNRAS.348..753S,ISR05}, to name a few examples. In
order not to have to repeatedly calculate the integrals over the spectrum,
they are tabulated as function of the optical depth at $\nu_{\rm th}$, the 
ionization threshold of hydrogen, which, together with the spectral shape 
completely determines these integrals. We refer the reader
to the above papers for further details on this method for solving the
non-equilibrium chemistry equations.

In the current context this approach for following the chemistry evolution
equations is particularly appealing since it provides us with an analytical
expression for the time-averaged ionization fraction
\begin{equation}
  \langle x \rangle = x_{\rm eq} +(x_0-x_{\rm eq}) (1-e^{-\Delta t /
  t_{\rm i}}){t_{\rm i}\over \Delta t}\,,
\label{mean_x}
\end{equation}
which can in turn be used to find the time-averaged optical depth of the cell
$\langle \Delta \tau \rangle$, as follows:
\be
\langle \Delta\tau_\nu \rangle=(1-\langle x\rangle)n_H\sigma_\nu\Delta r,
\label{meantau}
\ee
where $\langle x\rangle$ is given by Eq.~(\ref{mean_x}). 

\begin{figure}
\begin{center}
  \includegraphics[scale=1]{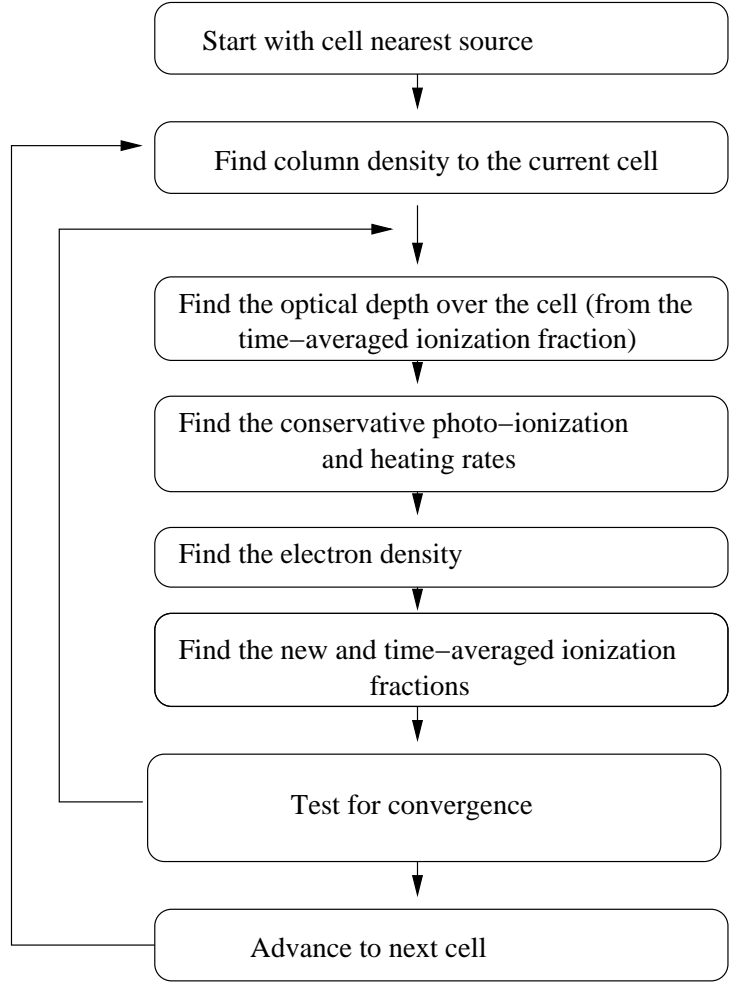}
\caption{Flow chart of our method for a given computational cell.
\label{chart}}
\end{center}
\end{figure}

In order to
  remain self-consistent, the value of the $\langle x \rangle$ should then be
  used to modify $\Gamma$ by changing both the optical depth and the neutral
  density:
\begin{equation}
  \Gamma=\int_{\nu_{\rm th}}^\infty {L_\nu e^{-\langle
        \tau_\nu \rangle} \over h\nu}
        {1-e^{-\langle\Delta\tau_\nu\rangle}\over \langle n_{\rm HI}
        \rangle V_{\rm shell}}{\rm d}\nu\,,
\label{photoncons_gamma}
\end{equation}
which leads to an iterative process. Furthermore, since $n_{\rm e}$ plays a
role similar to $\Gamma$, we use its time-averaged value 
$\langle n_{\rm e}\rangle$ when evaluating the ionization 
Eq.~(\ref{ionization}).  The
optical depth between the source and the cell edge $\tau_\nu$ is replaced by
its time-averaged value, by adding, in causal order, all the time-averaged
$\langle\Delta\tau_\nu\rangle$ of the cells lying between the source and the
cell under consideration. This total optical depth (evaluated at the
ionizing threshold of hydrogen, $\tau_0=\tau_{\nu=\nu_{\rm th}}$), is used 
to look up the values of the integral
\begin{equation}
  \Gamma'(\tau_0)\equiv
\int_{\nu_{\rm th}}^\infty \frac{L_\nu e^{-\langle
        \tau_\nu \rangle}}{h\nu\langle n_{\rm HI}
        \rangle V_{\rm shell}}{\rm d}\nu\,,
\label{photoncons_gamma1}
\end{equation}
in pre-calculated tables. 
Then the photoionization rate in Eq.~(\ref{photoncons_gamma}) is given by 
$\Gamma=\Gamma'(\tau_0)-\Gamma'(\tau_0+\Delta\tau_0)$, where 
$\Delta\tau_0=\Delta\tau_{\nu=\nu_{\rm th}}$.

In practice, the iteration for finding the ionization state of a given cell
proceeds as follows (see flowchart in Figure~\ref{chart}):
\begin{enumerate}
\item Initialize the mean ionization state to the initial values (given by the
  previous time step or the initial conditions).
\item Find the column density between the source and the cell
\item Iterate until convergence in the neutral fraction(s):
\begin{itemize}
\item[-] Calculate time-averaged column density of the cell
  [Eq.~(\ref{meantau})].
\item[-] Calculate the photoionization rate $\Gamma$
  [Eq.~(\ref{photoncons_gamma})]. 
\item[-] Calculate the mean electron number density, $\langle n_e\rangle$,
  based on the current mean ionization state.
\item[-] Calculate new and mean ionization states
  [Eqs.~(\ref{ionization})-(\ref{mean_x})].
\item[-] Check for convergence.
\end{itemize}
\end{enumerate}

\begin{figure}
  \includegraphics[scale=0.35]{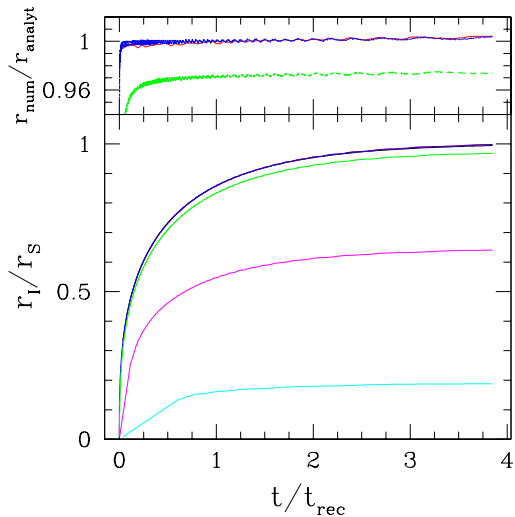}
\caption{Spherical I-front for point source in uniform gas: (a)(bottom panel) 
radius (in units of the Str\"omgren radius $r_S$) vs. time (in units of recombination
time $t_{\rm rec}$). 
Time-averaged photon-conserving photoionization rates from Eq.~(\ref{photoncons_gamma}) 
for $10^2$ time steps (top line) and the non-time-averaged (but otherwise also 
photon-conserving) rates from Eq.~(\ref{spatialgamma}) evaluated at the 
beginning of each time step, for $10^2, 10^3, 10^4$ and $10^5$ time steps (bottom to top 
lines). Note that the top curves (time-averaged and $10^5$ step non-time-averaged) 
are almost indistinguishable. (top panel) Ratios of the numerical solutions over the 
analytical one, $r_{\rm num}/r_{\rm analyt}$. Top lines, largely indistinguishable, 
are the time-averaged solution and non-time-averaged solution for $10^5$ time steps,
  bottom line is the non-time-averaged solution for $10^4$ time steps.
\label{photorates_fig2}}
\end{figure}

To illustrate the ability of our method to obtain the correct result with 
a small number of time steps, we consider the
problem of formation and evolution of an H~II region around a single source in
an uniform, initially-neutral density distribution, which has an exact analytical 
solution (this is Test~1 in \S~\ref{tests_sect} below, see there for details 
on the numerical parameters and the setup).  We use a one-dimensional grid of 
256 cells, gray opacities and uniform time steps. The optical depth per cell 
when it is fully-neutral is $\tau_{\rm cell}=11.5$. We solve this problem
two ways, using the time-averaged  photoionization rate in 
Eq.~(\ref{photoncons_gamma}) with 100 time steps, and using the 
non-time-averaged rates in Eq.~(\ref{spatialgamma}) evaluated at the 
beginning of each time step using $10^2, 10^3, 10^4$ and $10^5$ uniform time steps. 
Results are shown in Figure~\ref{photorates_fig2}. The time-averaged 
photoionization rates solution is essentially indistinguishable from the
analytical result, with errors much smaller than 1\%. To achieve a similar 
precision using the non-time-averaged photoionization rates we need up to
$\sim10^5$ time steps, or a factor of a 1000 more than in our method. The 
method with $\sim10^4$ time steps and no time-averaging still gives an 
acceptable, but clearly inferior solution, which is off by a few per cent 
from the analytical result. Using any smaller number of time steps and no 
time-averaging leads to a completely incorrect evolution, whereby the I-front 
initially propagates much more slowly than it should have, and producing a much 
smaller H~II region. It should be noted that ultimately even in those cases 
the correct Str\"omgren sphere is reached, but only at much later times.

The reason why the method without time-averaging of the optical depth would
need many more time steps is that it requires the time step, $\Delta t$, to be shorter (ideally much shorter) than the cell-crossing time, 
$t_{\rm cross}=\Delta x_{\rm cell}/v_I$, (i.e.\ that the I-front does not cross 
more than one cell per time step). When the I-front is very fast this condition
demands very short time steps. In particular problems, where the I-fronts
either propagate more slowly, or the fast-propagation phase is short-lasting, 
utilizing such short time steps could be feasible \citep[see e.g.][]{2004MNRAS.348..753S},
although still computationally-expensive. In more general situations, e.g.\ a
cosmological density field, I-fronts often propagate very fast at least somewhere 
in the computational domain. In such problems using time-averaging of the
optical depths is indispensable for this type of radiative transfer method. 

\section{Ray tracing}
\label{tracing_sect0}
The method described in \S~\ref{conserv_sect}
is one-dimensional, or in other words, along a ray. In order to use it in
three spatial dimensions, we need to trace rays across the computational
domain. Different ray-tracing methods exist and could be combined
with our method, as long as the rays are traversed in a causal order away
from the source. Here we use a method called 'short
characteristics', in which the rays for each point are constructed from
previously calculated neighboring points, instead of casting independent rays
to each cell ('long characteristics'). This is the approach used by
\citet{1999RMxAA..35..123R} and \citet{2003A&A...405..189L}, with minor
modifications. This method scales with the number of cells in
the computational mesh. We describe our method in detail in Appendix~\ref{tracing_sect}.

\begin{figure}
\begin{center}
  \includegraphics[scale=0.65]{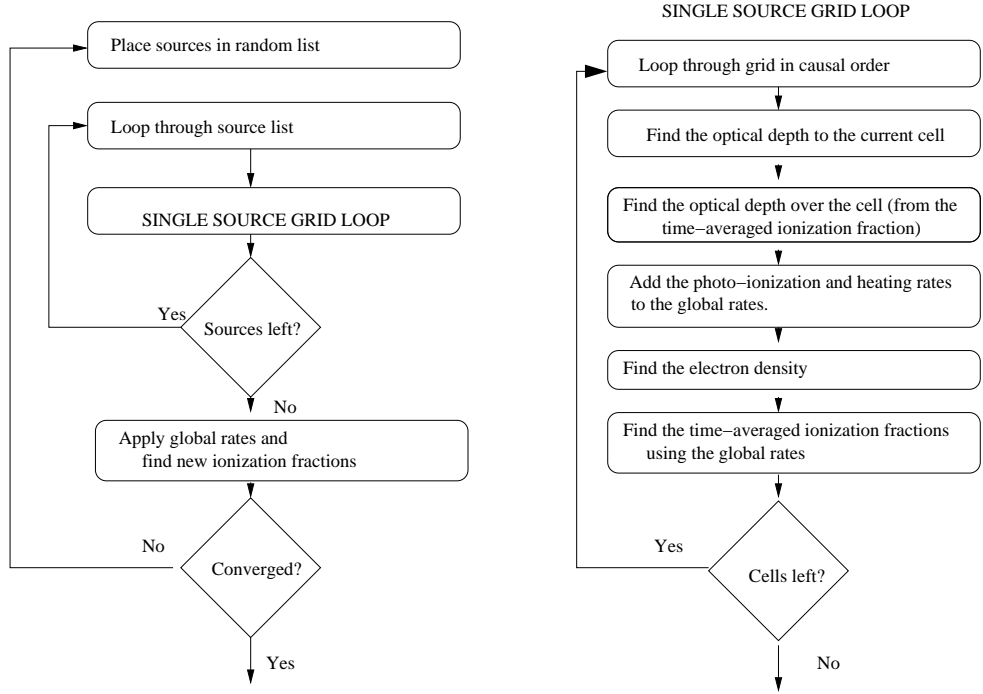}
\caption{Flow chart for the case of multiple sources.}
\label{multiple_sources_chart}
\end{center}
\end{figure}

\section{Treatment of Multiple Sources}
\label{multiple_sect}

When multiple sources of ionizing radiation are present, we start by
calculating the optical depths from each source outwards by casting
rays to each cell and using the method of short characteristics, as
described in Appendix~\ref{tracing_sect}. As we loop through the sources (in
random order) we add the contribution of each source to the total
photoionization rate, and we save the solution for the
(time-averaged) neutral fractions. The latter implicitly communicates
the presence of other sources to the current source. After having
treated all sources we apply the total photoionization rate to the
whole grid, this time without tracing any rays. This procedure is
repeated until the final neutral fraction satisfies our convergence
criterion. Note that in the multiple source case we do not test for
convergence when treating individual sources, but do so only at the
end, when we apply the total photoionization rates from all
sources. Figure~\ref{multiple_sources_chart} shows the above procedure
in the form of a flow chart.

Since ray-tracing is done for each individual source, the
method roughly scales with the number of sources times the typical
number of iterations needed for convergence. The latter number
depends on the density field and the amount of overlap between
the H~II regions, so it is hard to give a general number. Experience
shows it to be typically less than 10.

Source numbers, positions, luminosities, starting times and lifetimes could be
read in from a file or determined internally by the code.  Currently all
sources have the same spectrum, but within the framework of our method it is
easy to implement several types of sources (e.g.\  Pop.~II and III starbursts,
QSO's) by creating and using separate lookup tables for the photoionization
rates for each source type.

\section{Tests}
\label{tests_sect}

\subsection{Expansion of H~II regions about a single source}
\label{single_source_sect}
We first consider the simplest case of H~II region expansion, namely around 
a single source of ionizing radiation in a fixed density field. Consider a 
source emitting $\dot{N}_\gamma$ ionizing photons per unit time igniting in 
an initially-neutral medium consisting of pure hydrogen with number density 
$n_H$. The source propagates an I-front into the neutral gas. The I-front
propagation speed is determined by the balance at the I-front of the incoming 
flux of ionizing photons and the flux of neutral atoms flowing into the I-front 
and getting ionized. Assuming for simplicity that the I-front is ``sharp'', i.e.\ 
the width of the neutral-ionized transition region is small compared to the 
I-front radius, then this balance is expressed by the I-front jump
condition\footnote{It should be noted that strictly speaking the I-front jump
  condition in Eq.~(\ref{jump}) should be modified if the I-front is
  moving at relativistic speeds, see e.g.\ \citet{SIAS05}. The numerical
  radiative methods should also take special care not to allow
  superluminal motions. However, in most astrophysical problems this does not
  become a significant issue and thus for simplicity in this paper we limit
  ourselves to considering non-relativistic I-fronts.}:
\begin{equation} 
n_{H}\frac{{\rm d}r_I}{{\rm d}t} =F,
\label{jump}
\end{equation} 
where $n_H$ is the number density of hydrogen atoms and $F$ is the flux of
ionizing photons at the I-front. $F$ is equal to the photon output of the
source per unit time, $\dot{N}_\gamma$, minus the photons lost to recombinations
in the already ionized volume, $V_I= 4\pi r_I^3/3$:
\begin{equation} 
F=\frac{1}{4\pi r_I^2}\left(\dot{N}_{\gamma}-\int_{V_I}n_H^2
C\alpha_BdV\right),
\label{flux0}
\end{equation} 
where $C \equiv \langle n^2\rangle/\langle n\rangle^2$ is the volume-averaged
clumping factor of the gas (for a uniform gas $C=1$), which increases the
recombination rate of the ionized gas. Combining Eqs.~(\ref{jump}) and
(\ref{flux0}), we obtain
\begin{equation}
n_H\frac{dV_I}{{\rm d}t}={\dot{N}_\gamma-\int_{V_I}n_H^2 C\alpha_BdV}
\label{drdt_I_front}
\end{equation}
The width of an I-front is determined by the mean free path of the ionizing photons 
at the front. The assumption of ``sharp'' I-front is generally well-justified for 
soft ionizing spectra, but not necessarily for harder spectra, in which case the 
transition from neutral to ionized gas at the front could get fairly wide 
\citep[see][for further discussion and specific examples]{2004MNRAS.348..753S}. 

We performed four tests of single-source I-front propagation in several
density fields relevant to astrophysics, which are listed in Table~\ref{tests} 
and described in more detail below. In \S~\ref{1D_tests_sect} 
and \S~\ref{3D_tests_sect} we discuss the results in 1D spherical symmetry, and
3D, respectively, with the temperature kept fixed at $10^4$~K. For simplicity, 
we assume that the gas consists of pure hydrogen. While here we consider only the 
case of a pure hydrogen gas, our non-equilibrium chemistry solver can also easily 
accommodate multiple species \citep[see e.g.][]{1998A&A...331..335M,2004MNRAS.348..753S}. 
We also assume a gray opacity. In this limit the photoionization rate from Eq.~(\ref{photoncons_gamma})
becomes proportional to the number of ionizing photons 
produced by the ionizing source per unit time: 
\ba
\Gamma&\propto&\dot{N}_\gamma=\int_{\nu_{\rm th}}^\infty\frac{L_\nu}{h\nu}d\nu
\label{grey_rates}
\ea
In this simple setting there are exact analytical solutions for both the
I-front position and velocity for all density fields we consider in our tests.
To our knowledge, some of these solutions are derived here for the first time.

In all tests the computational domain size was chosen so that it is only slightly 
larger than the final H~II region size in order to keep the resolution roughly 
similar between the different tests. The source position for the 1D, spherical symmetry 
tests is at the origin, $r=0$, while for the 3D tests the source is in the center 
of the grid. We run each test four times in 1D and four times in 3D, all possible 
combinations of two different spatial resolutions (``coarse'' one with 16 cells, i.e.\ 
$\Delta x_{\rm cell}=x_{\rm box}/16$, and ``fine'' one with 128 cells in 1D, as well as the 
corresponding $32^3$ and $256^3$ cells in 3D), and two different choices of time steps, 
$\Delta t_{\rm coarse}=t_{\rm evol}/10$ and $\Delta t_{\rm fine}=t_{\rm evol}/100$, 
where $t_{\rm evol}$ is the total time for which we follow the I-front. For our 
coarse-resolution runs we chose these very low spatial and time resolutions in order 
to demonstrate the conservative properties of our method. There are many problems where 
such low resolutions are expected, e.g.\ during reionization when we try to follow the 
initial evolution of multiple H~II regions in a large simulation volume, so it is 
important to establish how reliable are the simulation results in such situations.
Our ``fine'' resolution runs also employ relatively modest resolutions in both time and 
space, which enables us to easily run the 1D and the 3D simulation results at equivalent 
resolutions in order to directly compare the results from the two. In all tests the 
parameters we picked are such that the individual cells start very optically-thick 
(see Table~\ref{tests}), so as to thoroughly test our code in such most demanding 
situations.

We define the I-front position to be at the point where the gas is 50\% neutral. 
Note that if the I-front width (which is typically 10-20 photon mean-free-paths, 
and is thus dependent on the type of ionizing spectrum and the spectrum hardening
ahead of the I-front) is unresolved (the front is ``sharp''), then the exact value 
of the neutral fraction used to define the I-front position is irrelevant. However, 
when the I-front structure is resolved (the front is ``thick''), then its 
position (but not its velocity) depends on the neutral fraction at which we define it. 
Therefore, in such cases the I-front position obtained numerically could be offset from 
the position given by the analytical solutions, which all assume ``sharp'' I-fronts for 
simplicity. Inside a cell the I-front position is found by linear interpolation. The 
numerical value for the I-front velocity, $v_I$, was obtained by simple 
finite-differencing of its position and the time, according to 
\be 
v_i(t_{i,ave})=\frac{r_i-r_{i-1}}{t_i-t_{i-1}},
\label{vel_finite_diff}
\ee 
where $t_{i,ave}=(t_i+t_{i-1})/2$. Whenever the time for the I-front to cross
a cell is larger than the time step, the average speed of the front over a
single time step is numerically poorly defined by
Eq.~(\ref{vel_finite_diff}). In that case, the speed was evaluated over a
larger interval of time, comparable or greater than the cell-crossing time.  

\begin{table}
\caption{Parameters of the test problems described in the
text. The box sizes listed are for the 3D tests. For the corresponding 1D 
tests we used a box size which was one-half of the values listed. $\tau_{0,cell,min}$
and $\tau_{0,cell,max}$ are the minimum and maximum optical depth per cell (when 
neutral) at the Lyman limit of hydrogen at the corresponding resolution 
(lr=low resolution run, hr=high-resolution run). 
\label{tests}}
\begin{tabular}{lllll}
\hline Test & 1 & 2 &3 &4\\ \hline 
$\dot{N}_{\gamma}$(s$^{-1}$)&$10^{54}$ &$10^{51}$ &$10^{51}$ & $10^{54}$\\ 
$n_{\rm HI}$ (cm$^{-3}$)&$1.87\times 10^{-4}$ &0.015$\left(\frac{5~\mathrm{
kpc}}{r}\right)$&$3.2\left(\frac{91.5~\mathrm{pc}}{r}\right)^2$&$1.87\times
10^{-4}(1+z)_{10}^3$\\ 
$C$ & 5 &1 &1 &5 \\ 
$t_{\rm evol}$ (Myr) & $500$ &15 &1 &$500$ \\
$x_{\rm box}$ (cm) &$5\times10^{24}$ &$1.4\times10^{22}$ & $6\times10^{21}$ &
$7\times10^{25}$ (comov.)\\
\hline $t_{\rm rec}$ (Myr)& $130$&$8.2\left(\frac{r}{5\,\rm kpc}\right)$
&$0.04\left(\frac{r}{91.5\,\rm pc}\right)^{2}$&
$130(1+z)_{10}^{-3}$\\ 
$r_S$ (kpc) & $563$ & $1.86$ & ... & ...\\
$t_{\rm evol}/t_{\rm rec}$& 3.85&$1.84\left(\frac{r}{5\,\rm
kpc}\right)^{-1}$&$26\left(\frac{r}{91.5\,\rm pc}\right)^{-2}$
&$3.85(1+z)_{10}^{3}$\\
$\tau_{0,cell,min}$,lr  &184 &91   & 34  &50\\
$\tau_{0,cell,min}$,hr &23  &11   & 4   &6\\
$\tau_{0,cell,max}$,lr  &184 &2200 & 3800&184\\
$\tau_{0,cell,max}$,hr &23  &2200 & 470 &23\\
\hline
\end{tabular}
\end{table}

Test 1 involves the simplest situation, an I-front propagating in a uniform,
constant gas density $n_H$ with a constant clumping factor $C$. There is a
well-known analytical solution for the expansion of such an I-front, which is
determined by two parameters, the Str\"omgren radius $r_{\rm S}$ and the
recombination time $t_{\rm rec}$ \citep[see e.g.][]{2003adu..book.....D}, which
are defined as
\begin{equation}
r_{\rm S}=\left[{3\dot{N}_\gamma\over 4\pi \alpha_B(T)C n_{\rm
  H}^2}\right]^{1/3}\,,
\end{equation}
and
\begin{equation}
t_{\rm rec}=\left[C\alpha_B(T) n_H\right]^{-1}\,.
\end{equation}
The analytical
expressions for the I-front position, $r_I$ and velocity, $\rm v_I$, as a 
function of time are then
\ba
r_I&=&r_{\rm S}\left[1-\exp(-t/t_{\rm rec})\right]^{1/3}\,.\\
\rm v_I&=&\frac{r_{\rm S}}{3t_{\rm rec}}\frac{\exp{(-t/t_{\rm rec})}}
{\left[1-\exp(-t/t_{\rm rec})\right]^{2/3}}\,,
\label{strom}
\ea  
i.e.\ the H~II region reaches a finite radius, $r_S$, and zero velocity at
$t\rightarrow\infty$ (in practice, after a few recombination times), at
which point the recombinations in the ionized volume balance the new photons 
arriving from the source. 
In physical units we choose typical values for cosmological I-fronts
propagating during reionization, with the gas density equal to the mean
density of the universe at redshift $z=9$ and a source with ionizing photon
production rate $\dot{N}_{ph}=10^{54}\,\rm s^{-1}$ (Table~\ref{tests}).

In Test 2 we study the propagation of an I-front from a source in the center
of a singular, decreasing density profile $n_H=n_0(r_0/r)$, where
$n_0=0.015\,\rm cm^{-3}$ is the gas number density at the characteristic
radius $r_0=5\,\rm kpc$. This test is related to the problem of an I-front
propagating outward from a source in the center of a galactic halo with a Navarro, Frenk
\& White profile \citep{1997ApJ...490..493N} (assuming the gas follows the dark 
matter density profile). 
For this density profile Eq.~(\ref{drdt_I_front}) reduces to 
\be 
\frac{{\rm d}r_I}{{\rm d}t}=\frac{L}{r_I}-K,
\label{inv1}
\ee 
where we defined $L\equiv{\dot{N}_\gamma}/{(4\pi n_0r_0)}$,
$K\equiv{n_0r_0C\alpha_B}=r_0/t_{\rm rec,0}$, where $t_{\rm
  rec,0}\equiv(n_0C\alpha_B)^{-1}$ is the recombination time at the
characteristic density $n_0$. Eq.~(\ref{inv1}) has an analytical
solution, which for initial condition $r(0)=0$ is given by
\begin{equation} 
r_I(t)=r_S\left\{1+{\rm LambertW}\left[-\exp\left(-\frac{r_0t}{r
_St_{\rm rec,0}}-1\right)\right]\right\},
\label{test2_soln}
\end{equation}
where $r_S=L/K$ is the Str\"omgren radius for this test and ${\rm LambertW}(x)$
is the solution of the algebraic equation $y(x)e^{y(x)}=x$, which can be
calculated e.g.\ using readily available public software.

In Test 3 we follow the propagation of an I-front in a density profile
$n_H=n_0(r_0/r)^{2}$, with a flat core of gas number density $n_0$ and radius
$r_0$. This density profile is steeper than the one we consider in Test 2, and
the H~II region evolution is qualitatively different, as we show below 
\citep[see also][for detailed discussion of I-front propagation in power-law 
density profiles]{1990ApJ...349..126F,SIAS05}. This
test, with the dimensional parameters we have chosen, $n_0=3.2\,\rm cm^{-3}$
and $r_0=91.5\,\rm pc$, (Table~\ref{tests}) resembles the problem of
inside-out ionization of a dwarf galaxy formed at redshift $z=9$ by
an ionizing source at its center. In this case Eq.~(\ref{drdt_I_front})
reduces to
\begin{equation} 
\frac{{\rm d}r_I}{{\rm d}t}=L+\frac{K}{r_I},
\label{inv2}
\end{equation} 
where we defined
\begin{equation} 
L=v_{\rm lim}\equiv\frac{\dot{N}_\gamma}{4\pi n_0r_0^2}-\frac43n_0r_0C\alpha_B, 
\end{equation} 
which physically is the terminal velocity $v_{\rm lim}$ of the I-front for
$r\rightarrow\infty$, and $K\equiv n_0r_0^2C\alpha_B=r_0^2/t_{\rm rec,core}$.
Assuming that $\dot{N}_\gamma>4\pi r_0^3n_0^2C\alpha_B/3$ i.e.\ source is
strong enough to ionize more than just the core,
it is clear that $v_I>0$ for all radii and thus the I-front will never stop, 
eventually reaching the constant terminal velocity $v_{\rm lim}$. 
Equation~(\ref{inv2}) for arbitrary values of the parameters $K$ and $L$ has 
a complex analytical solution for the initial
condition $r(0)>r_0$. However, one particularly simple solution is obtained
when $L=0$, i.e.\ $\dot{N}_\gamma=16\pi r_0^3n_0^2C\alpha_B/3$, and $r(0)=r_0$,
in which case the radius of the I-front is given by
\begin{equation}
r_I=r_0(1+2t/t_{\rm rec,core})^{1/2}.
\end{equation}
This is the case we will use in our Test 3. In calculating the column
densities for this test we use the weightings in Eq.~(\ref{weightings})
with $\tau_0=\epsilon>0$, where $\epsilon\ll1$.

Finally, our Test 4 is the same as Test 1, but for a cosmological I-front
propagating in a uniform-density medium with a density equal to the
time-evolving mean background density of the universe\footnote{The
  cosmological parameters we use for the cosmological tests in this paper
are $H_0=70\,\rm km\,s^{-1}Mpc^{-1}$, baryon density $\Omega_b=0.043$ and
$\Omega_0=1-\Omega_\Lambda=0.27$.}. The ionizing source is
switched on at redshift $z=9$. The I-front evolution of cosmological I-front
in an IGM with mean volume-averaged clumping factor $C$ has an exact analytical
solution \citep{1987ApJ...321L.107S}, 
given by 
\be
y(x)={\lambda}e^{\lambda/x} \left[x Ei(2,{\lambda}/{x})-Ei(2,\lambda)\right].
\label{y_exact}
\ee 
Here $y\equiv[r_{I,c}(t)/r_{S,i}]^3$, $r_{I,c}=r_I(1+z)$ is the comoving
radius of the I-front at time $t'=t_i+t$,
$r_{S,i}=[3\dot{N}_\gamma/(C\alpha_B(n_{H,c})^2)]^{1/3}$ is the initial
Str\"omgren radius, $n_{\rm H,c}$ is the mean comoving density of hydrogen
(defined at epoch of source turn-on, i.e.\ the scale factor is normalized
$a_i=1$),
\be 
\lambda\equiv \frac{t_i}{t_{\rm rec,i}}= t_i C\alpha_B n_{H,c}, 
\ee 
is the ratio of recombination time at present to the age of the
universe when the source turned on,
and $Ei(2,x)\equiv\int_1^\infty \frac{e^{-xt}}{t^2}{\rm d}t$ is the
Exponential integral of second order. Note that the exponent in the analytical
solution in Eq.~(\ref{y_exact}), $e^{\lambda t_i/t}$, is generally very
large, which could easily lead to  numerical overflow problems. There is a similar
exponent, but with negative sign, in the Exponential integrals and it is
therefore better to numerically evaluate them together, as $e^{\lambda
  t_i/t}Ei(2,\eta\frac{t_i}{t})$, rather than separately.

\subsubsection{1D Ionization Fronts}
\label{1D_tests_sect}

\begin{figure}
  \includegraphics[scale=0.35]{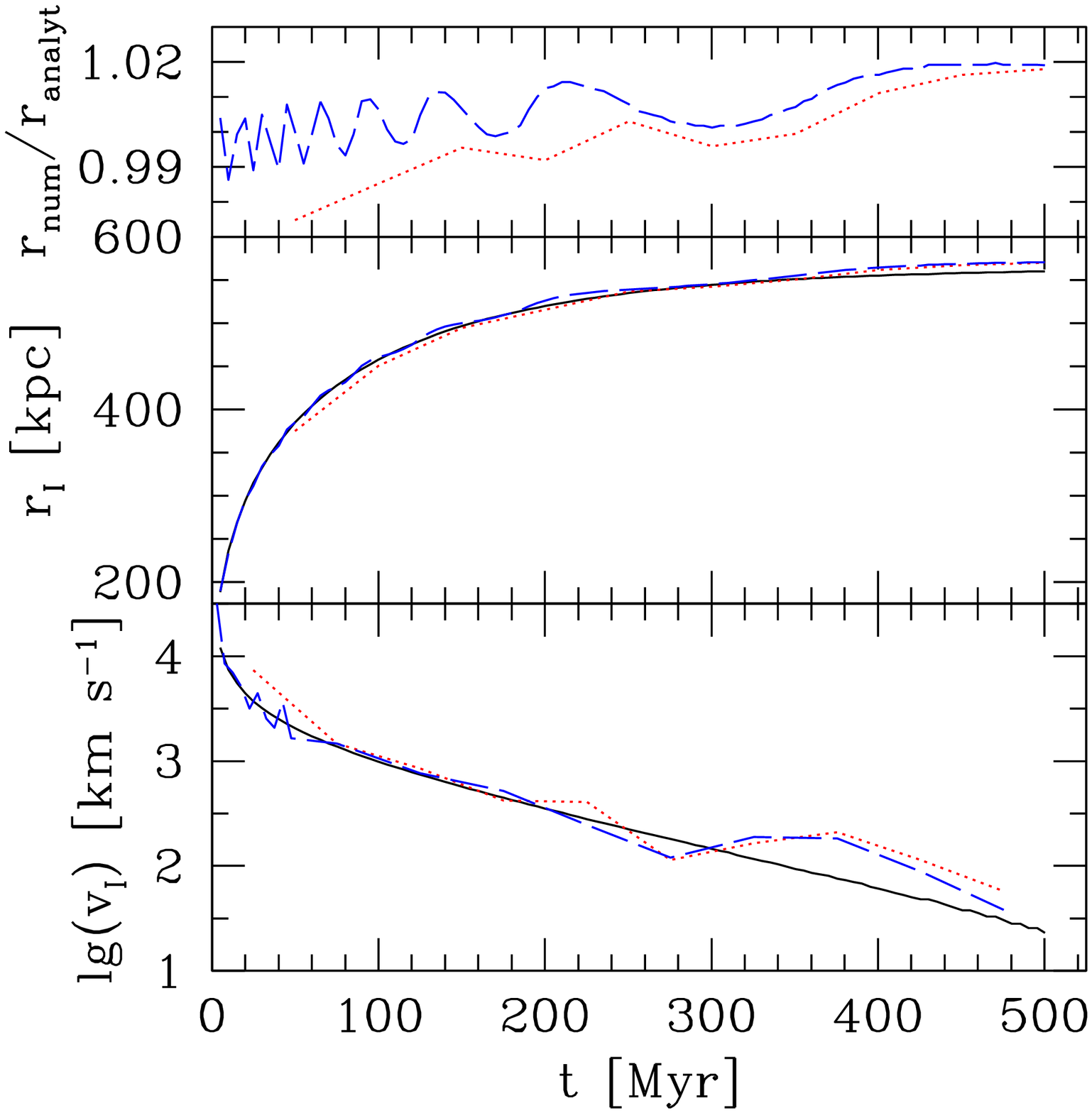}
  \includegraphics[scale=0.35]{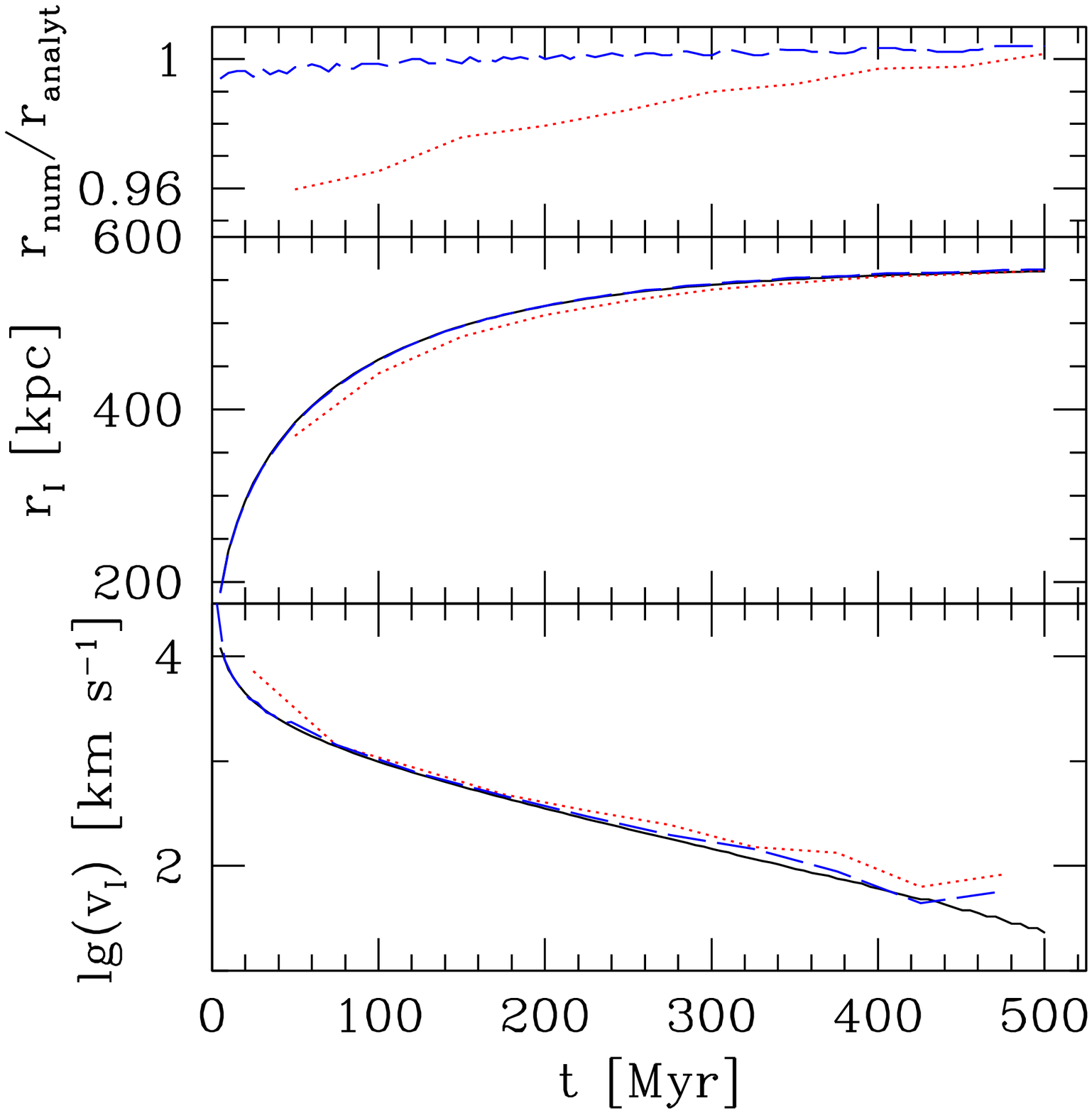}
\caption{Ionization front evolution for Test 1 (H~II region expansion
  in an uniform gas) in 1D spherical symmetry, 
  using coarse grid (left panels) and fine grid (right panels). Shown are the
  analytical solution (solid, black), and the numerical solutions for $\Delta
  t_{\rm coarse}$ (dotted, red) and $\Delta t_{\rm fine}$ (dashed,
  blue) for the I-front position (middle panels), ratio of the numerical to
  the exact analytical position (top panels) and the I-front velocity (bottom
  panels).
\label{test1_1D_fig}}
\end{figure}

Figure~\ref{test1_1D_fig} shows our results for Test 1 (constant density).
The numerical results match the exact analytical solution quite well, even at
very low spatial and temporal resolutions. The I-front position is never
off by more than a fraction of a cell size at the corresponding spatial
resolution. The only exception is the low temporal, but high spatial resolution 
case where the H~II region size is initially underestimated by a few per 
cent (or $\sim1.25$ cell-sizes). However, even in this case the agreement quickly 
improves and the final Str\"omgren radius is excellently reproduced. The numerical
I-front velocity is also in excellent agreement with the analytical solution,
regardless of the time- or spatial resolution, although the coarse spatial
resolution runs tend to overestimate the I-front speed  slightly when the I-front
slows down below $v_I\approx 100\,\rm km\,s^{-1}$. 
This could be expected, since the I-front cell-crossing time for this velocity is 
$t_{\rm cross}\approx 500$ Myr, i.e.\ the I-front has essentially stalled and 
remains inside a single cell, making the velocity estimate unreliable. This 
is confirmed by our higher spacial resolution runs shown in Figure~\ref{test1_1D_fig} 
(right) in which case the cell-crossing time for the same I-front velocity
is 8 times shorter and hence the velocity is correctly reproduced down to
few tens of $\rm km\,s^{-1}$. 
\begin{figure}
  \includegraphics[scale=0.35]{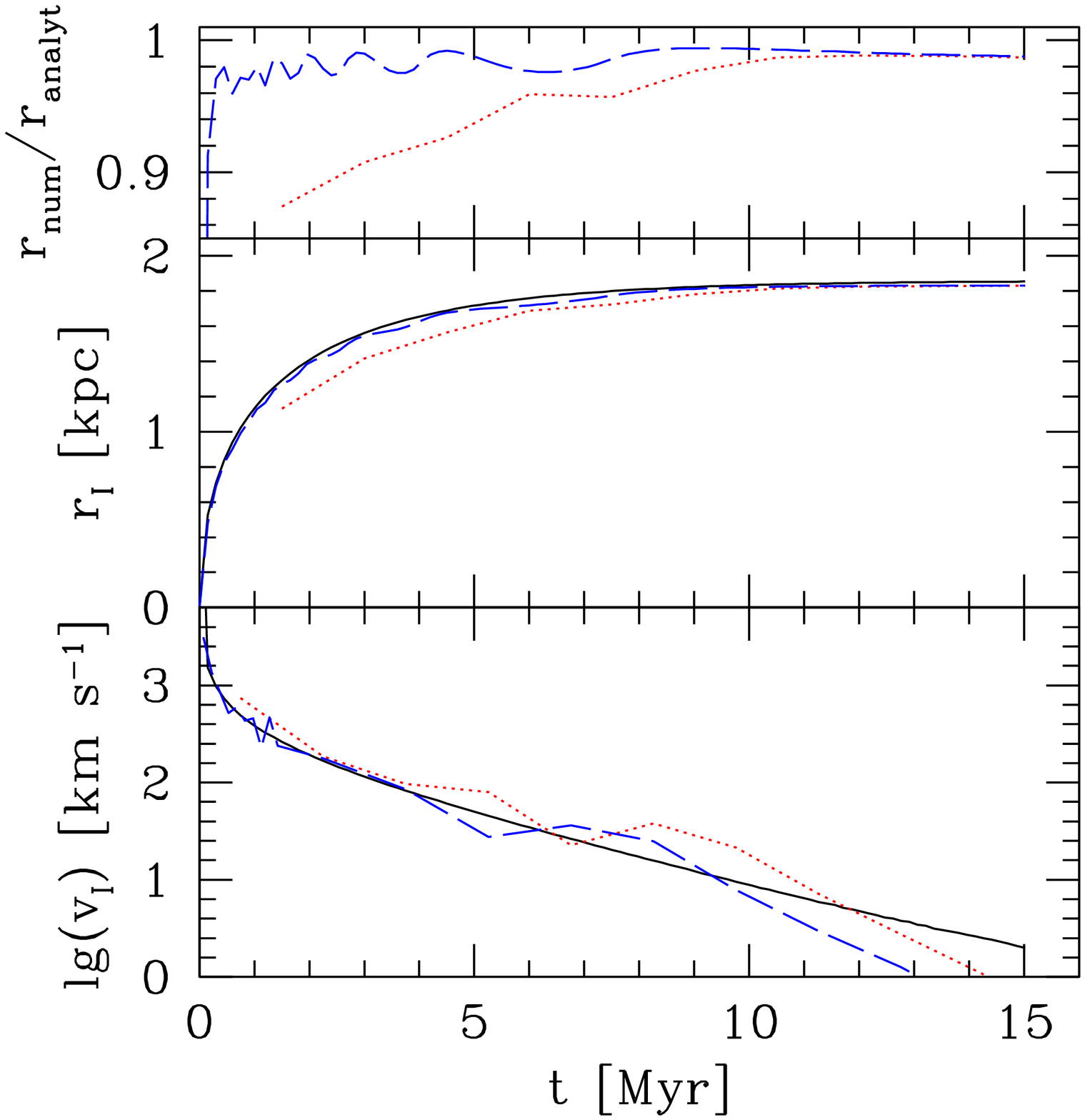}
  \includegraphics[scale=0.35]{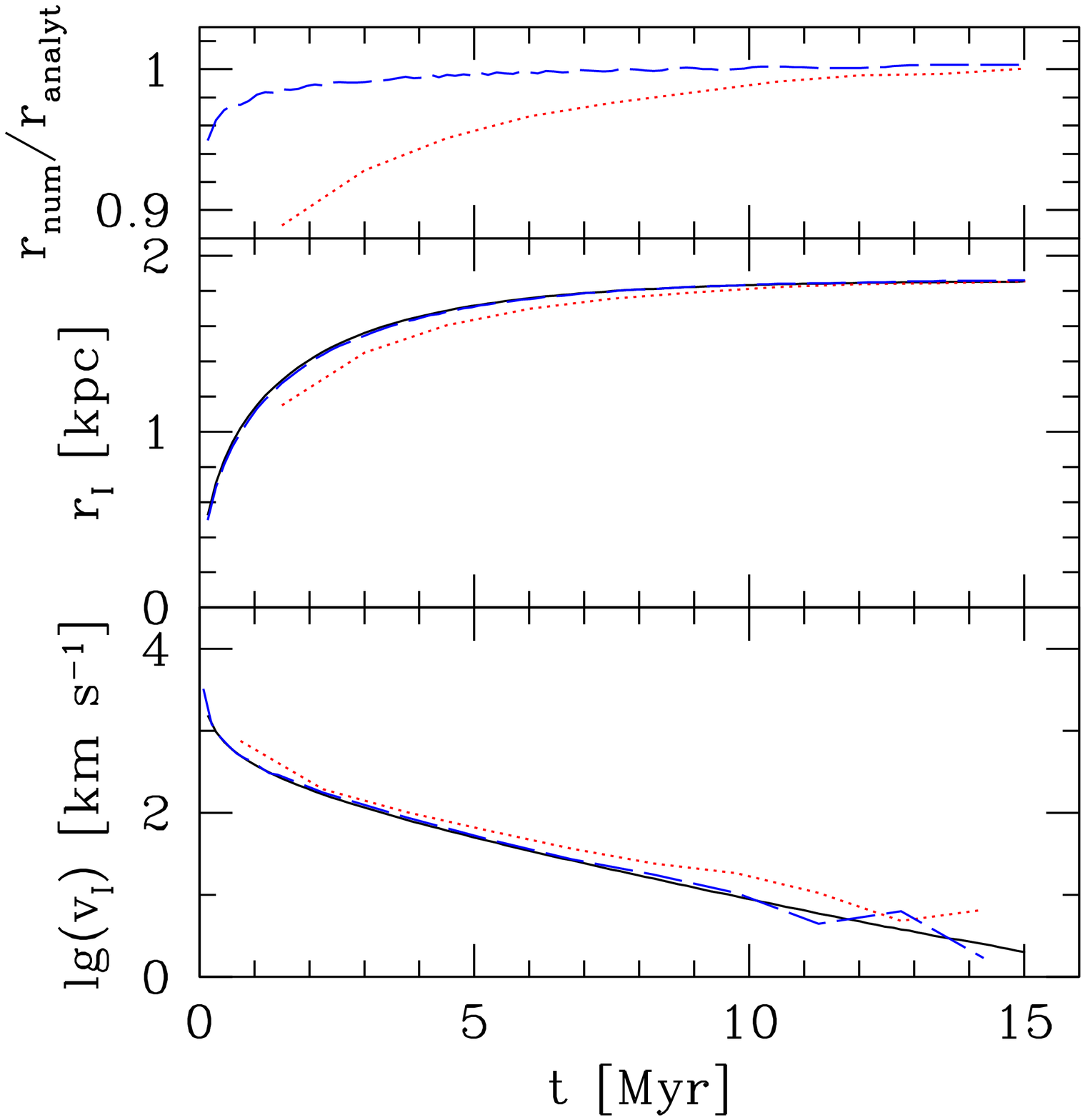}
\caption{I-front evolution for Test 2 (H~II region expansion
  in $r^{-1}$ density profile) in 1D spherical symmetry. 
  Same notation as in Fig.~\ref{test1_1D_fig}.
\label{test2_1D_fig}}
\end{figure}
\begin{figure}
  \includegraphics[scale=0.35]{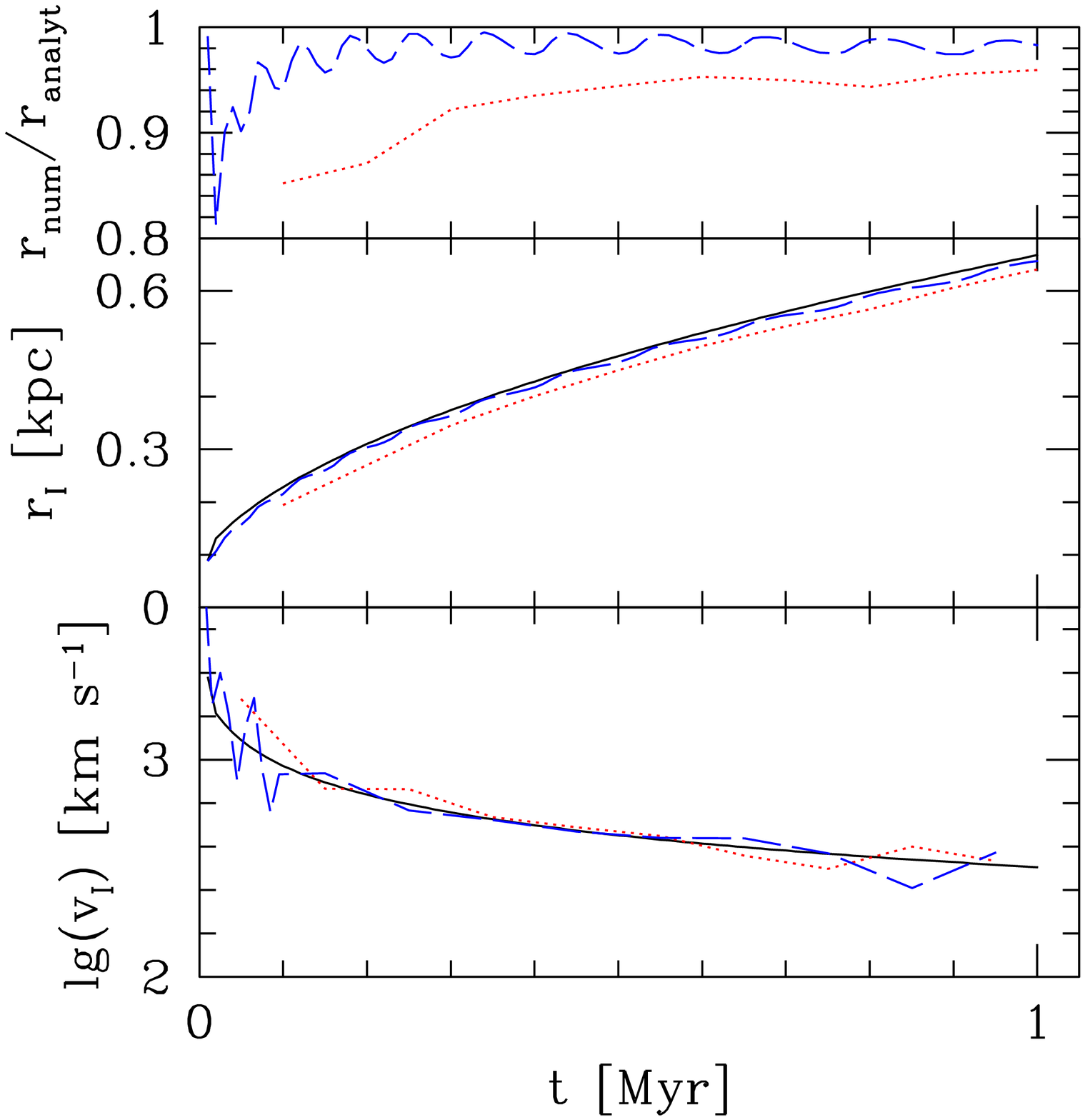}
  \includegraphics[scale=0.35]{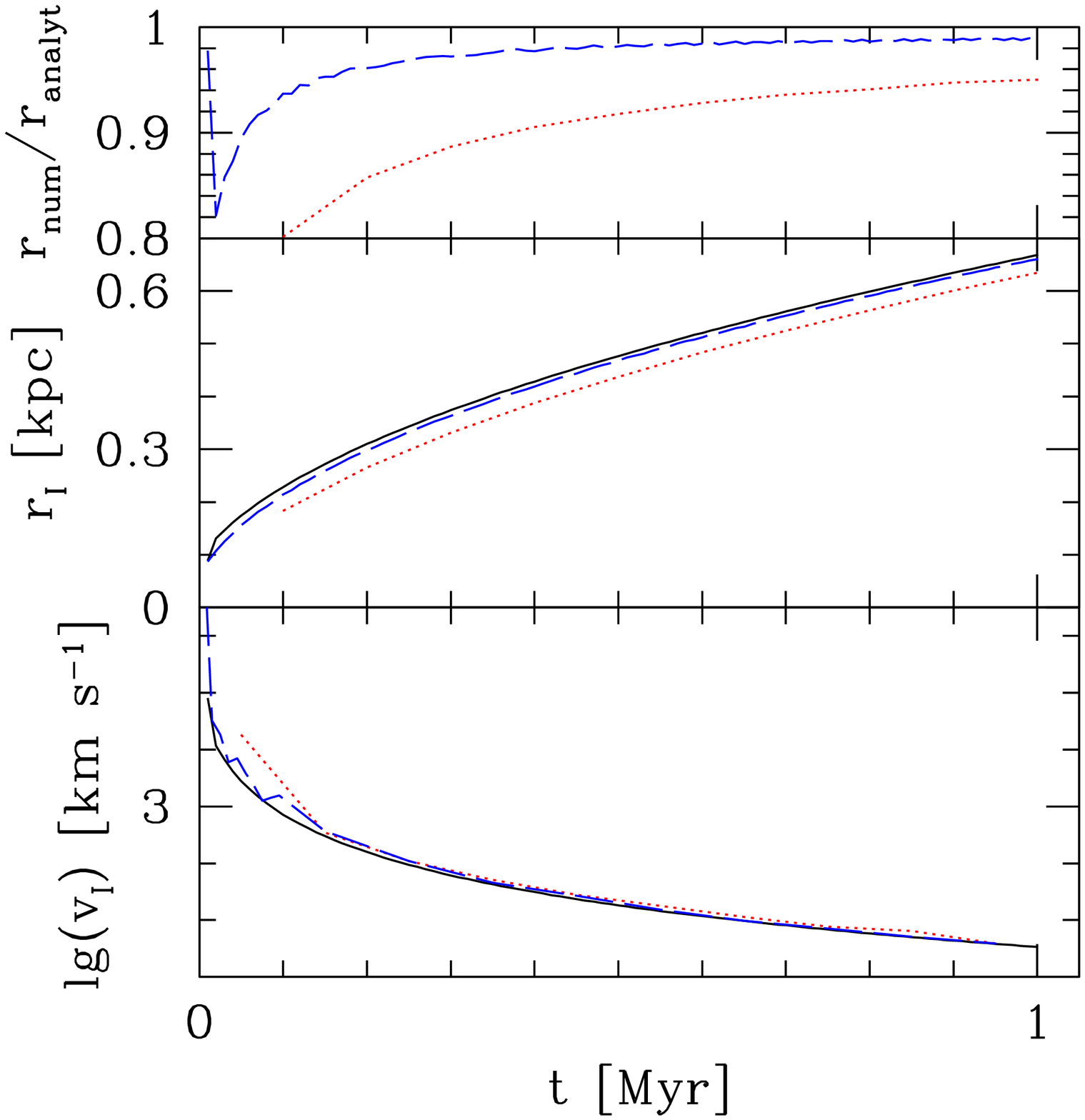}
\caption{I-front evolution for Test 3 (H~II region expansion
  in $r^{-2}$ density profile) in 1D spherical symmetry. 
  Same notation as in Fig.~\ref{test1_1D_fig}.
\label{test3_1D_fig}}
\end{figure}

In summary, for this test, the good temporal resolution is more important
in achieving good agreement with the analytical solution than is high spatial
resolution, although the results from our numerical method are quite acceptable 
in all cases.

The reason for this behavior is that for time steps of order of the 
recombination time, our method overestimates the number of recombinations. 
In this test problem this happens when, within one time step, the H~II region almost 
reaches its Str\"omgren sphere and the 
I-front essentially stalls. In this case, some cells near the front remain
neutral for a large part of the time step. Using the time-averaged electron
density in this case overestimates the number of recombinations. A better
result is obtained if the time step, $\Delta t$ is significantly shorter than the 
recombination time, $t_{\rm rec}$, while here 
$\Delta t_{\rm coarse}=50\,\rm Myr=0.4t_{\rm rec}$.

\begin{figure}
  \includegraphics[scale=0.35]{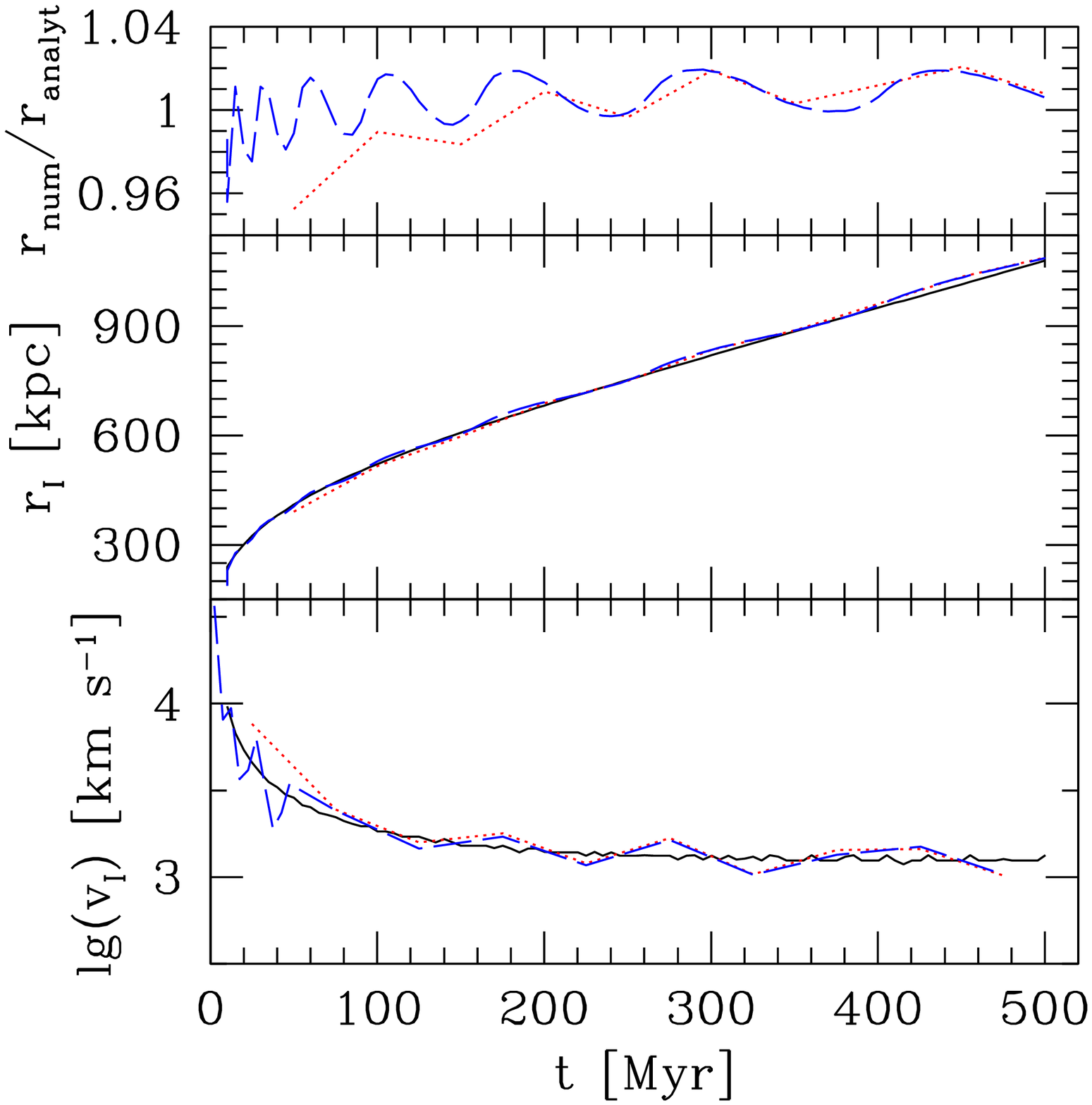}
  \includegraphics[scale=0.35]{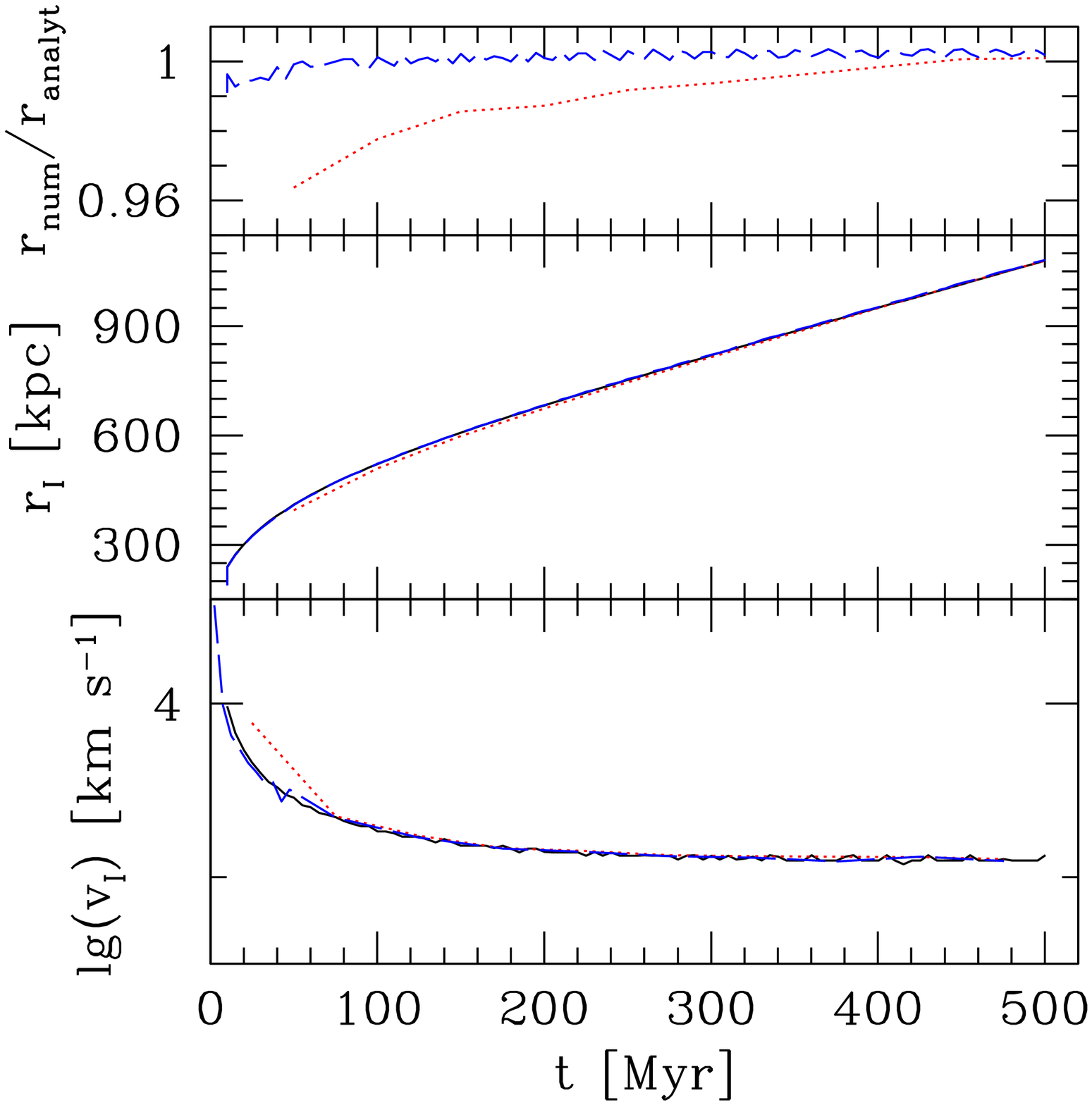}
\caption{I-front evolution for Test 4 (cosmological H~II region expansion) 
  in 1D spherical symmetry. Same notation as in Fig.~\ref{test1_1D_fig}.
\label{test4_1D_fig}}
\end{figure}
Our results for Test 2 ($1/r$ density) are shown in Figure~\ref{test2_1D_fig}. In this case the
density profile is singular, and hence the evolution is dominated by recombinations 
from the start, rather than just at late times as was in Test 1. The H~II region 
reaches the correct Str\"omgren radius, with errors $\sim1\%$ in all cases 
regardless of the resolution, but once again the early evolution of the H~II 
region is much better followed (generally within 1-2\% or better) in the runs 
with higher temporal resolution.  
However, even for very poor time resolution the I-front position is correct 
within $\sim10\%$, and to better than 4\% after the first few time steps. The 
reason for this small initial discrepancy is that the recombination time in the 
inner, high-density cells is very short, shorter than (or at most of order) 
$\Delta t_{\rm coarse}=1.5$ Myr, and hence the numerical I-front initially propagates 
slightly slower than it should. Nonetheless, thereafter the I-front velocity is 
correctly reproduced in all cases up to the Str\"omgren sphere.

\begin{figure}
  \includegraphics[scale=0.35]{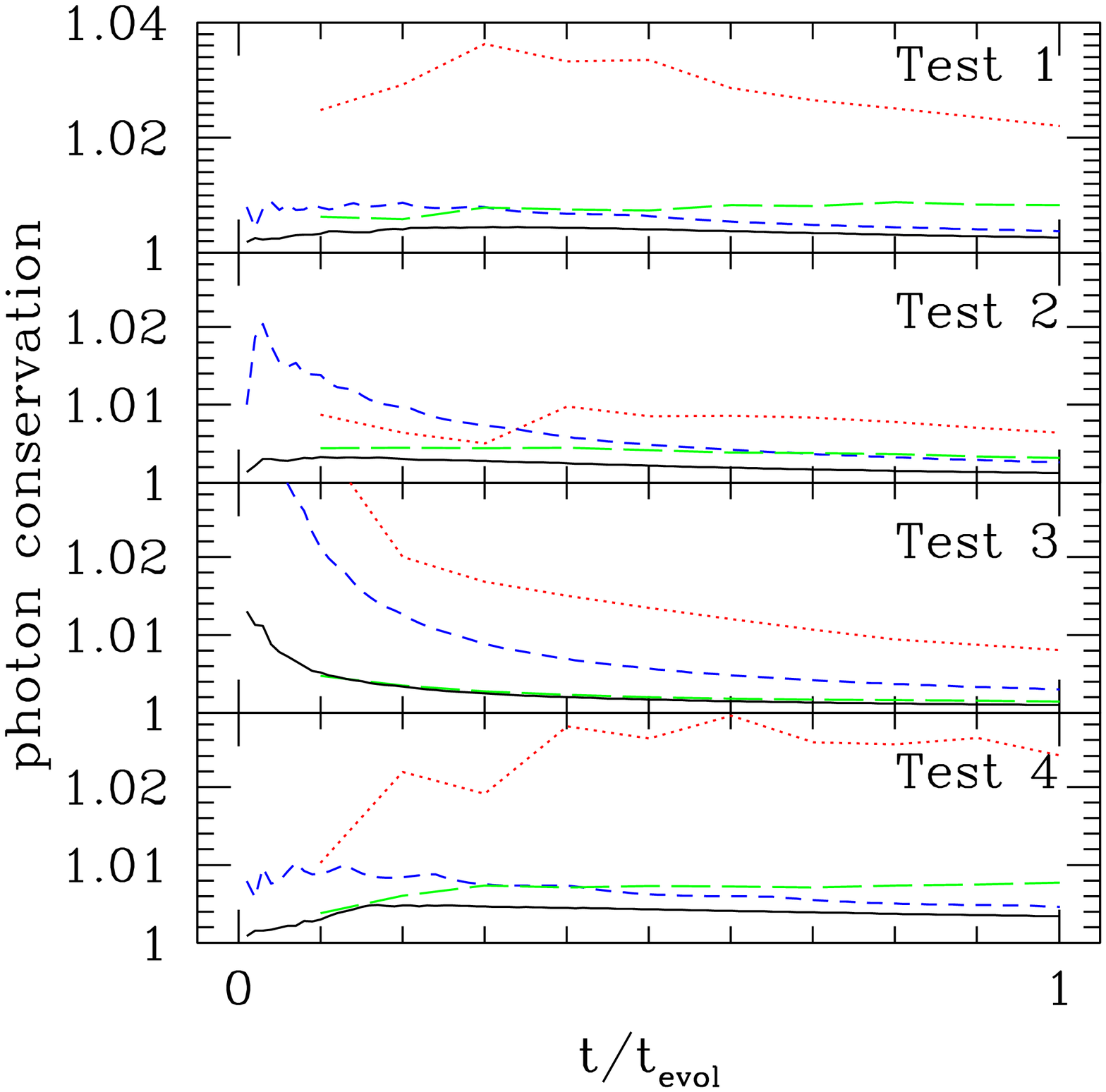}
  \includegraphics[scale=0.35]{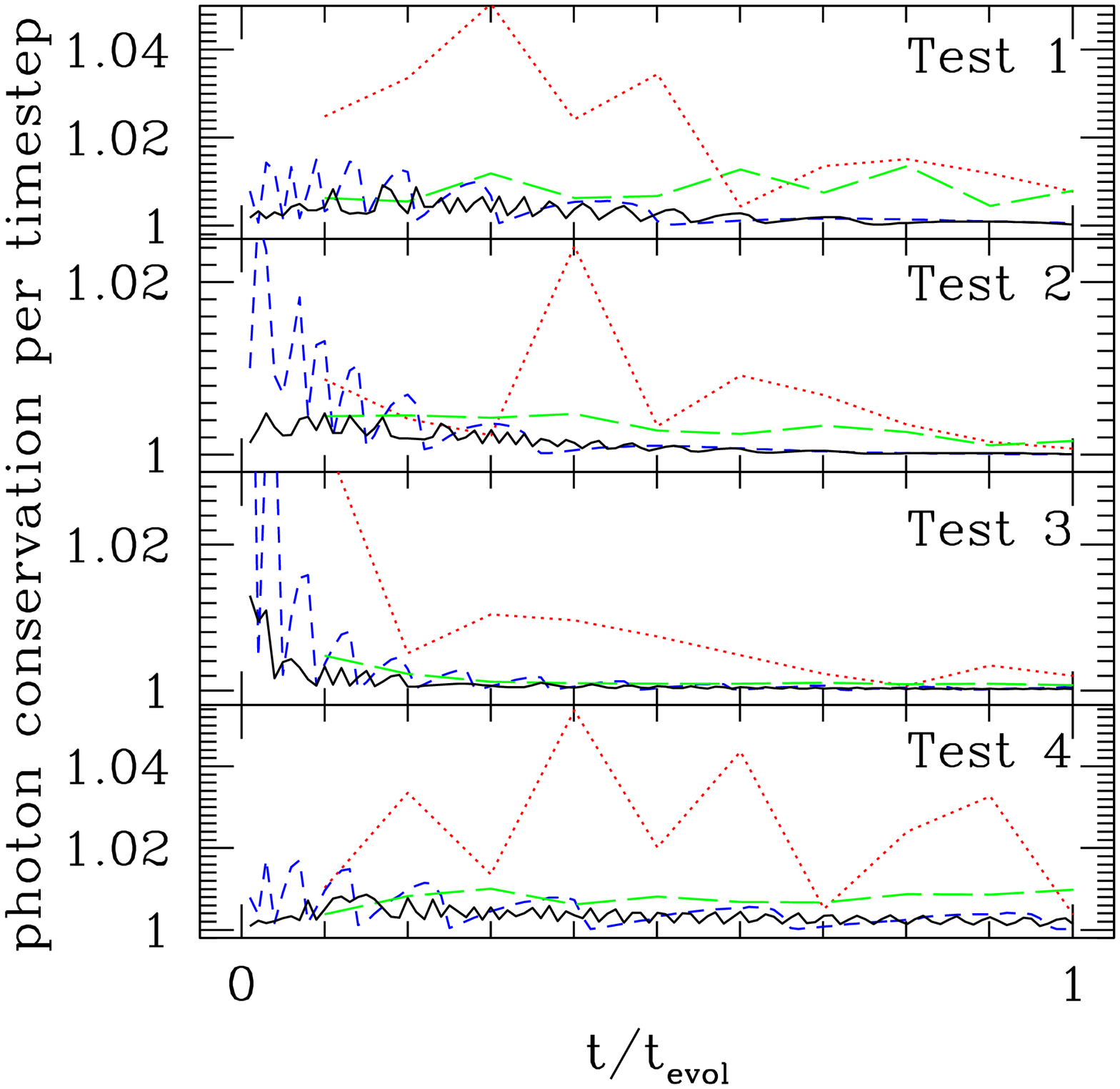}
\caption{Photon conservation for Tests 1-4 in 1D. Cases plotted are: 16 cells, 
  10 time steps (red, dotted), 16 cells, 100 time steps (blue, short-dashed),
  128 cells, 10 time steps (green, long-dashed), 128 cells, 100 time steps
  (black, solid).
\label{conserv_1D_fig}}
\end{figure}
Test 3 ($1/r^2$ density) proved most difficult for our numerical
method to follow at low temporal resolutions (Figure~\ref{test3_1D_fig}). The H~II region is smaller than
the exact, analytical one by 5-10\% percent (and up to 20\% at the very
beginning). Within one such relatively long time step ($\Delta t_{\rm coarse}=0.1$ Myr)
the I-front crosses the density profile core and advances significantly down the
steep density gradient. The recombination time in the core is 
$t_{\rm rec,core}=0.04$~Myr and $\Delta t_{\rm coarse}=2.5t_{\rm rec,core}$. Our 
time-averaging procedure slightly overestimates 
the mean optical depth over the time step in such conditions. Interestingly, the
I-front exact analytical velocity is nevertheless perfectly reproduced. Thus for
coarse time steps the I-front propagates at the correct speed, but with slightly 
offset position. In the high temporal resolution cases the analytical I-front 
radius is matched to better than a few percent after the first few steps, regardless
of the spatial resolution employed. 

Test 4 (expanding universe) resulted in excellent agreement between our 
numerical solution and the exact one, for all spatial and temporal
resolutions (Figure~\ref{test4_1D_fig}). This was in fact expected, since in this case the recombination time
is always longer than even our coarse time step, i.e.\ 
$t_{\rm rec}/\Delta t_{\rm coarse}=2.6/(1+z)_{10}^3$, especially at later
times as the background expands with the Hubble flow. We note that, as a
consequence of this continuous decrease of the gas density, the Str\"omgren
sphere is never reached and the I-front velocity remains very fast at all times, 
always above $10^3\rm km\,s^{-1}$, i.e.\ the I-front is of fast, R-type. This is a 
generic property of global cosmological I-fronts, as was first shown by 
\citet{1987ApJ...321L.107S}.

\begin{figure}
  \includegraphics[scale=0.35]{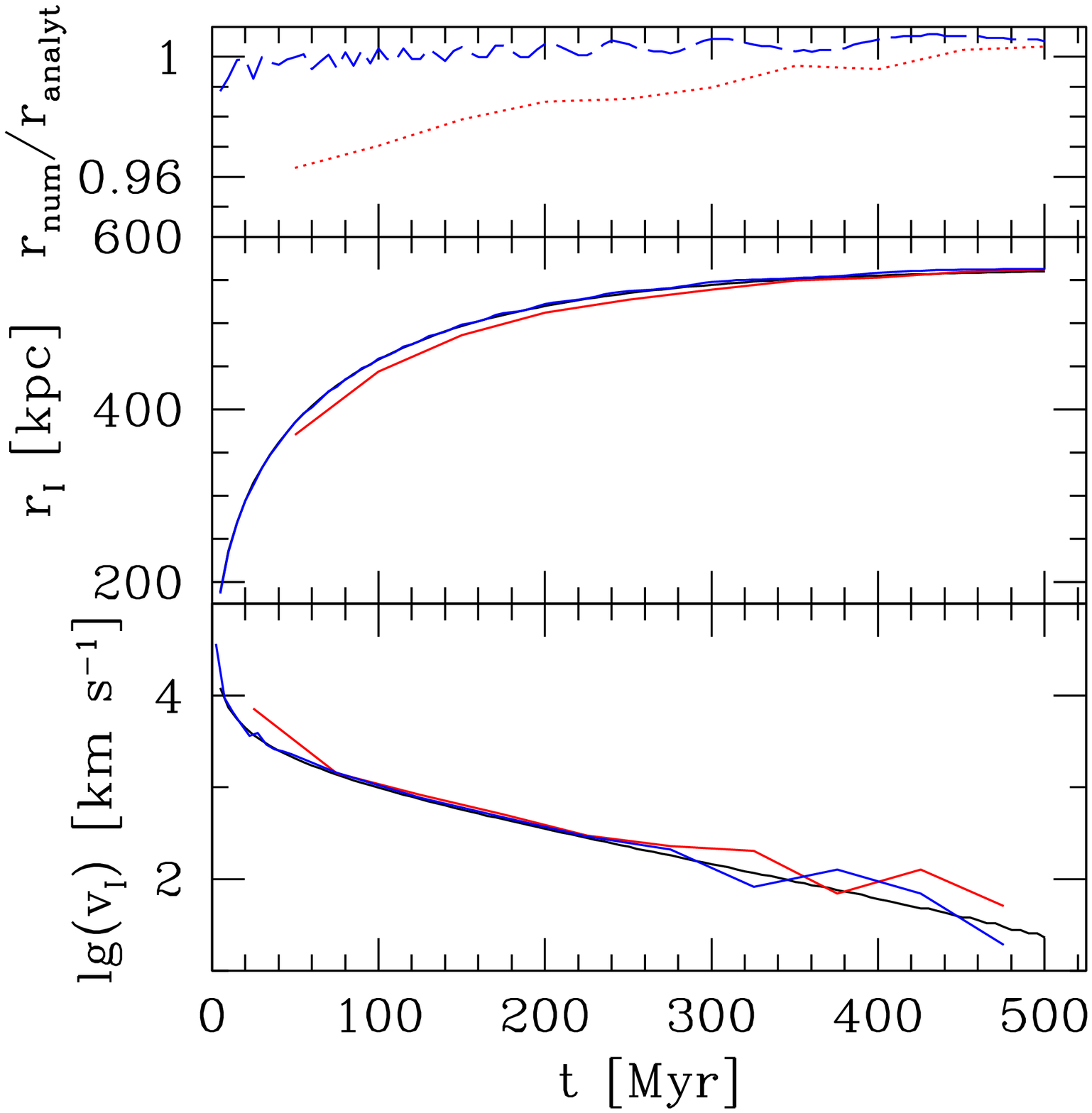}
  \includegraphics[scale=0.35]{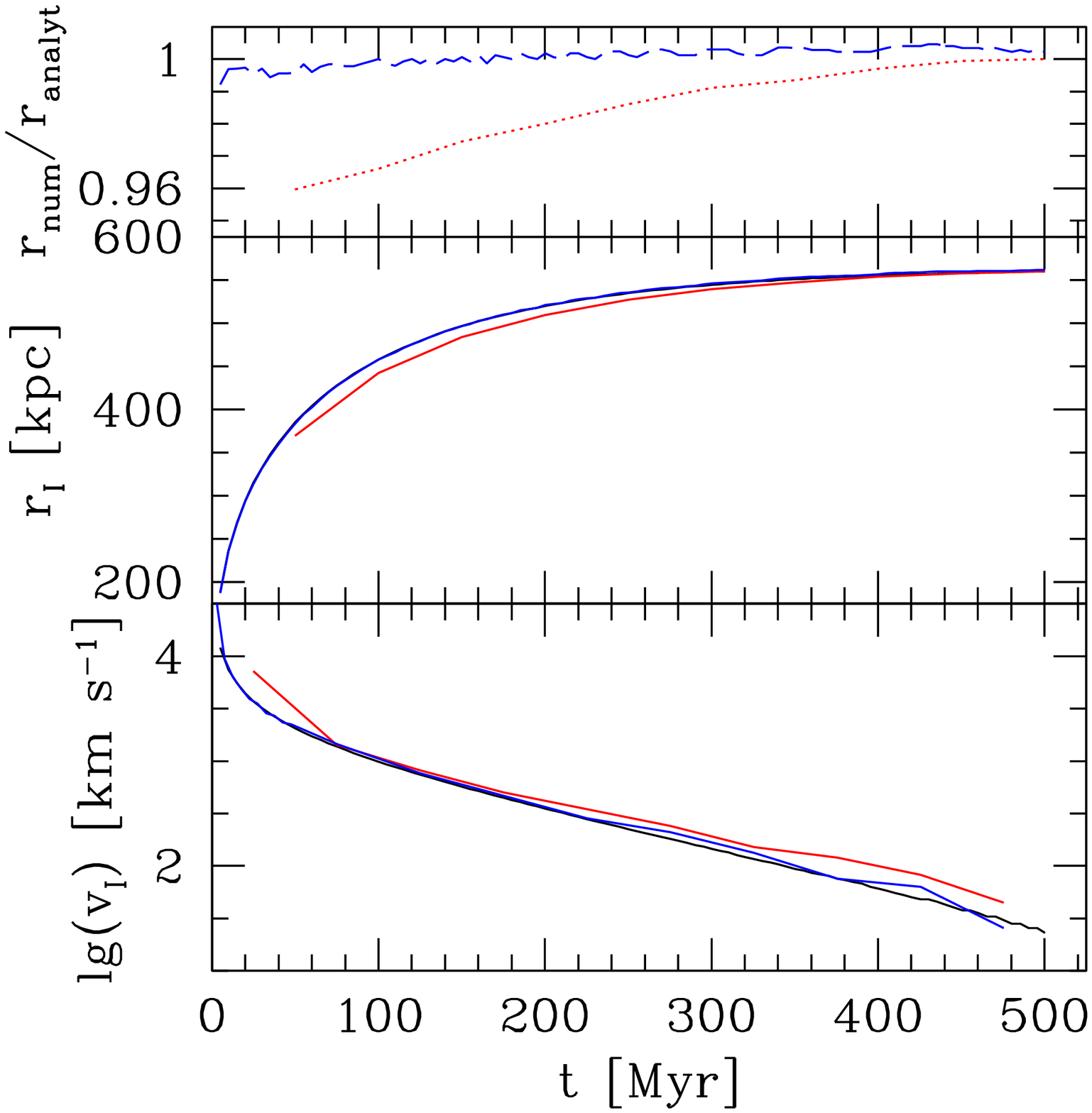}
\caption{I-front evolution for Test 1 (H~II region expansion
  in an uniform gas) in 3D, using $32^3$ cells (left panels) and 
  $256^3$ cells (right panels), otherwise same notation as in
  Fig.~\ref{test1_1D_fig}.
\label{test1_3D_fig}}
\end{figure}

\begin{figure}
  \includegraphics[scale=0.35]{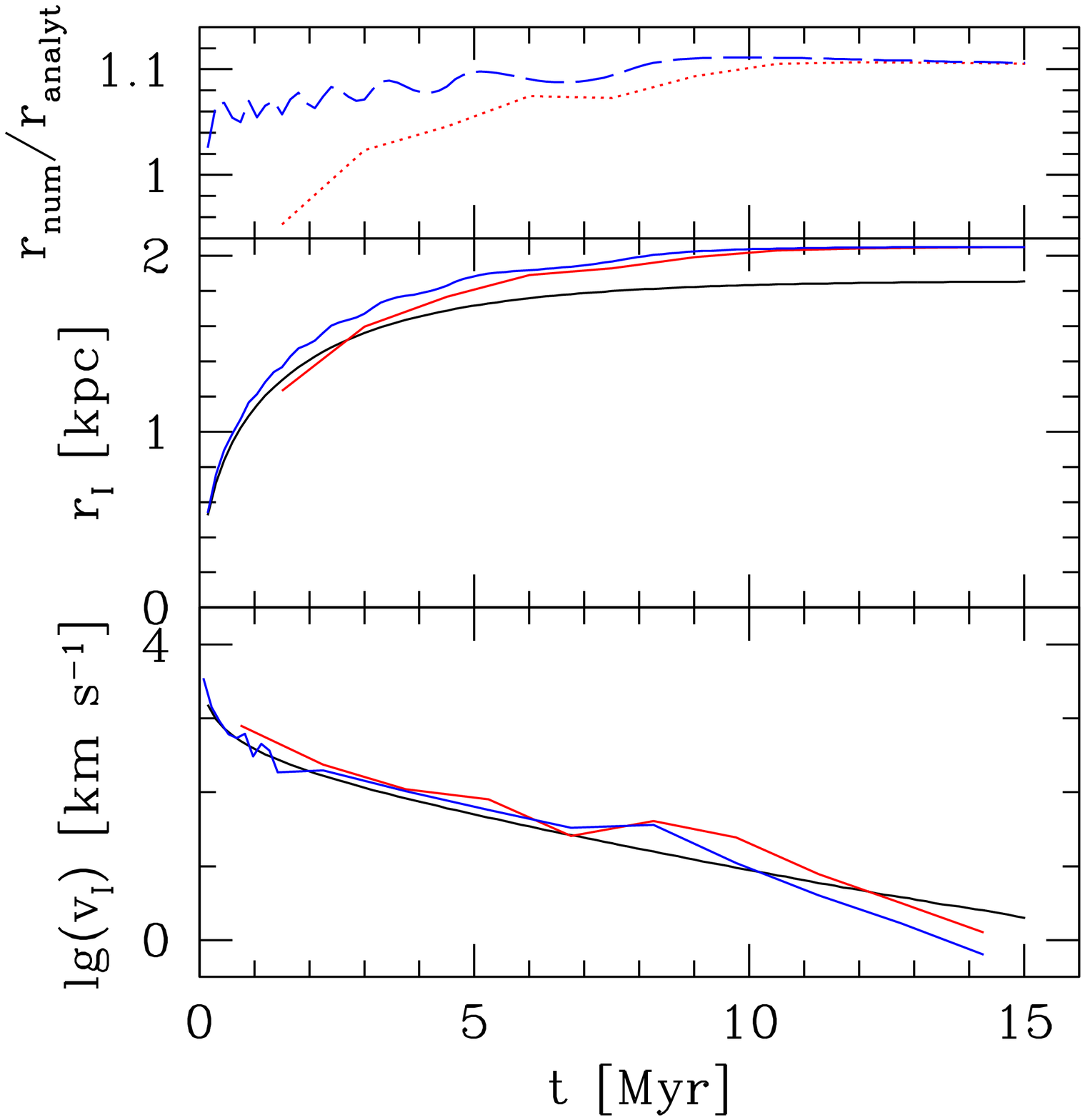}
  \includegraphics[scale=0.35]{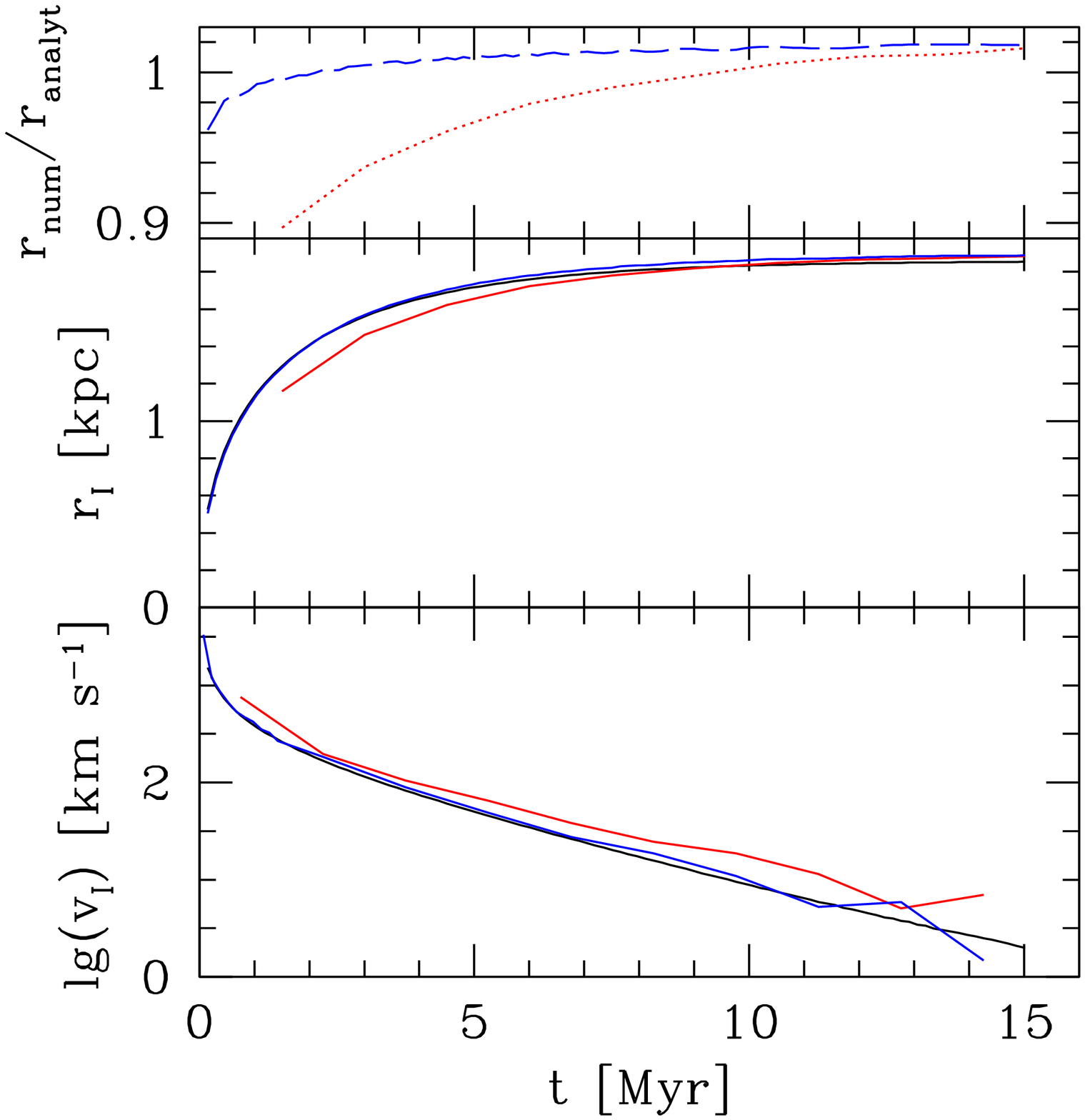}
\caption{I-front evolution for Test 2 (H~II region expansion
  in $r^{-1}$ density profile) in 3D. Same notation as 
  in Fig.~\ref{test1_3D_fig}.
\label{test2_3D_fig}}
\end{figure}

Finally, in Figure~\ref{conserv_1D_fig} we show the photon conservation
numbers, which we define as the ratio of the total of all new ionizations 
and recombinations divided by the number of photons provided by the ionizing 
source per time step  (right), or integrated over time (left) for Tests 1-4 
in 1D spherical symmetry. We see that, even at the coarsest space- and 
time-discretizations, photons are generally conserved to within a few per
cent. When either the spatial- or the time-resolution is not the coarse one 
the photon conservation of our method is generally better than 1\%.

\begin{figure}
  \includegraphics[scale=0.35]{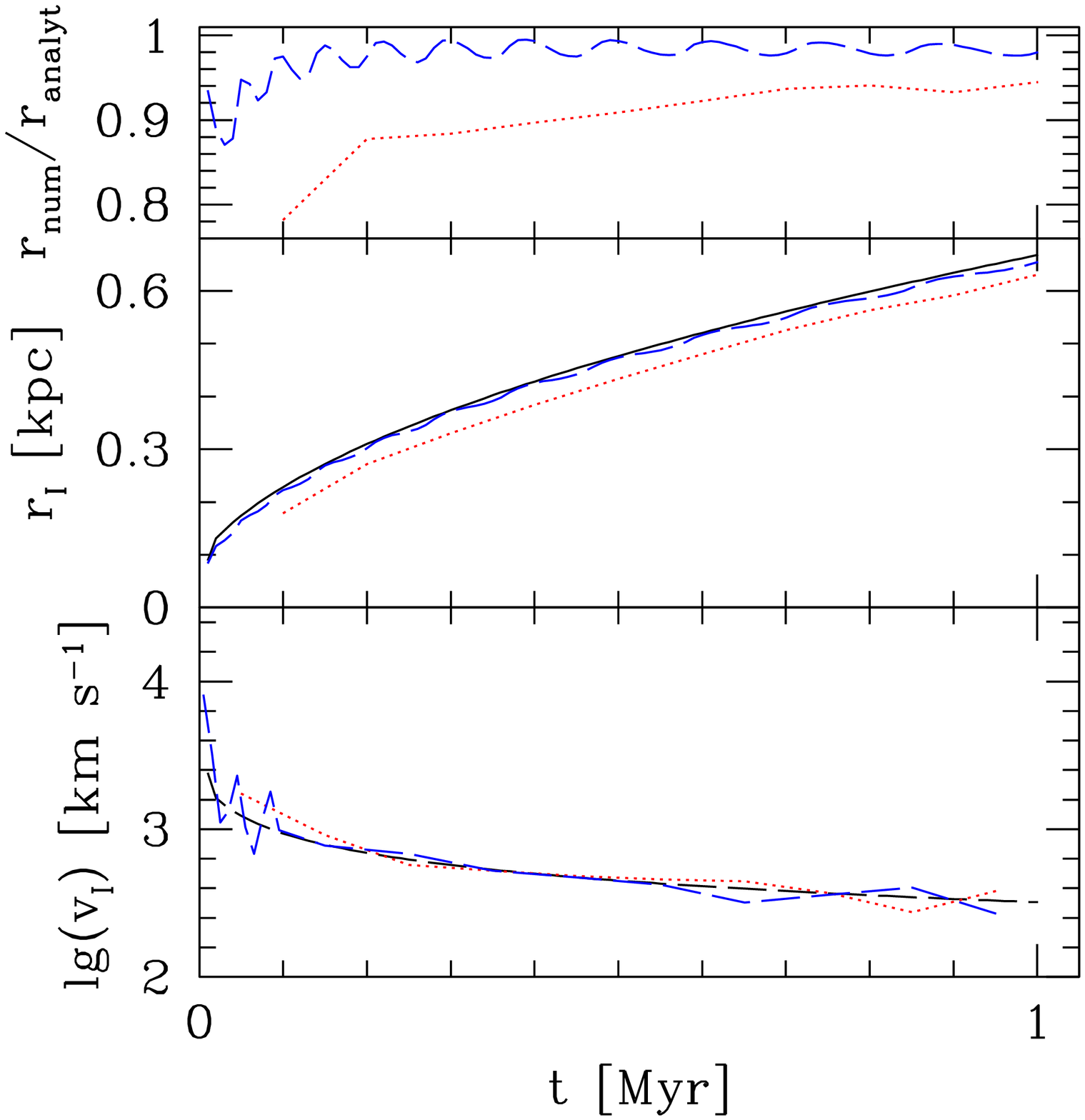}
  \includegraphics[scale=0.35]{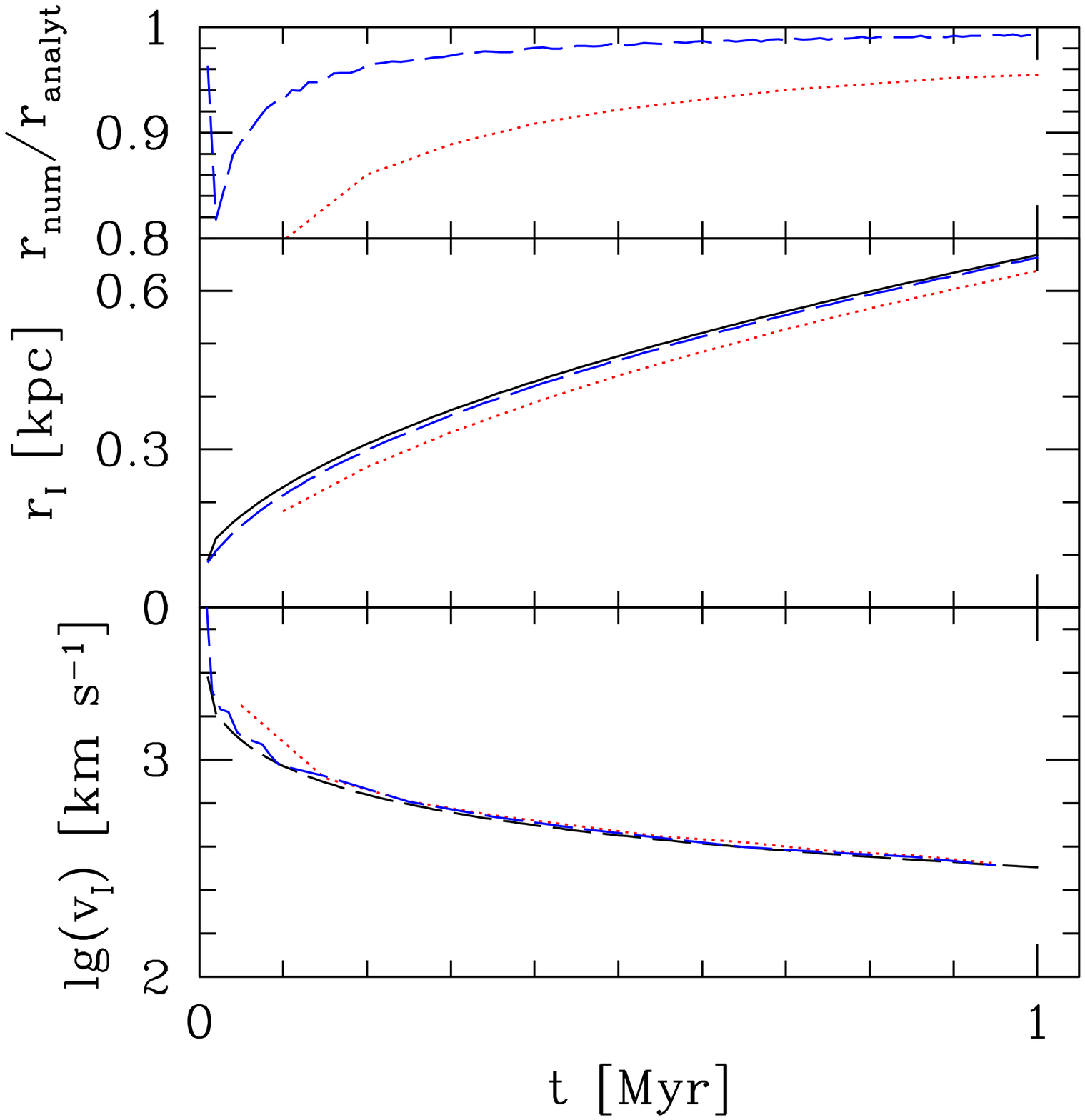}
\caption{I-front evolution for Test 3 (H~II region expansion
  in $r^{-2}$ density profile) in 3D. Same notation as 
  in Fig.~\ref{test1_3D_fig}.
\label{test3_3D_fig}}
\end{figure}

\begin{figure}
  \includegraphics[scale=0.35]{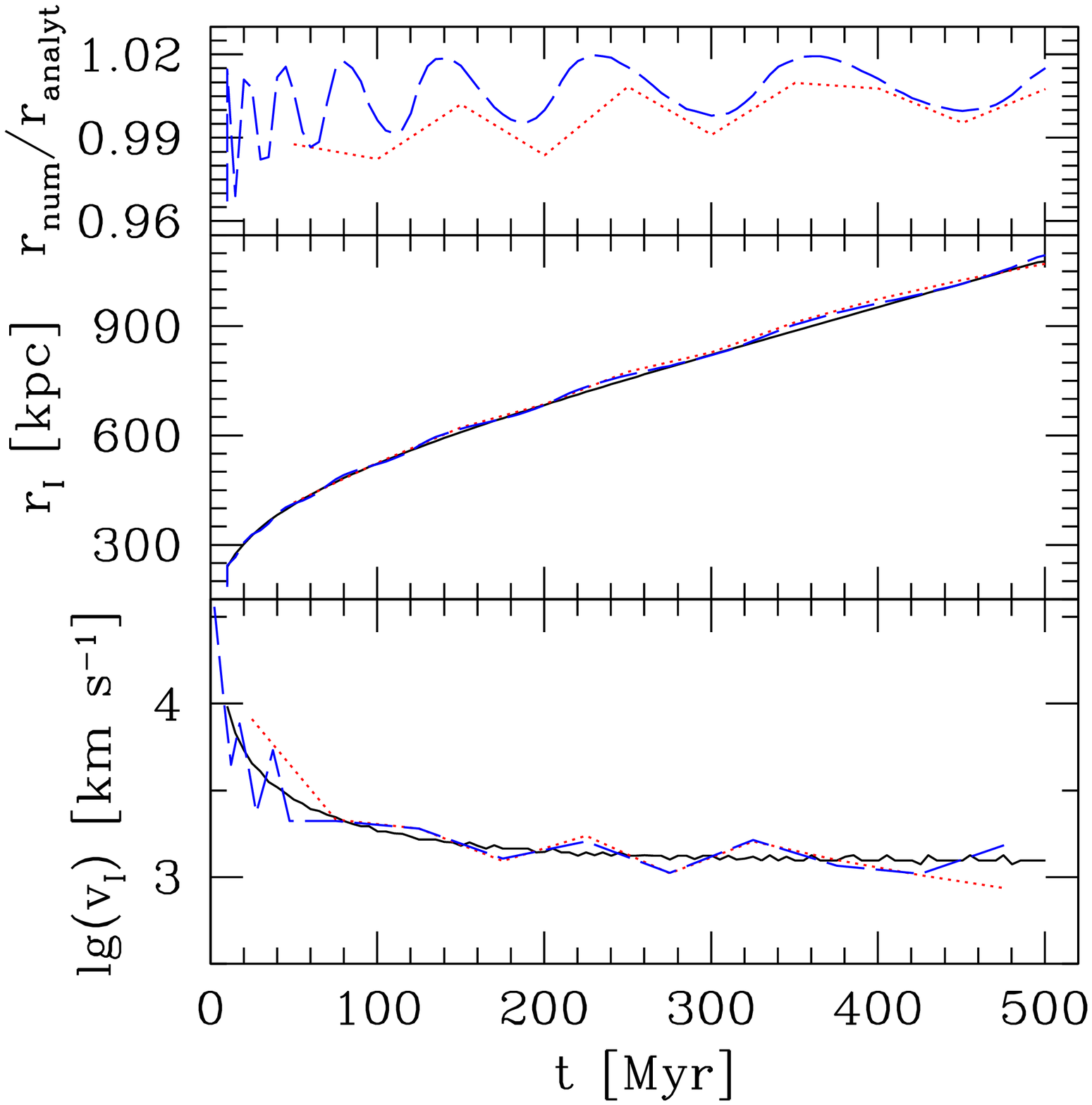}
  \includegraphics[scale=0.35]{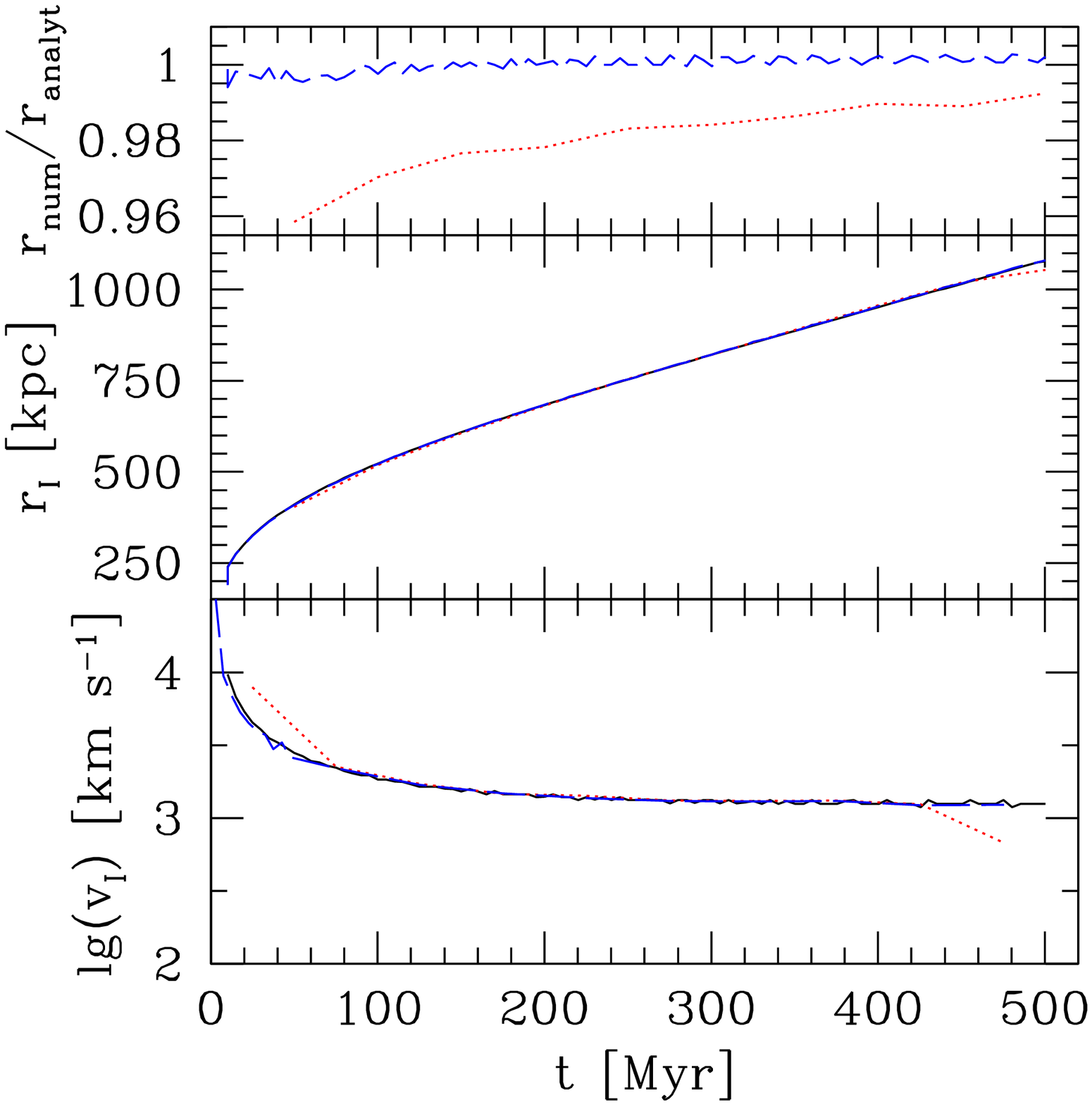}
\caption{Ionization front evolution for Test 4 (cosmological H~II region expansion) 
in 3D. Same notation as in Fig.~\ref{test1_3D_fig}.
\label{test4_3D_fig}}
\end{figure}

\begin{figure}
  \includegraphics[scale=0.35]{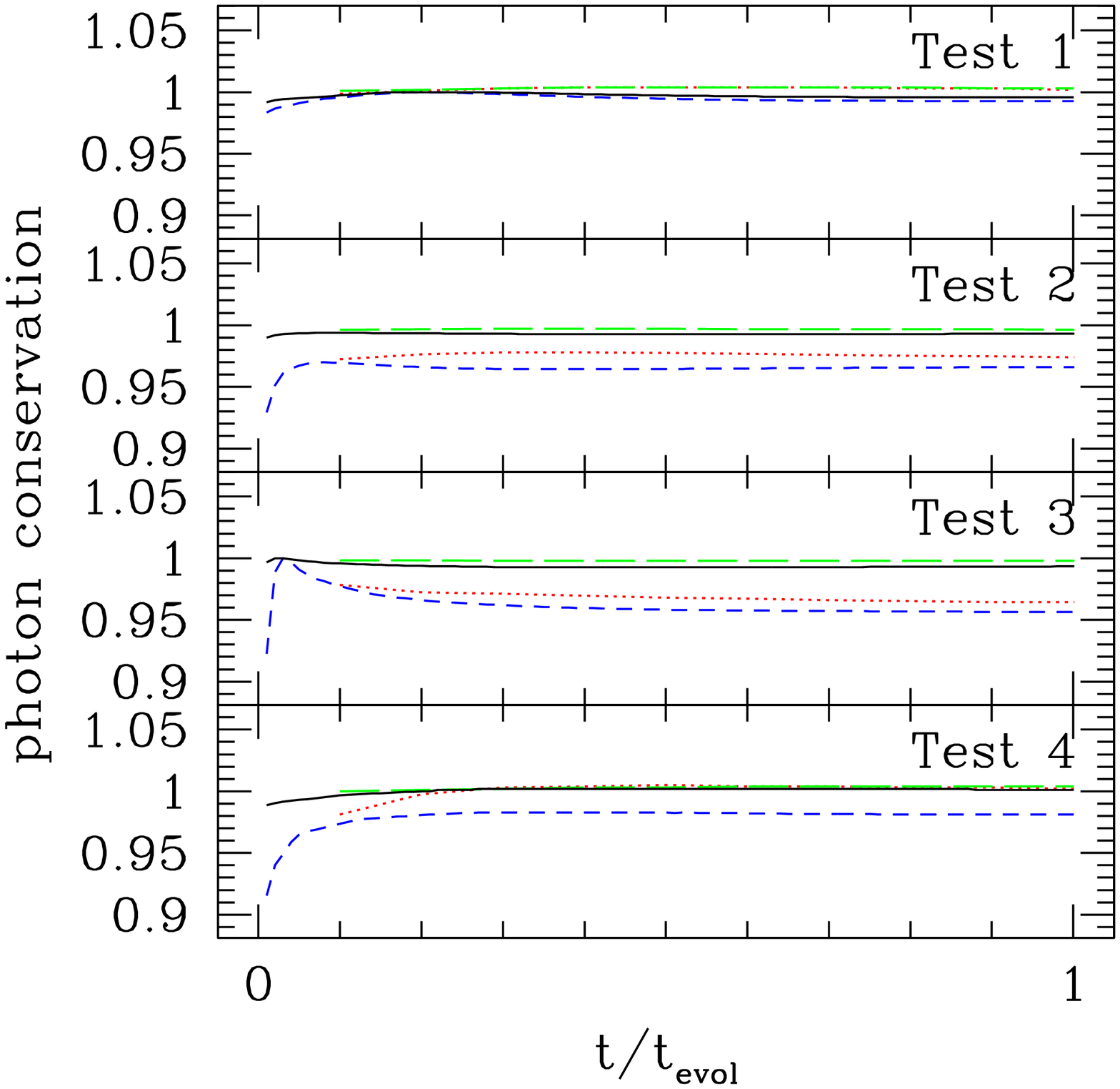}
  \includegraphics[scale=0.35]{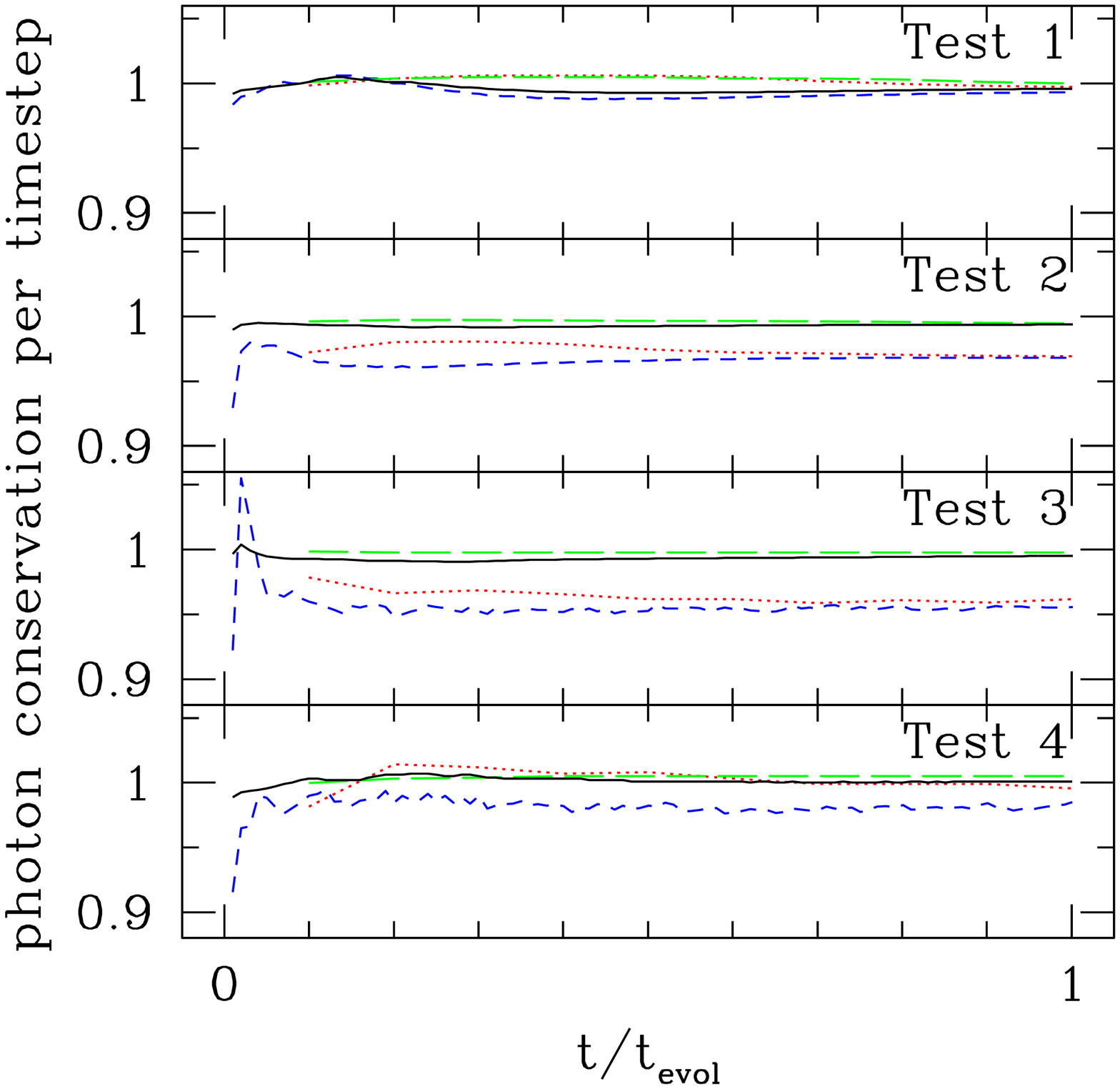}
\caption{Photon conservation for Tests 1-4 in 3D. Cases plotted are: $32^3$ cells, 
  10 time steps (red, dotted), $32^3$ cells, 100 time steps (blue,
  short-dashed), $256^3$ cells, 10 time steps (green, long-dashed), $256^3$
  cells, 100 time steps (black, solid).
\label{conserv_3D_fig}}
\end{figure}

\subsubsection{3D Ionization Fronts} 
\label{3D_tests_sect}
\begin{figure}
  \includegraphics[scale=0.35]{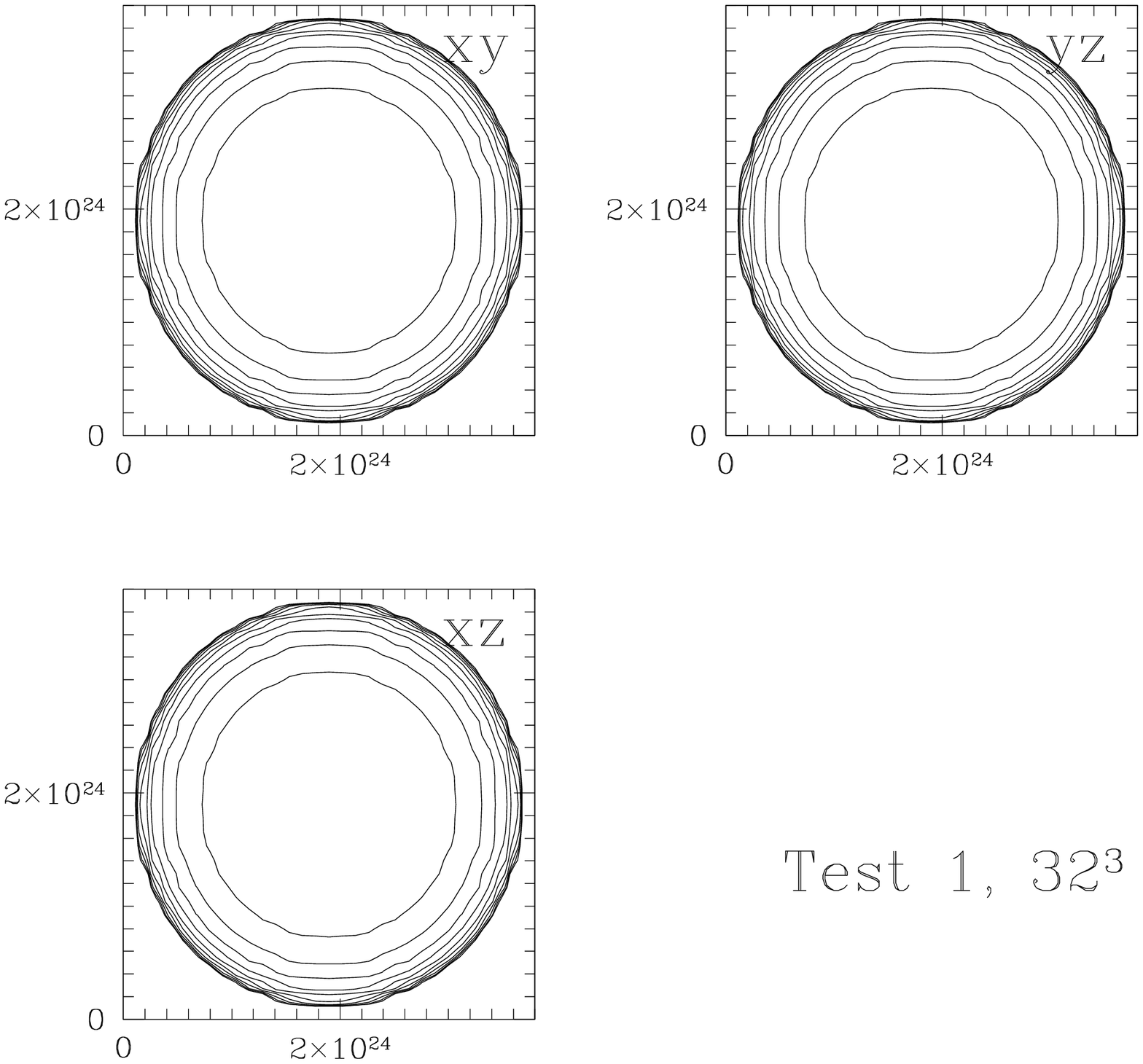}
  \includegraphics[scale=0.35]{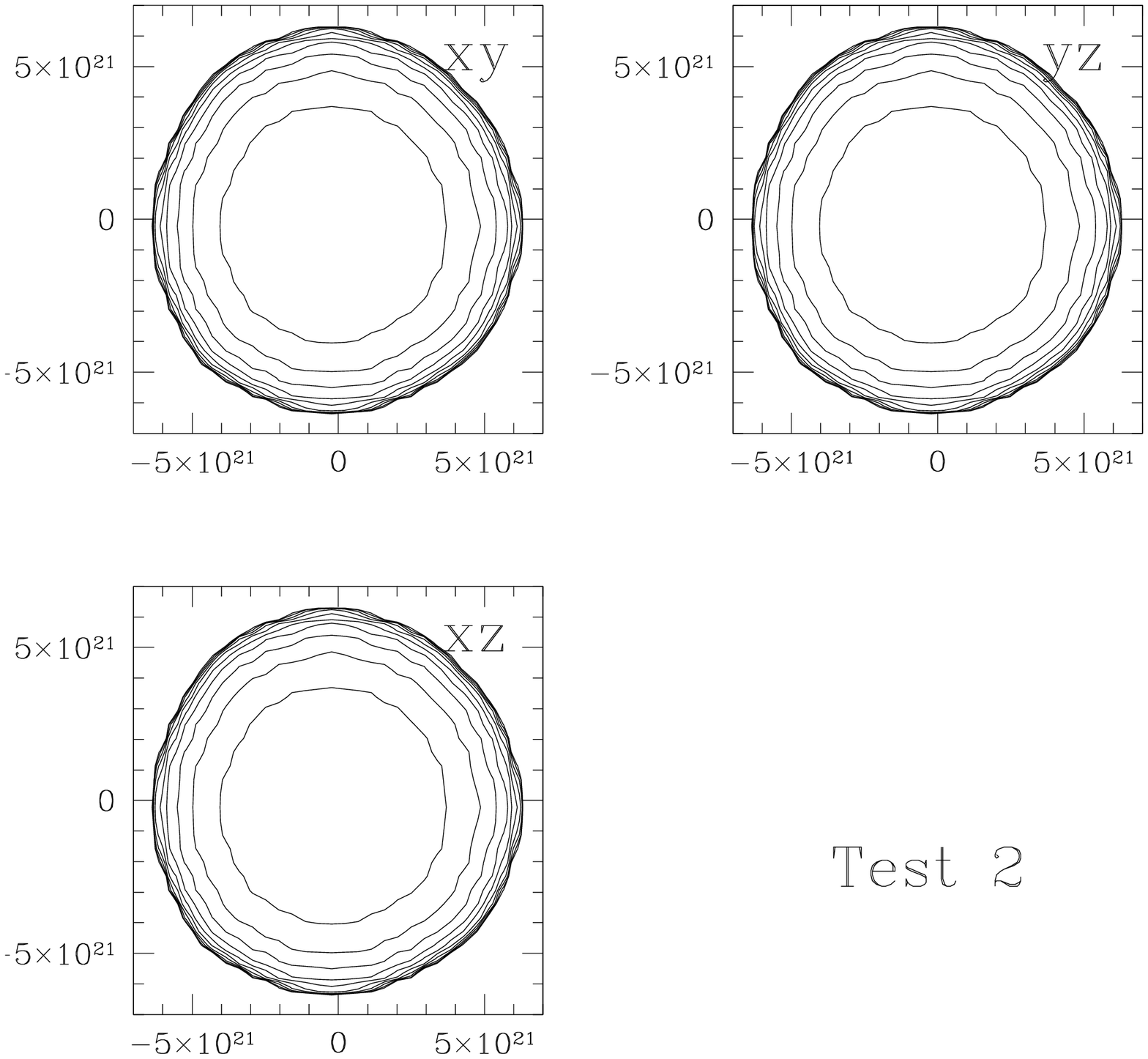}
  \includegraphics[scale=0.35]{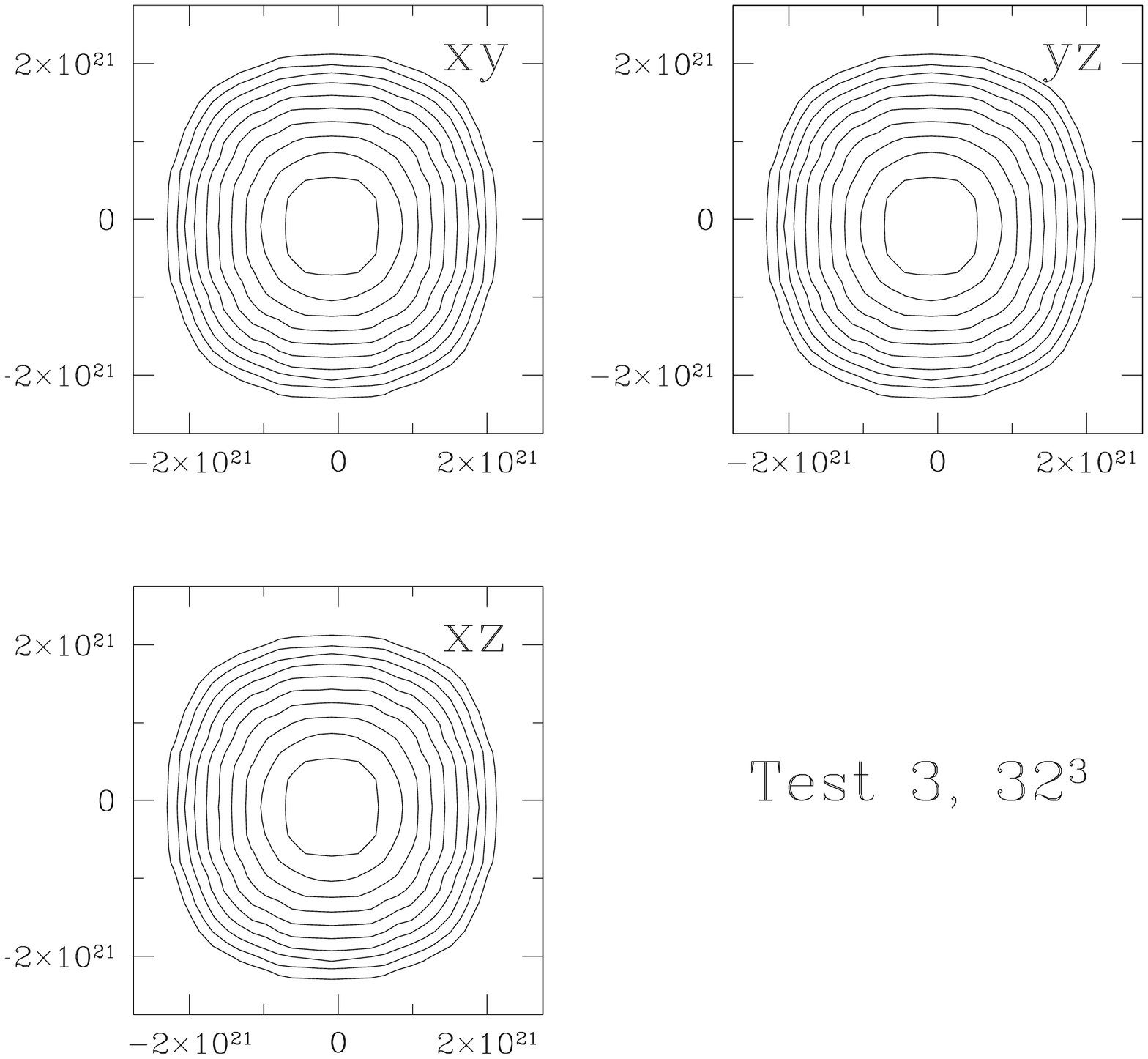}
  \includegraphics[scale=0.35]{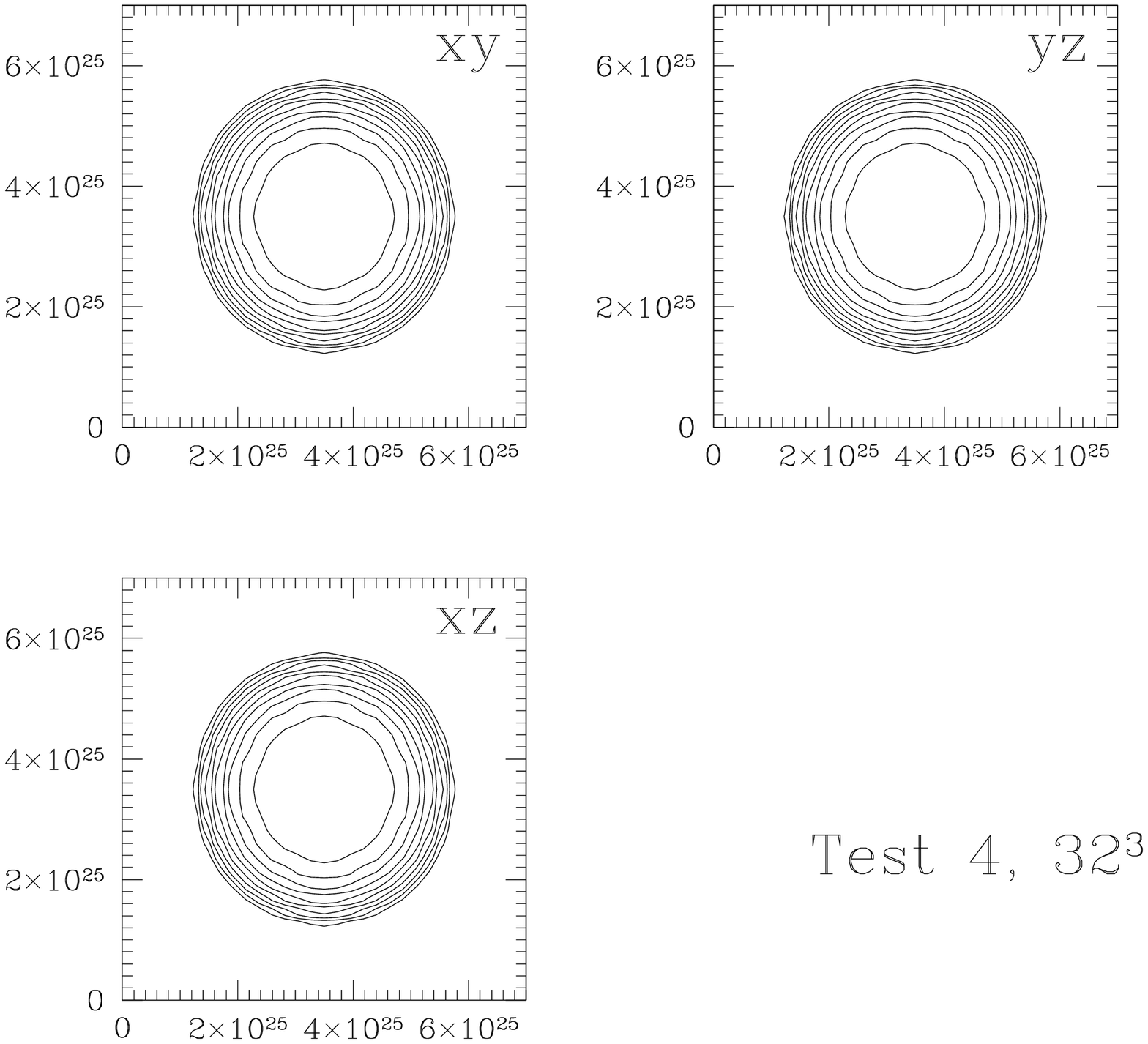}
\caption{I-front evolution. Shown is the 0.5 ionization level contour in 
  time-sequence from $t=0.1t_{\rm evol}$ to $t=t_{\rm evol}$, every $0.1t_{\rm
    evol}$ for Tests 1-4, as labeled. All plots are for the case with 
    coarsest-resolution in both space and time. On all occasions the I-front 
remains spherical at all times except 
for slight non-sphericity in Test~3 ($1/r$ density). In all higher-resolution cases the
    I-front remains perfectly spherical at all times.  
\label{contours}}
\end{figure}
For the 3D tests we use the same approach as in \S~\ref{1D_tests_sect}, i.e.\
constant temperature at $T=10^4$ K and gray opacities. The resolutions used 
are a coarse one at $32^3$ cells and a fine one at $256^3$ cells. We placed 
the source in the center of the grid. Hence, the radial profiles are calculated 
at radial resolution of 128 and 16 cells, respectively, and are thus directly 
comparable to the 1D results in \S~\ref{1D_tests_sect}. Again, for each case we 
use two temporal resolutions, $\Delta t_{\rm coarse}=t_{\rm evol}/10$ and 
$\Delta t_{\rm fine}=t_{\rm evol}/100$.  
Results are shown in Figs.~\ref{test1_3D_fig}-\ref{conserv_3D_fig}. The radial 
cuts are along the positive $x$-axis. The results are largely identical to the 
corresponding ones in 1D spherical symmetry. The numerical solution slightly
overestimates the initial I-front velocity for low temporal resolution cases in 
Tests 1 (uniform density) and 2 ($1/r$ density). The final Str\"omgren spheres in the same Tests 1 and 2 have 
the correct sizes, to within few per cent, except for low spatial resolution 
runs of Test 2, where it is off by about 10\%, or slightly more than 1 cell size. 
Once again the low temporal resolution runs in Test 3 ($1/r^2$ density) somewhat underestimate the 
H~II region radius (although the match improves markedly at later times, to better 
than 5\%). The I-front velocity in all cases is in good agreement with the
exact result. Similarly to 1D runs, Test 4 (uniform, expanding universe) exhibits the best agreement with
the analytical solution, regardless of resolution.

The level of photon conservation, either per time step or integrated over time
in the 3D tests is similar to the one we observed
in the 1D spherical runs, typically within few percent (Fig.~\ref{conserv_3D_fig}). The conservation is
worst for the low spatial resolution runs of Test 3 ($1/r^2$ density), largely due to a 
slight non-sphericity of the I-front (see Figure~\ref{contours}) caused by the 
steep gradient of the density profile. Nevertheless, even in these cases the photon
conservation holds to within $\sim4\%$. The high spacial resolution runs conserve
photons to better than a fraction of a percent in all cases.

\subsubsection{Testing Against Explicit Ionization Front Tracking for a 
Cosmological Density Field}

As we have described, the method presented here calculates the evolution of
the ionized fraction at each point in space by solving the ionization rate
equations.  As a way to check our implementation of this method for a single
source surrounded by a static inhomogeneous density field, we have used a
different, simpler approach which makes approximations similar to the ones 
used in obtaining the analytical solution in Eq.~(\ref{strom}).  
The approximations are as follows. The I-front speed in any direction is 
given by the I-front jump condition in Eq.~(\ref{jump}), using the flux
on the ionized side and the density of neutral atoms on the neutral side of
the front at that location. This flux is calculated according to
Eq.~(\ref{flux0}) which assumes ionization equilibrium on the
ionized side of the front. This allows us to attenuate the flux by integrating 
the recombination rate per unit area between the source and the position of
the front along any direction, instead of
calculating the optical depth. Along any given direction, the only thing that
makes it necessary for us to solve these equations numerically, rather than
analytically, is the fact that the gas density varies along the ray with
distance from the source and so we must do a quadrature to get the integrated
recombination rate along the ray to the position of the front.
The position of the front versus time along each ray (independent of
every other ray) is determined by finite time-differencing the integration of
the I-front velocity obtained from the jump condition as described above.  
Here we briefly describe our numerical implementation of I-front tracking.  

We use a set of isotropically distributed rays emanating from the source,
and discretize each of the rays radially into segments, so that segment $i$
covers all radii $r$ such that $r_{i}<r<r_{i+1}$, and the center of the segment
is given by $r_{i+1/2}^3 \equiv (r_i^3+r_{i+1}^3)/2$. This is a version of
``long-characteristics'' approach to ray tracing, different from the 
short-characteristics employed by C$^2$-Ray (see Appendix~\ref{tracing_sect}). The density $n_i\equiv
n(r_{i+1/2})$ at the center of the ray segment is determined by tri-linear 
interpolation from the eight nearest cell centers on the mesh.  Along each ray,
the I-front jump condition implies that
\beq 
\frac{{\rm d}r}{{\rm d}t}=\frac{{\dot{N}_\gamma}(r,t)}{4\pi r^2n_H(r)},
\label{diffeq}
\eeq
where ${\dot{N}_\gamma}(r)$ is the number of ionizing photons per unit time 
arriving at the I-front,
\beq 
\dot{N}_\gamma(r)=\dot{N}_{\gamma,0}-4\pi \int_0^r \alpha_Bn_H^2(r)r^2{\rm d}r,
\label{reccorrect}
\eeq
where $\dot{N}_{\gamma,0}={\dot{N}_\gamma}(r=0)$. The discretization of 
Eq.~(\ref{reccorrect}) yields
\beq \dot{N}_i\equiv \dot{N}_{\gamma,0}-\sum_j^{i-1} \alpha_B n_{H,j}^2\Delta V_i,
\eeq
where $\Delta V_i \equiv 4\pi(r_{i+1}^3-r_{i}^3)/3$.
Equation (\ref{diffeq}) is solved numerically to obtain the evolution of
the I-front along each ray.  At any given time, the ionized fraction on the
original grid is set to zero if the distance from the source is smaller than
the radius of the I-front along that direction, and is set to one otherwise.

\begin{figure}
  \includegraphics[scale=0.35]{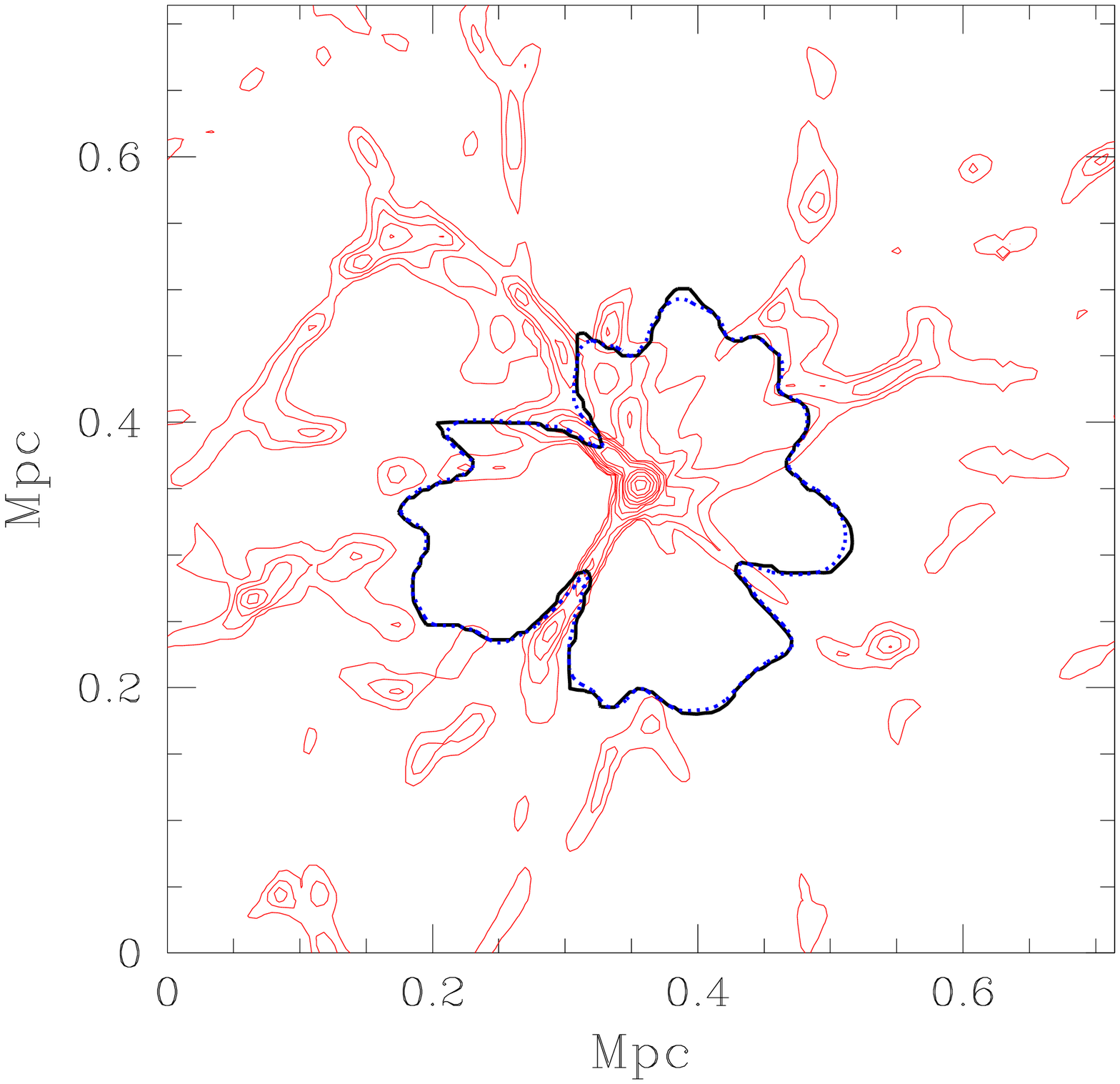}
  \includegraphics[scale=0.35]{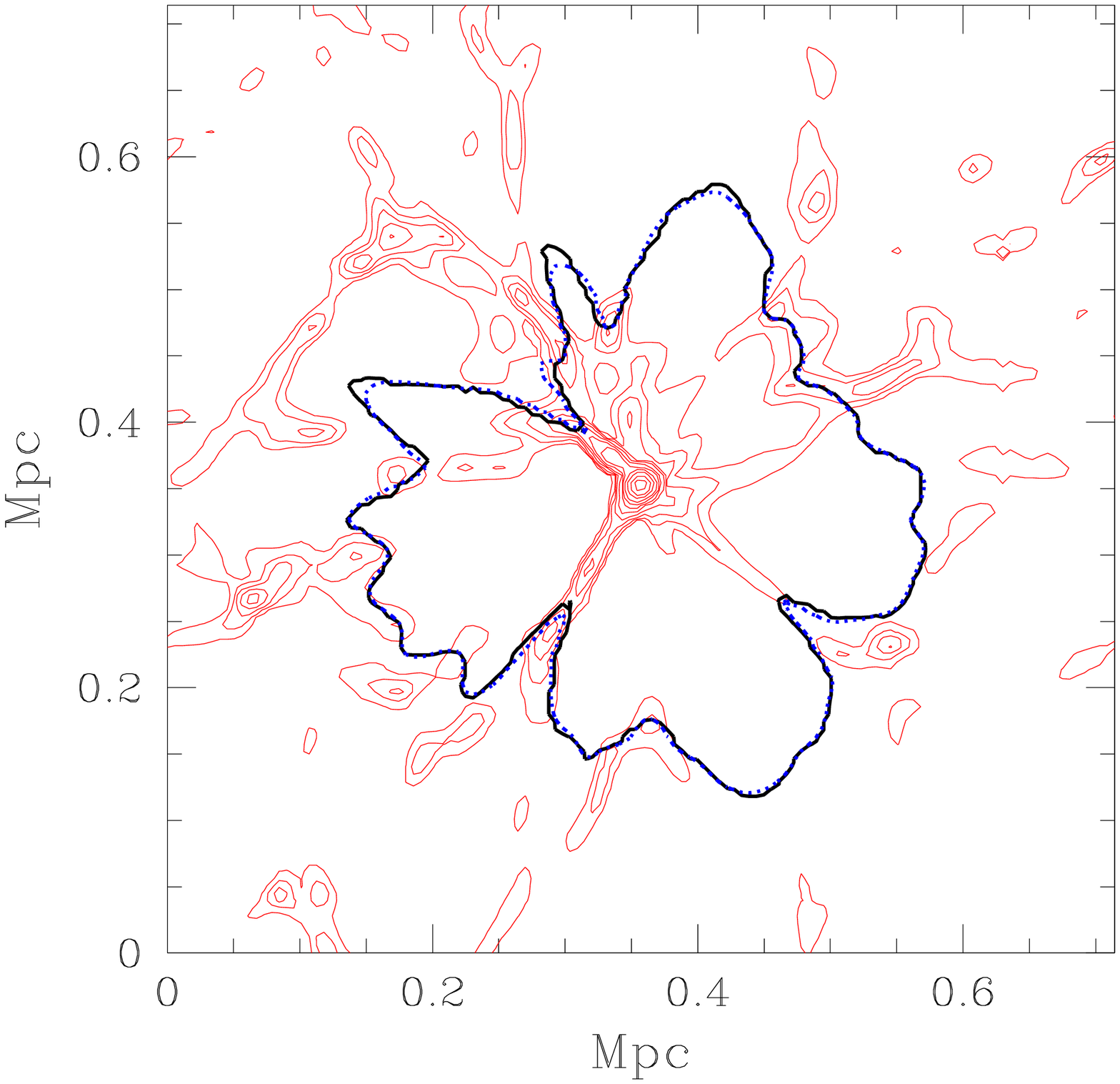}
\caption{Full radiative transfer results from our method (thick dotted blue line) 
vs. explicit I-front tracking (thick solid black line) at times $t=0.5$ Myr
(left) and $t=1$ Myr (right). Also shown are the density contours of the 
underlying density field (thin red lines).
\label{cosmo_fig2}}
\end{figure}
We compare the results from the two methods as applied to the case of a single
ionizing source in a cosmological density field (see \S~\ref{cosmo_appl} for 
more details on the simulations). Since the explicit I-front tracking assumes 
sharp I-fronts and fixed temperature, for closest direct comparison with the 
results of our method for the same test we assumed an ionizing source with a soft 
black-body spectrum with effective temperature $T_{\rm eff}=20,000$~K. The 
gas temperature is fixed to $T=10^4$~K everywhere. In Fig.~\ref{cosmo_fig2} we 
show the positions of the I-fronts obtained by the two methods at times $t=0.5$~Myr 
and 1~Myr. The dotted blue lines show the I-front calculated with full radiative 
transfer method, while the
solid black lines show the I-front position according to the explicit I-front 
tracking. The thin red contours show the underlying density field. The H~II regions
assume a characteristic ``butterfly'' shape since the I-fronts, after they escape from the dense vicinity of the ionizing source, propagate faster
in the voids and slower along the denser filaments. The results from the two methods 
show excellent agreement, thus verifying our ray-tracing procedure.

\section{Some simple illustrative applications}
\label{applic_sect}

\begin{figure}
  \includegraphics[scale=0.6]{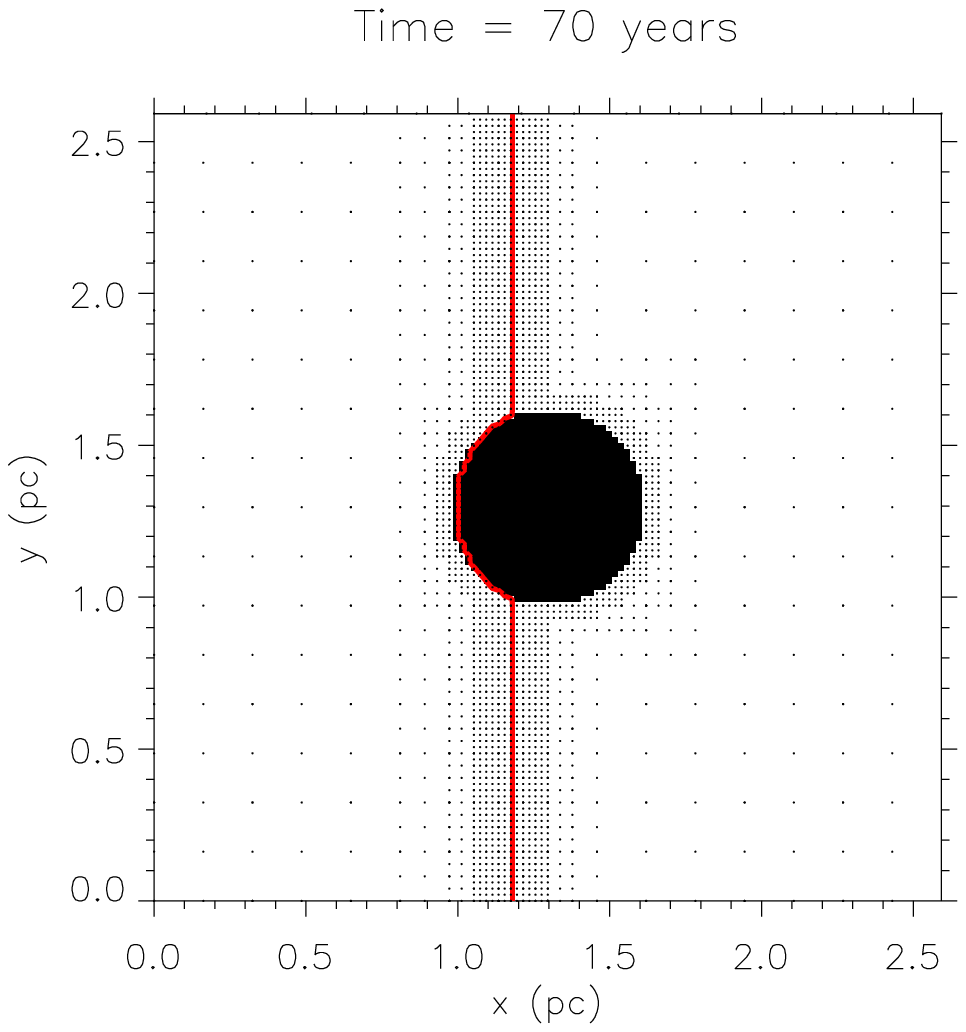}
  \includegraphics[scale=0.6]{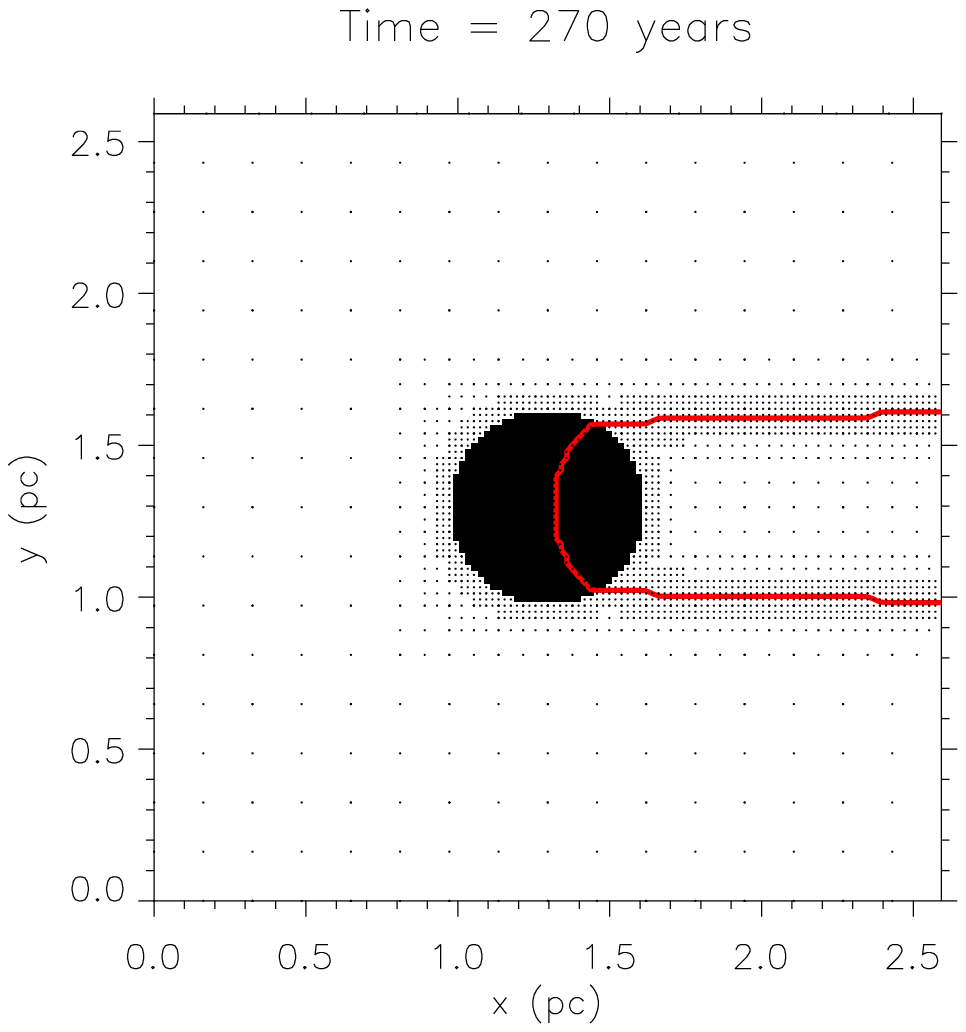}
  \includegraphics[scale=0.6]{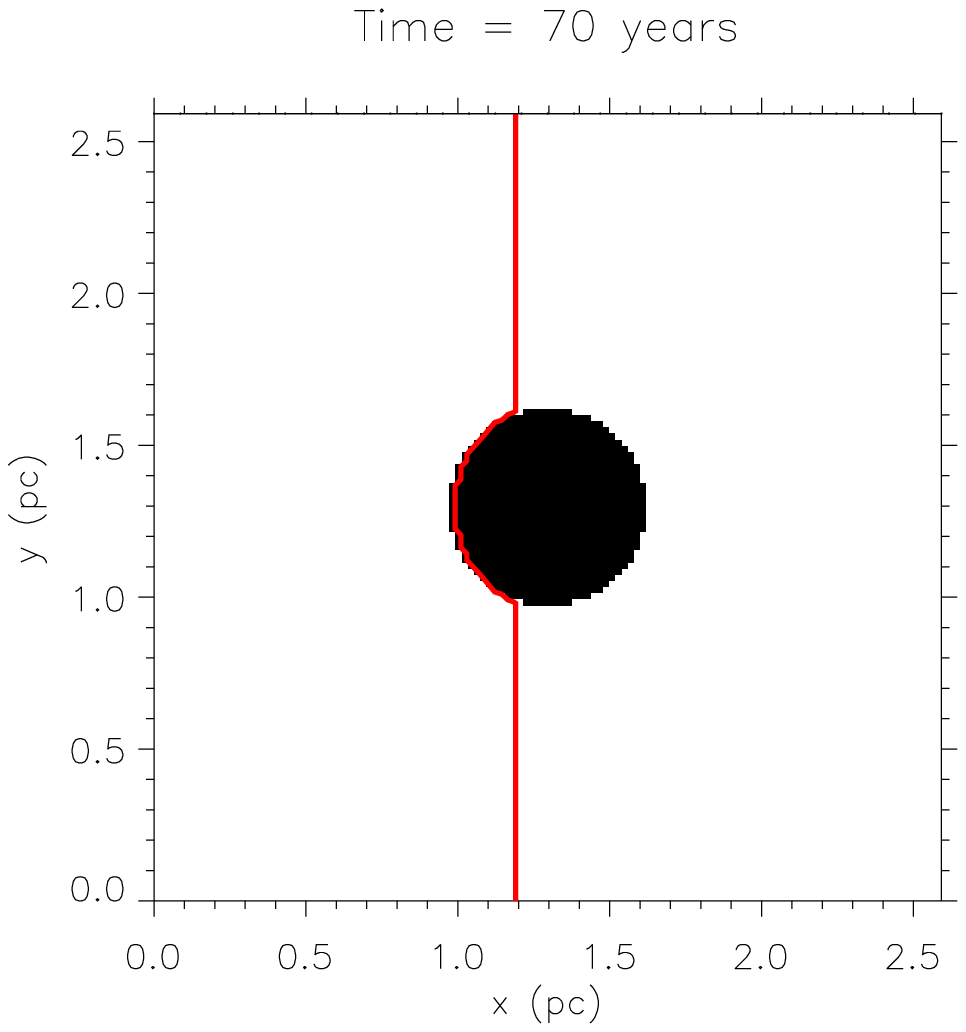}
  \includegraphics[scale=0.6]{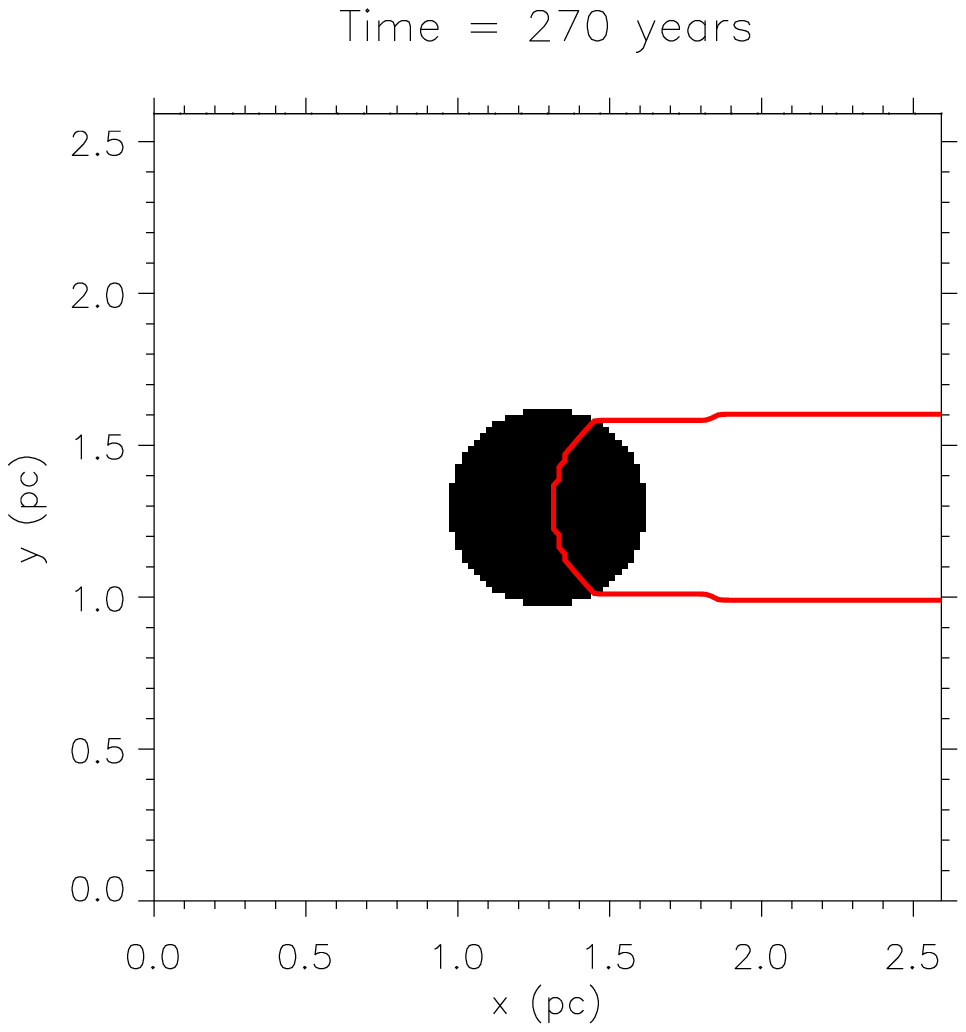}
\caption{Plane parallel ionization front crossing a high density region.
  Top row: calculated on a 5-level adaptive mesh; Bottom row: calculated on a
  fixed mesh. The snapshots show the density in the central $xy$ plane
  (black=high density, white=low density), with the red contour showing the
  ionization front ($x=0.5$). In the AMR simulation the dark points indicate
  the computational mesh points.
\label{amr_fig}}
\end{figure}

\subsection{Plane parallel ionization front on an adaptive mesh}

Our radiative transfer and ray-tracing method are directly applicable to any
type of rectangular mesh, not just a uniform one, but also nested and
adaptively-refined (AMR) meshes. We have currently implemented it for 
a plane-parallel I-front on an AMR mesh, in addition to the fixed, uniform 
mesh case.
  
To illustrate the performance of our method on an adaptive mesh, we ran a
simulation of a plane-parallel ionization front overrunning a dense spherical
cloud.  This problem was calculated both with a fixed mesh version, and with
an adaptive mesh. The AMR method used is the one implemented in the {\it Yguazu}\/
code \citep{2000RMxAA..36...67R}. It refines on a point by point
basis, based on a given set of quantities, determined by the user (these could
include e.g.\ pressure, density, ionized fractions of various species, etc.)
and employs refinement steps of a factor of two. For this particular 
simulation we used a 3D box with sides of $8\times 10^{18}$~cm, filled with 
uniform gas of 5~cm$^{-3}$ and containing a ten times denser cloud of 
radius $10^{18}$~cm positioned in the center of the computational volume. We 
used five refinement levels, and the maximum resolution (at finest refinement 
level) is $128^3$. We refined on the gradients of the ionization fraction 
and density. For comparison we also ran the same problem on a fixed, uniform 
mesh of $128^3$ cells. 

In Figure~\ref{amr_fig} we show the comparison between the AMR and the fixed
mesh simulation results. Our AMR implementation correctly identifies both the
I-front and the clump and refines around them. The I-front position
and shape are identical at both times in the AMR and fixed-mesh results. This
test demonstrates that our method interacts correctly with the adaptive mesh
and properly tracks the I-front. For this particular problem the AMR simulation 
was completed approximately 10 times faster than the fixed mesh simulation,
demonstrating the power of mesh adaptivity to speed up codes without loss of
accuracy. Further tests and full description of the AMR implementation of C$^2$-Ray
will be presented in a subsequent paper which will discuss the direct coupling of 
our radiative transfer method with hydrodynamics \citep{hydropaper}.
\begin{figure}
  \includegraphics[scale=0.5]{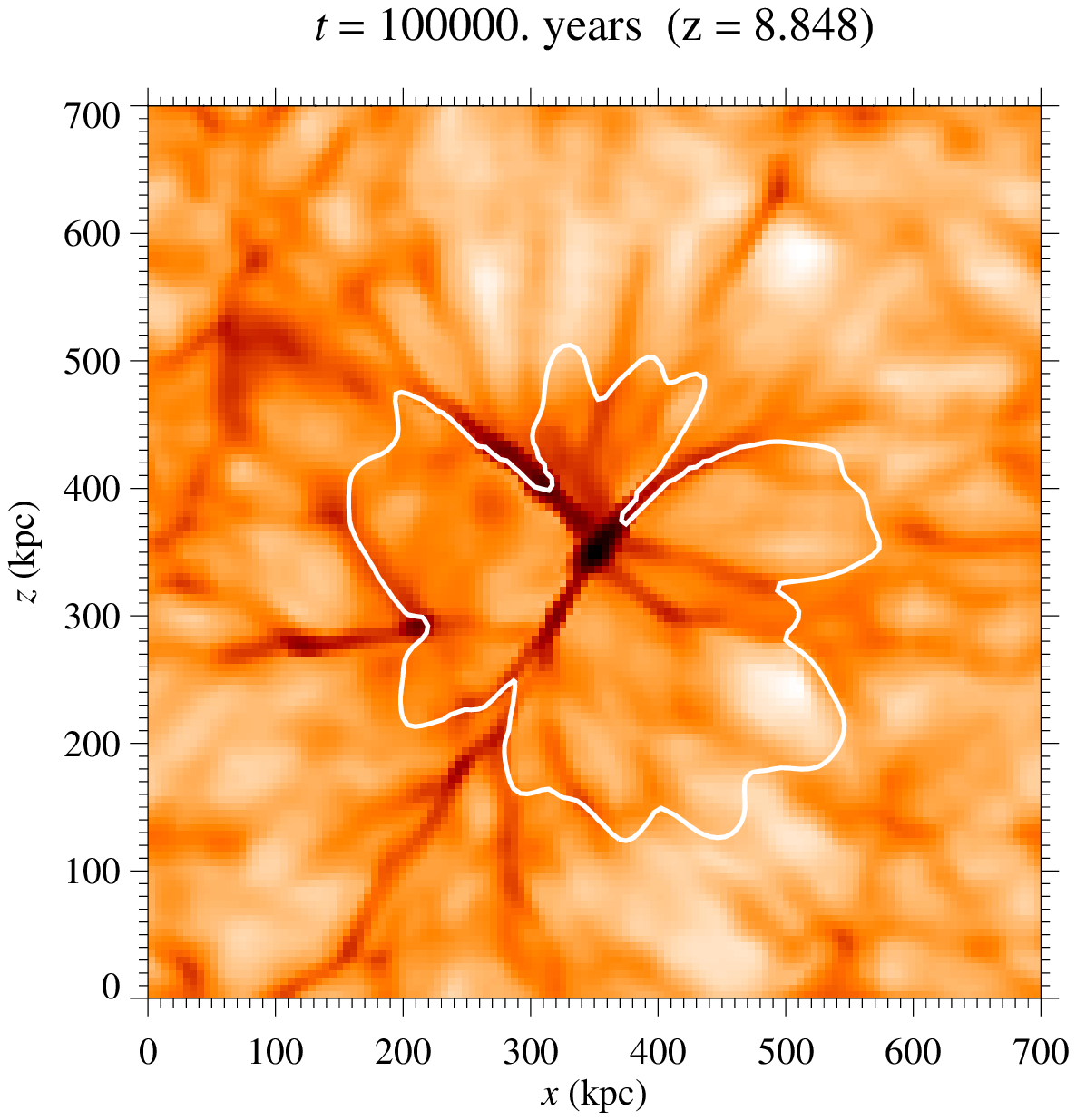}
  \includegraphics[scale=0.5]{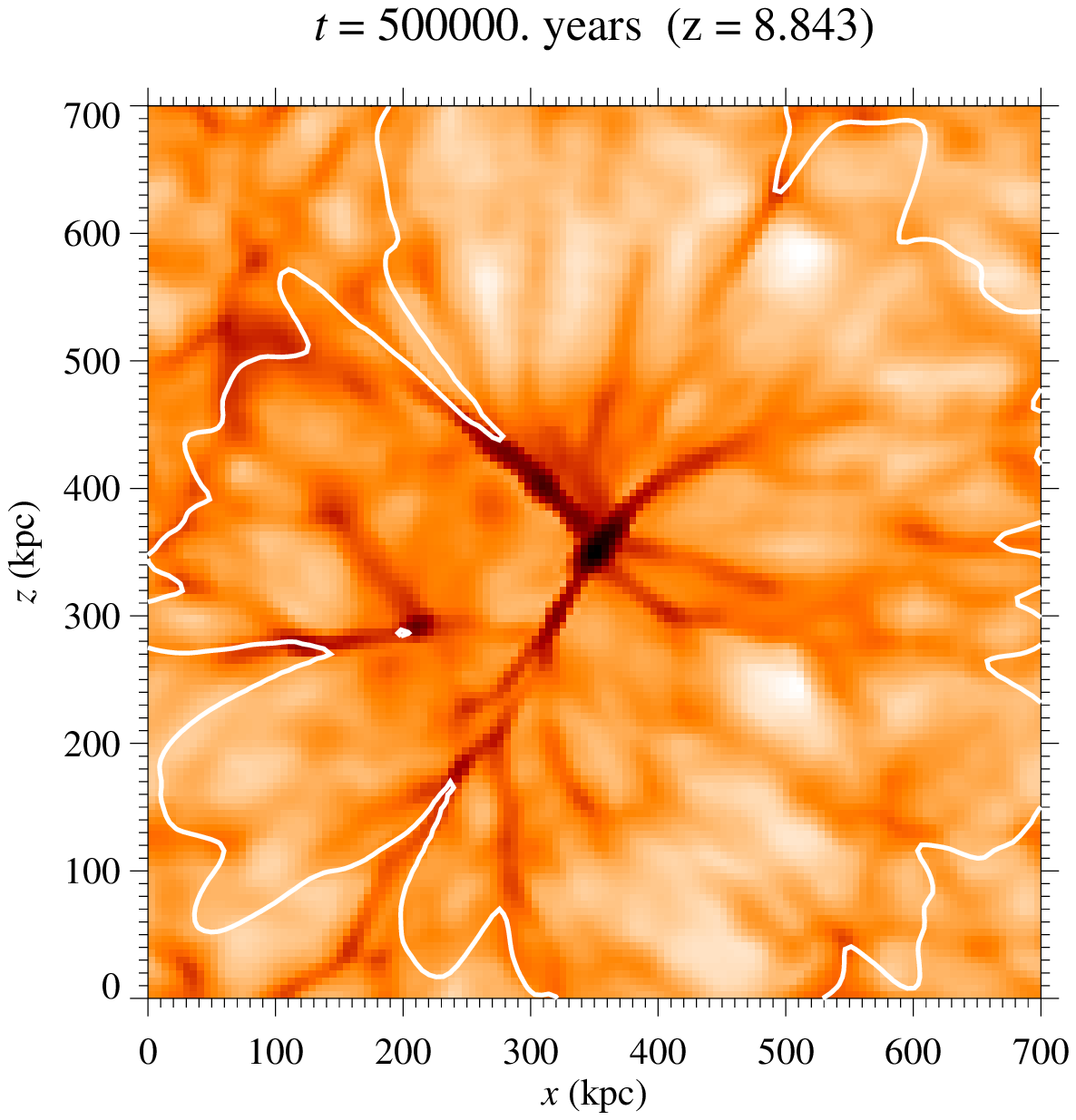}
\caption{I-front propagation around a single ionizing source in a cosmological 
density field at times $t=0.1$ Myr (left) and $t=0.5$ Myr (right).
   Image is a cut through the source $y$-plane and shows the color-coded gas
   density field (the darker regions are
   denser). The current position of the I-front is indicated by the white line.
   See Movies 1--3 for an animated view of the growth of these ionization 
   fronts in three planes.
\label{cosmo_fig_single}}
\end{figure}

\subsection{Reionization of a cosmological density field}
\label{cosmo_appl}
As a further illustration and testing of our code, we have applied our method
the simulation of reionization of a fixed cosmological density field at
redshift $z=8.85$.  The gas distribution was obtained from a PM+TVD N-body and
gasdynamical simulation \citep{SAAIMR05} performed using 
the cosmological dynamics code described in \citet{1993ApJ...414....1R}.
The simulation box size is $0.5\,h^{-1}$ Mpc, the resolution is $128^3$ cells,
$2\times64^3$ particles. 
 
\begin{figure}
  \includegraphics[scale=0.5]{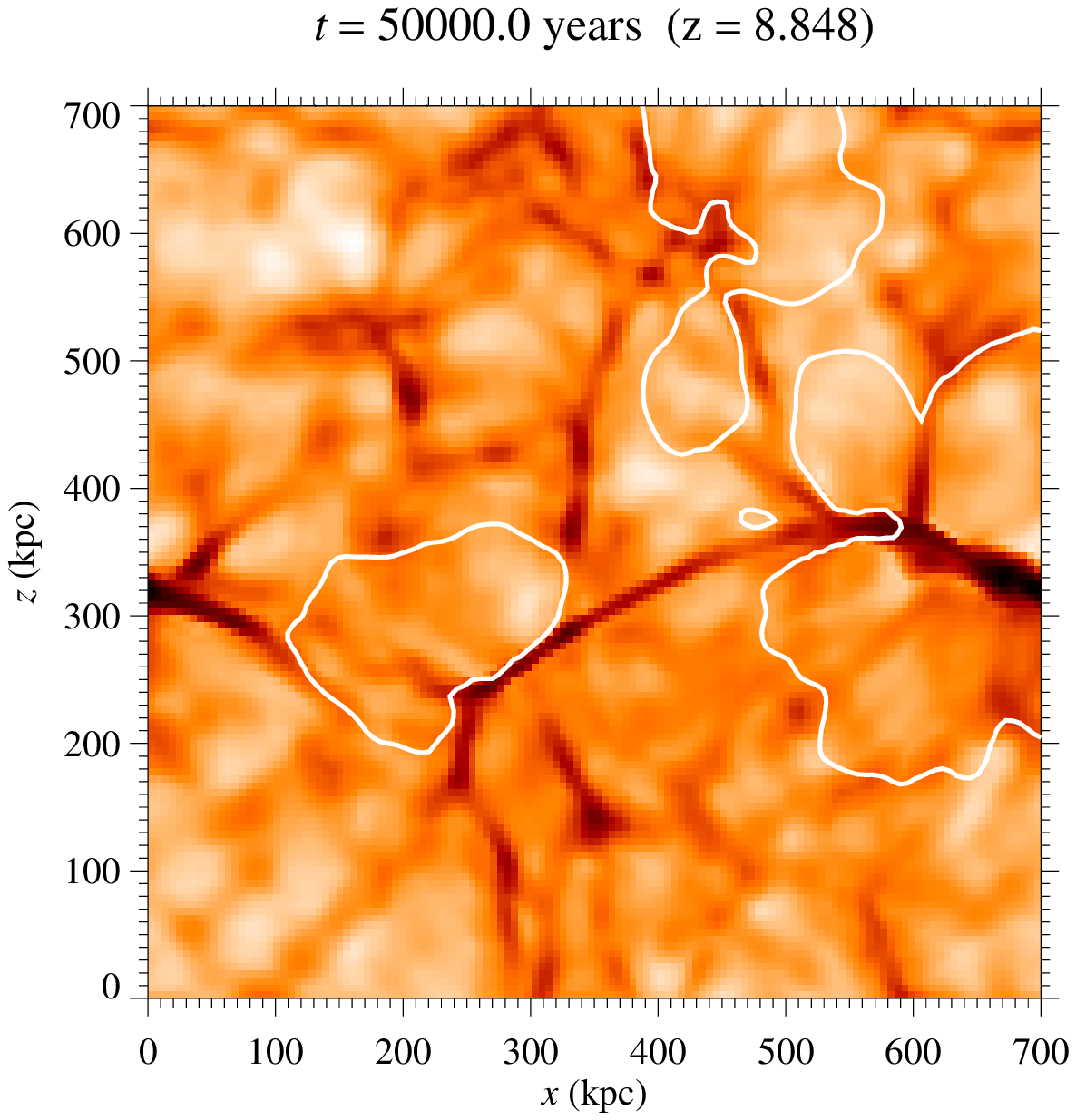}
  \includegraphics[scale=0.5]{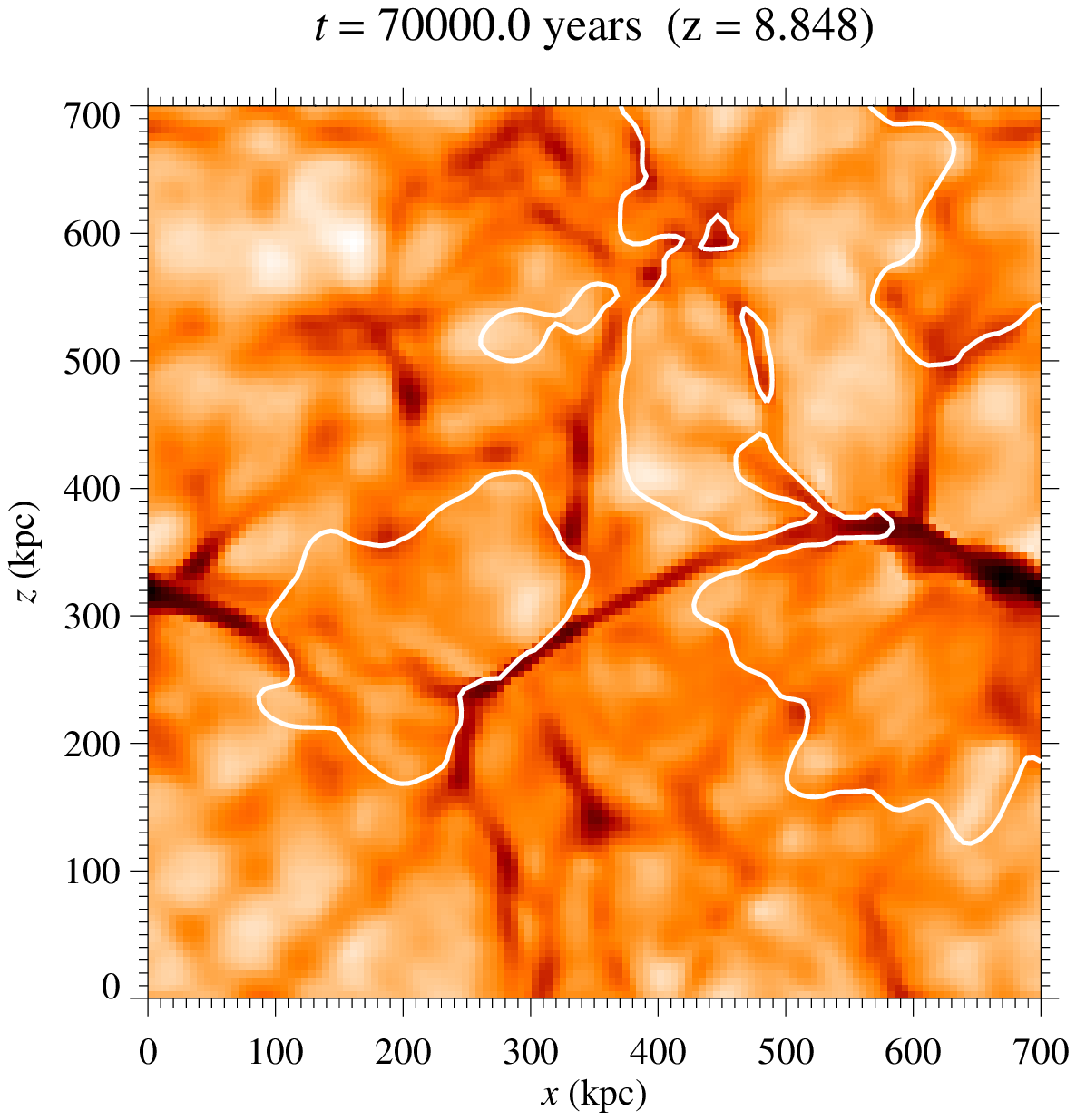}
  \includegraphics[scale=0.5]{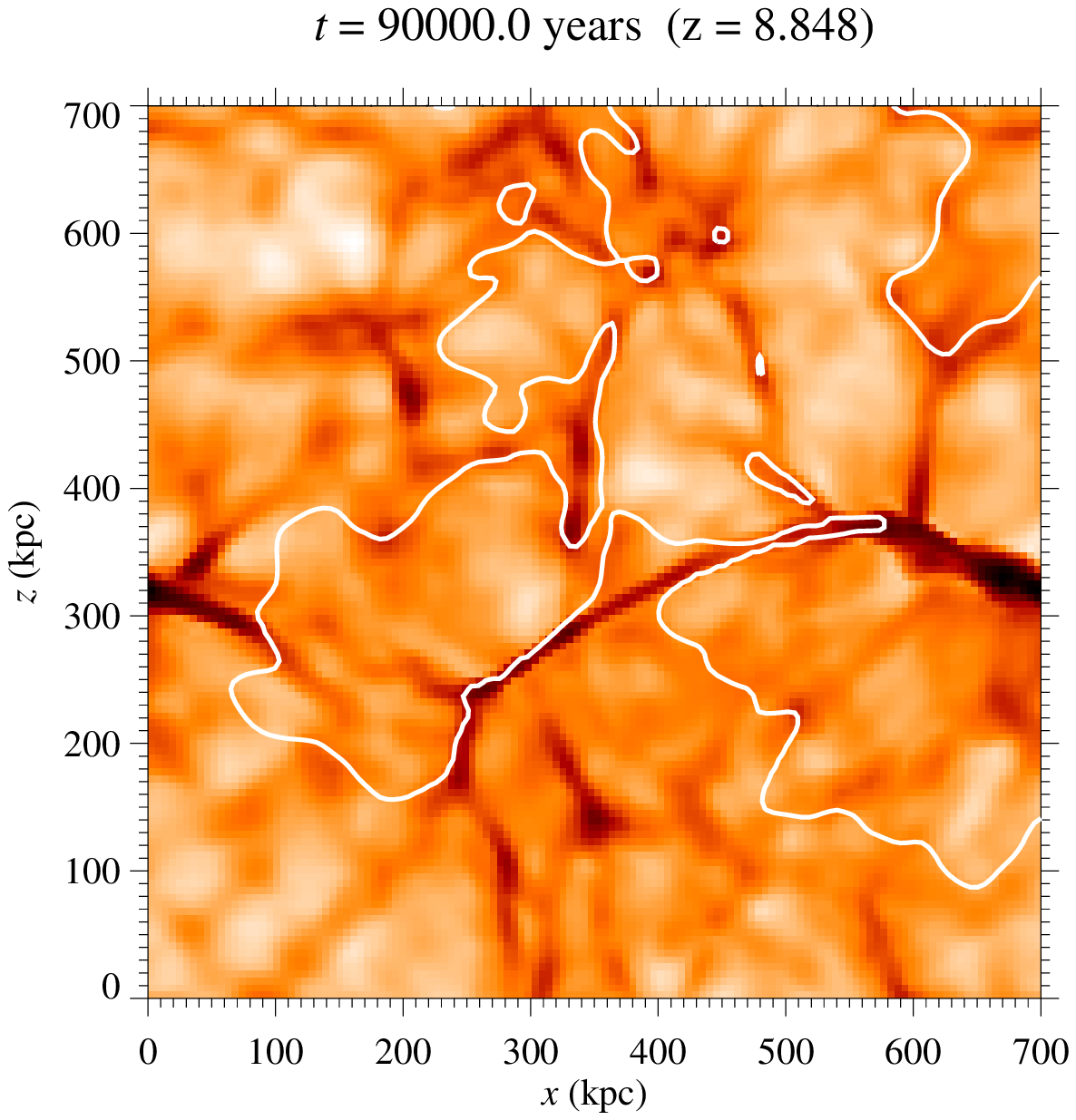}
  \includegraphics[scale=0.5]{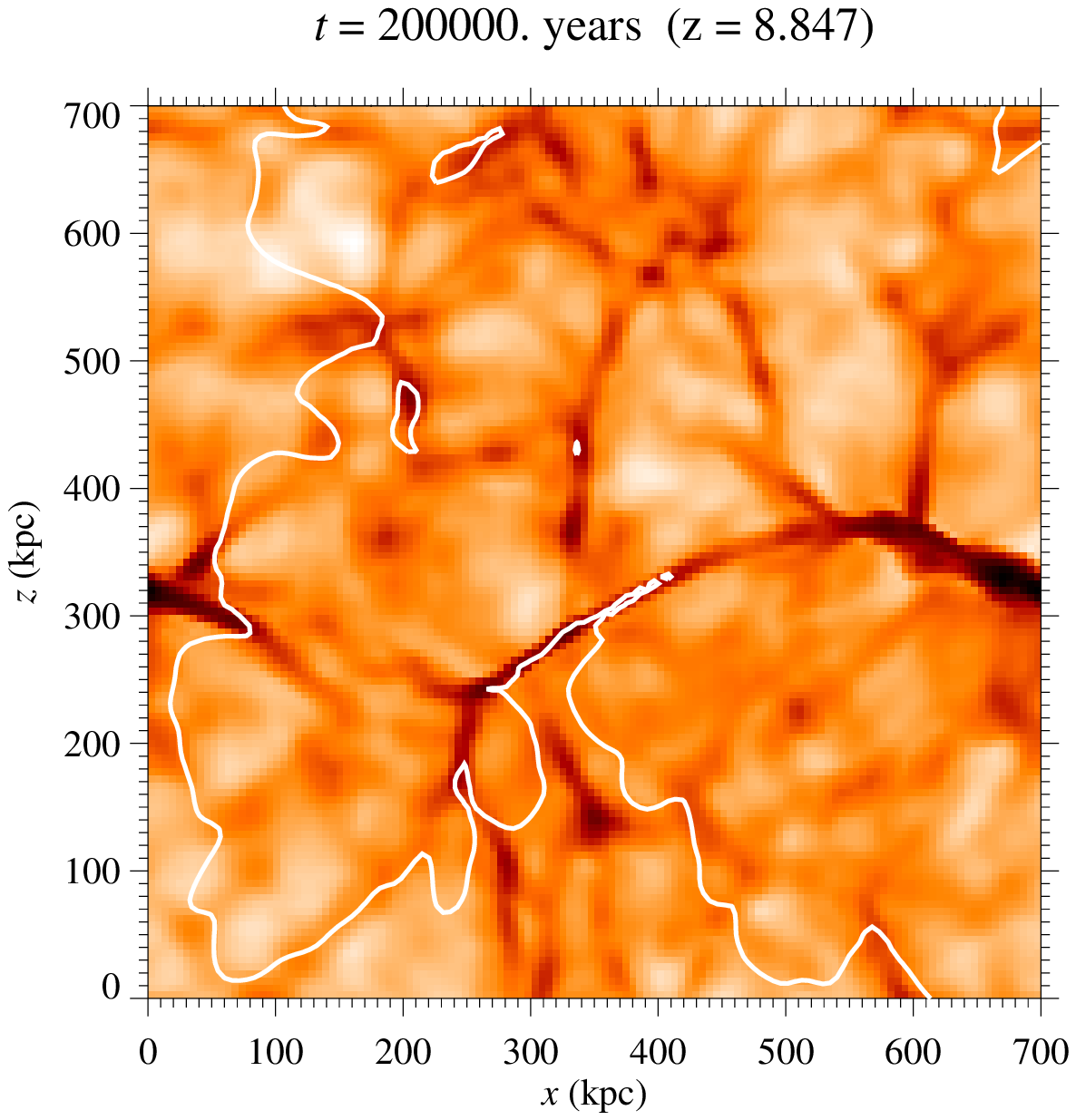}
\caption{Ionization of a cosmological gas density distribution by  
  multiple sources at time $t=0.05$ Myrs (top left) $t=0.07$ Myrs (top right),
  $t=0.09$ Myrs (bottom left), and $t=0.2$ Myrs (bottom right). Image shows
  the color-coded gas density field, cut through the middle of the box ($y$-plane). The
  current position of the I-fronts is
  indicated by the white lines.  Movies 4--6 show the growth of 
  these ionization in three planes through the center of the volume.
\label{cosmo_fig_mult}}
\end{figure}

We ran two simulations, the first one with a single ionizing source in the
computational volume, the other with multiple sources. For the single-source run
the ionizing source is positioned at the highest-density point of the grid
(which for visualization purposes is moved to the middle of the grid using the
periodic boundary conditions), with luminosity $L=10^9 L_\odot$ and black-body
spectrum of effective temperature $T_{\rm eff}=50,000$ K, as appropriate for
a massive Pop~II star, corresponding to an
ionizing photon production rate of $\dot{N}_\gamma=6.822\times10^{52}\,\rm
s^{-1}$.  For the multiple-source run the ionizing sources are chosen to be
the 6 most massive halos found in the box, with masses above $4\times10^7
M_\odot$, which mass at that redshift corresponds to a halo virial temperature 
 $10^4$ K \citep{2001MNRAS.325..468I}, below which temperature the gas cannot 
cool efficiently and form stars easily. The halos in the simulation box were 
found using a friends-of-friends halo finder with a linking length of 0.25. The 
ionizing photon production rate for each source is  constant in time and 
is assigned assuming that each source lives $t_s=3$ Myr and has a constant 
mass-to-light ratio, emitting a total of $f_\gamma=250$ ionizing photons
per atom during its lifetime, as is appropriate for massive stars
\citep{2005ApJ...624..491I}: 
\be 
\dot{N}_\gamma=f_\gamma \frac{M\Omega_b}{\Omega_0 m_pt_s}, 
\ee
where $M$ is the total halo mass and $m_p$ is the baryon mass. The 
source cell is assigned to the densest cell inside the $3^3$ cells centered on 
the halo position. For simplicity all the sources are assumed to ignite at the
same time.

In the single-source case, the the H~II region is initially confined to the
dense region around the source. Once most of the gas there is ionized, the
I-front quickly escapes into the lower-density voids, while the denser
filaments temporarily trap it and become ionized only slowly, creating a
characteristic ``butterfly'' shape (Fig.~\ref{cosmo_fig_single}). Later on 
the H~II region becomes somewhat more spherical and eventually most of the 
computational box becomes ionized at $t\sim 1$ Myr.

\begin{figure}
  \includegraphics[scale=0.35]{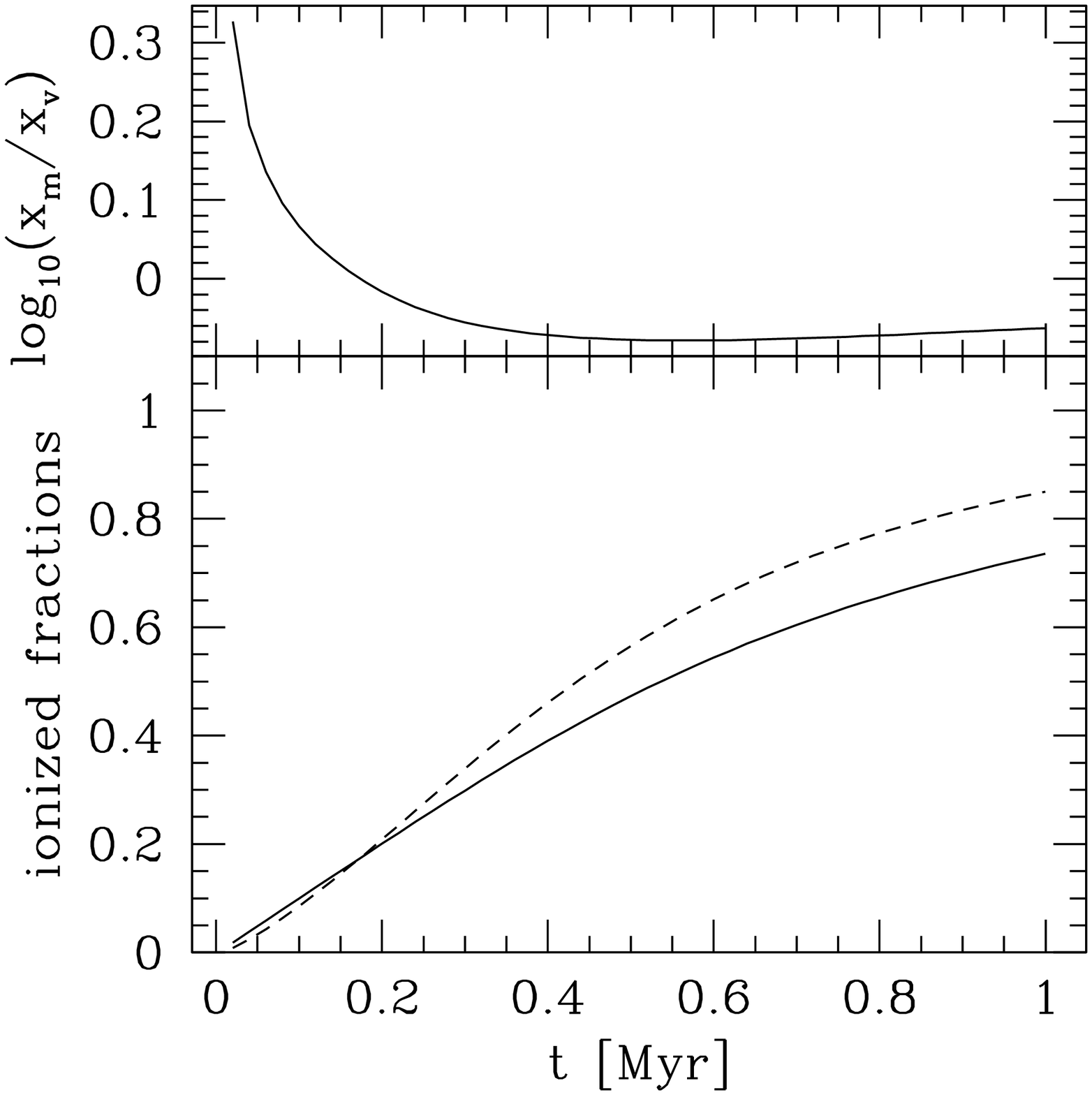}
  \includegraphics[scale=0.35]{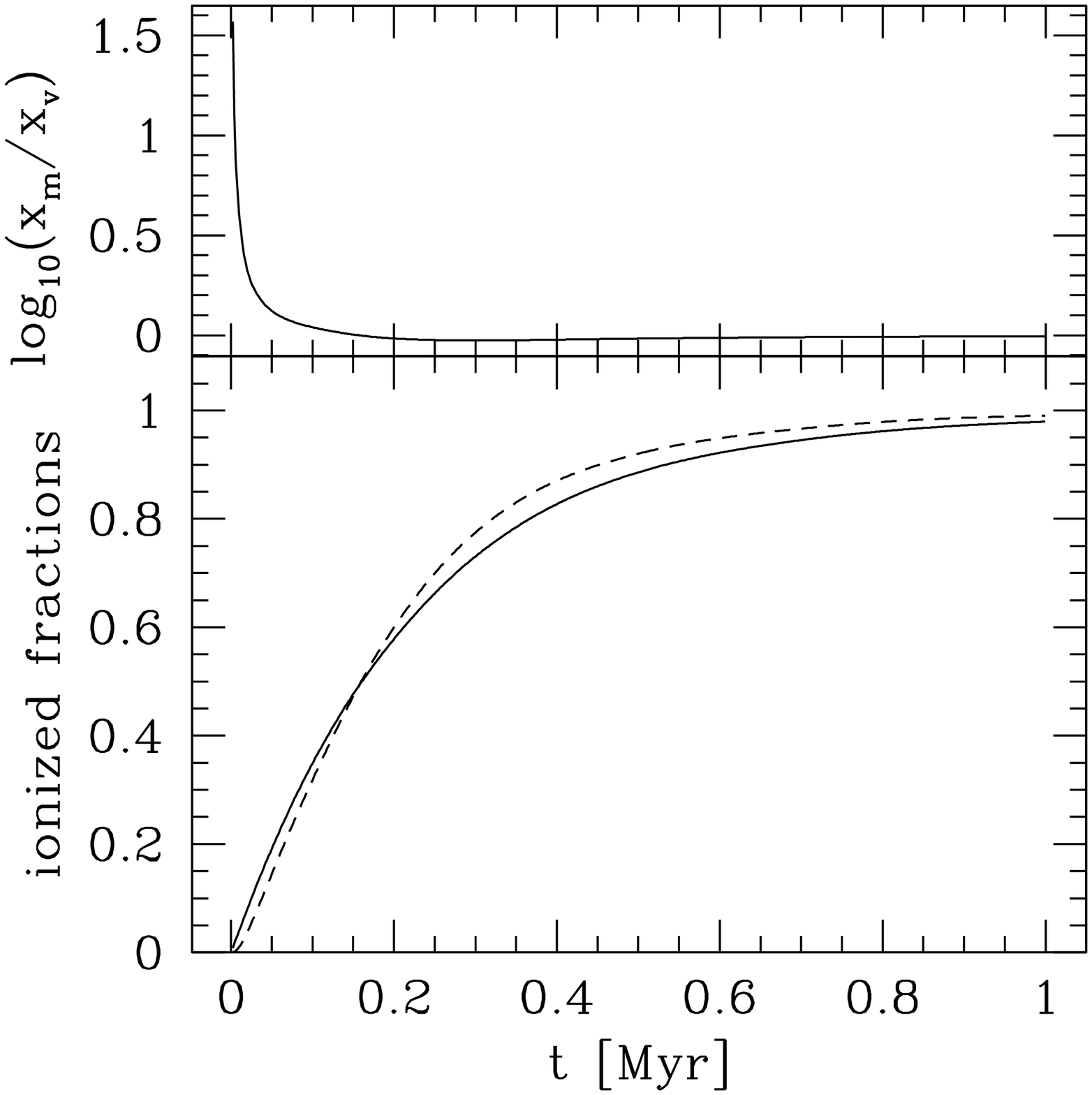}
\caption{Ionized gas fraction for (a)(left) single-source and (b)(right)
  multiple ionizing sources on a cosmological density field: mass-weighted,
  $x_m$, (solid) and volume-weighted, $x_v$ (dashed) (bottom panels) and their
  ratio (top panels).
\label{cosmo_ion_frac}}
\end{figure}

When multiple ionizing sources are present in the computational volume 
(Fig.~\ref{cosmo_fig_mult}), the topology of the H~II regions changes and 
becomes much more complex. Initially, individual H~II regions form around 
each source, but due to the significant source clustering at high redshift 
these small isolated H~II regions quickly merge into a few larger ones, with 
some remaining neutral pockets of denser gas along the filaments 
inside them which temporarily self-shield and trap the propagating I-front. 
The shape of the H~II regions is quite non-spherical at all times, dictated 
by the interplay between the source clustering and the underlying density 
distribution. The computational domain becomes completely ionized within 
$\sim1$ Myr.

This shows that reionization is not as simple as either ``inside-out''-- dense
regions ionized first, low-density regions last -- or ``outside-in'' --
low-density regions, or voids, ionized first, dense regions last, claimed by
\citet{2000ApJ...530....1M}. The dense regions in the immediate vicinity of
the sources were ionized first, but then the I-fronts quickly escaped into the
voids, and thereafter the overdense and under-dense regions were being ionized
simultaneously. This point is further illustrated in Figure~\ref{cosmo_ion_frac}, 
which shows the mass- and volume-weighted ionized fractions for both the single- 
and the multiple-source runs. Initially, in both cases the mass-weighted ionized
fraction is significantly larger than the volume-weighted one. However, the ratio 
quickly, within 0.1-0.2 Myr, drops to $\sim1$. The evolution is somewhat different 
thereafter. In the single-source case the ratio becomes somewhat less than one,
as the voids become ionized faster than the dense regions. In the multiple-source 
case, on the other hand, the ratio stays very close to 1, indicating that while the 
voids (which take most of the volume) are ionized relatively quickly, the dense knots
and filaments (which contain most of the gas mass) are also being ionized at the
same time, which keeps the ratio of the mass- to volume-ionized fraction similar. 
With this simple example we thus confirm the results 
of e.g.\ \citet{2002ApJ...572..695R}.

\section{Conclusions}

We presented a new method, called C$^2$-Ray, for photoionization 
calculations and radiative transfer in optically-thick media. 
Our method is explicitly photon-cons\-erving, which allows us to relax
the often-required condition for the computational cells to be 
optically-thin, and thus be able to obtain accurate results even for
very coarse spatial discretizations. Furthermore, we introduced 
time-averaged optical depths, which we obtain from a relaxation 
solution of the non-equilibrium chemistry equations, which allow us 
to use similarly coarse time-discretizations without any significant 
loss of accuracy. These developments result in a very fast, efficient 
and accurate calculation of the evolution of the ionization state of a gas 
distribution regardless of the spatial- and time-resolution. 

We tested our method systematically, and in significant detail, in several 
astro\-physically relevant situations in both 1D (in order to test its basic 
functionality) and in full 3D, at various resolutions in time and space. We
compared the results against the corresponding exact analytical solutions, 
some of which we derived here for the first time. The results of these tests
show that our method follows the analytical solutions to within a few percent,
and conserves photons with the same accuracy or better, even when using 
extremely coarse grids and very long time steps. We have also shown that
our radiative transfer method can be combined with adaptive mesh refinement 
(AMR), which leads to substantial increase in the calculation speed and much 
lower memory requirements.
These tests demonstrate the importance of extensive testing of RT
codes in a wide range of density fields - both for code validation and
for possible adjustments of free parameters for better handling of
special cases, e.g.\ steep density gradients. A project to thoroughly
compare different codes used for cosmological radiative transfer
methods is underway (Iliev et al.~2005, in preparation).

Our radiative transfer method imposes no limitations on the computational 
cell size and optical depth, and imposes no hard limits on the time step 
sizes. The only requirements we found are related to the desired accuracy 
of the solution obtained, for high accuracy the time step should be 
significantly shorter than the local recombination time. Longer time steps, 
comparable to the recombination time, lead to gradual loss of accuracy, but 
no incorrect or unstable solutions.

As a first simple application, we studied the reionization of a cosmological 
density field by both a single- and multiple-sources confirmed the findings 
of e.g.\ \citet{2002ApJ...572..695R} that neither inside-out, nor outside-in 
models describe the progress of cosmological I-fronts correctly. We showed that, 
instead, the mass- and volume-weighted ionized gas fractions are generally 
very similar, indicating that the dense and underdense regions are ionized 
at the same time.

The accurate I-front tracking over long time steps and coarse spatial 
resolutions makes our code ideal for direct dynamical coupling to 
multi-dimensional gasdynamics and N-body codes, where the evolution 
time-scales are generally much longer than the ionization and I-front 
crossing times. Allowing for larger, optically-thick cells further 
relaxes the time step limits imposed on the gasdynamic evolution through 
the Courant condition. In a follow-up paper we will address how we combine 
our C$^2$-Ray method with 3D gas- and N-body dynamics, including more details 
on the coupling to AMR and calculating the effects of photo-ionization 
heating.

The C$^2$-Ray's high computational efficiency and correct I-front
evolution over long time steps make it also a very attractive tool for
studying reionization using high-resolution precomputed density fields. As 
a ray-tracing method, its performance scales linearly with the number of 
ionizing sources and the number of computational cells, but our tests show 
that on the available hardware we can perform such post-processing calculations 
on large computational meshes of $400^3$ to $800^3$ cells and for thousands of 
ionizing sources. In a future paper we will describe the results of these 
calculations and show how we use them to make detailed observational predictions
e.g.\ for the large-scale 21-cm signal from the Epoch of Reionization.

GM is grateful to CITA and the LKBF for providing support for visits to CITA
where significant part of this work was completed. The work of GM is partly 
supported by the Royal Netherlands Academy of Arts and Sciences. During the
initial stages of this work ITI was supported in part by the Research and 
Training Network ``The Physics of the Intergalactic Medium'' established by 
the European Community under the contract HPRN-CT2000-00126. This work was 
partially supported by NASA grants NAG5-10825 and NNG04G177G and Texas
Advanced Research Program grant 3658-0624-1999 to PRS. MAA is grateful for the 
support of the US Department of Energy Graduate Fellowship in Computational
Sciences. 

\bibliographystyle{apj}

\appendix

\section{Ray tracing}
\label{tracing_sect}

\begin{figure}
  \includegraphics[scale=0.6]{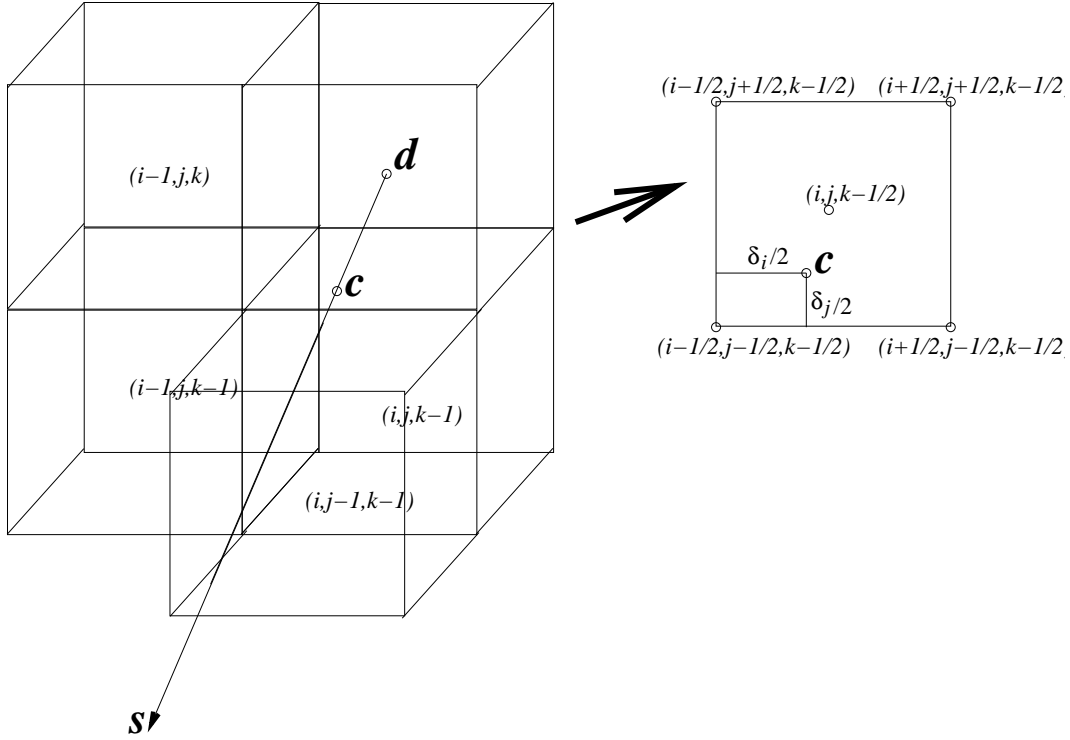}
\caption{Short-characteristics ray tracing. 
\label{raytrace_fig}}
\end{figure}

Here we describe how the column densities for a given cell are
constructed using the short-characteristics method. For this we
consider a point $d$ with position $(x,y,z)$ at mesh position
$(i,j,k)$, see Fig.~\ref{raytrace_fig}. The source point $s$ in physical coordinates is
$(x_{\rm s},y_{\rm s},z_{\rm s})$ or in mesh coordinates $(i_{\rm
s},j_{\rm s},k_{\rm s})$. The point where the ray enters the cell is
called $c$. Assuming that the cells are cubical, the ray
calculation is most easily done in mesh coordinates, which is what we
will use.

For our method we need two column densities for each cell, the column
density {\it to} the cell, called $N_c$ 
and the column density {\it over} the cell, denoted $\Delta N$.
$N_c$ corresponds to the optical depth $\tau_\nu$, and $\Delta N$
corresponds to $\Delta \tau_\nu$ in Eq.~(\ref{spatialgamma}).

$\Delta N$ is simply given by the $n_{\rm HI}{\rm d}s$, with ${\rm d}s$ 
the path length ${\rm d}s$ through the cell, which is twice
the distance between $c$ and $d$ and can be shown to be
\begin{equation}
  {\rm d}s=\sqrt{1 + {(i-i_{\rm s})^2 + (j-j_{\rm s})^2 \over
      (k-k_{\rm s})^2}}
\end{equation}
in units of cell size.  The appropriate value for $V_{\rm shell}$ in
Eq.~(\ref{spatialgamma}) is then given by $4\pi r_{sd}^2 {\rm d}s$, where
$r_{sd}$ is the distance between the source and the point $d$.

The column density to the cell $N_c$ is constructed from the column
densities of the neighboring cells. In order to pick the correct
neighbors, we need to establish where the ray enters the cell around
$d$. This can be done using the differences $\Delta i=i-i_{\rm s}$,
$\Delta j=j-j_{\rm s}$, $\Delta k=k-k_{\rm s}$ which show whether the
ray from the source enters the cell through the $x$, $y$, or $z$
plane. E.g.\ if $|\Delta k| > |\Delta j|$ and $|\Delta k| > |\Delta
i|$ the crossing is on the constant $z$-plane. Considering this case
we calculate the point $c$ where the ray enters the cell, with mesh
coordinates $(i_c,j_c,k_c)$, where $k_c=k-{1 \over 2}\sigma_k$ and
$\sigma_k={|\Delta k| \over \Delta k}$.

The neighbors closer to the source are the four cells with mesh positions
$e1$: $(i,j,k-\sigma_k)$, $e2$: $(i,j-\sigma_j,k-\sigma_k)$, $e3$:
$(i-\sigma_i,j,k-\sigma_k)$, and $e4$: $(i-\sigma_i,j-\sigma_j,k-\sigma_k$),
with $\sigma_{i,j}={|\Delta i,j| \over \Delta i,j}$. From these we construct
the column density at the crossing point $c$:
\begin{equation}
  N_c = w_1 N_{e1} + w_2 N_{e2} + w_3 N_{e3} +w_4 N_{e4}\,,
\label{coldens_weighted}
\end{equation}
with $w_i$ a set of normalized weights. These weights can be constructed in
different ways. The most straightforward one is taking the $x$ and $y$
distances from point $c$ to the corner points: \be
\begin{array}{l}
  {\delta}_i=2|i_c-i+\sigma_i/2|\\
  {\delta}_j=2|j_c-j+\sigma_j/2|
\end{array}
\ee and using \be
\begin{array}{l}
  w_1=(1-\delta_i)(1-\delta_j)\\
  w_2=\delta_i(1-\delta_j)\\
  w_3=(1-\delta_i)\delta_j\\
  w_4=\delta_i\delta_j
\end{array}
\ee With this weighting $N_c=N_{e1}$ if the ray is parallel to the $z$-axis,
and $N_c=N_{e4}$ if the ray travels diagonally over the mesh.

However, we found that this simple geometric weighting in case of a very
clumpy medium gives too much spreading of shadows. As discussed
previously in \citet{1999RMxAA..35..123R}, we suppress this by adding an extra
optical-depth-dependent factor in the weights $w_1-w_4$.  \be
\begin{array}{l}
  w_1={(1-\delta_i)(1-\delta_j)\over {\rm max}(\tau_0, \tau_{e1})\sum_n w_n}\\
  w_2={\delta_i(1-\delta_j) \over {\rm max}(\tau_0,\tau_{e2})\sum_n w_n}\\
  w_3={(1-\delta_i)\delta_j \over {\rm max}(\tau_0,\tau_{e3})\sum_n w_n}\\
  w_4={\delta_i\delta_j \over {\rm max}(\tau_0, \tau_{e4})\sum_n w_n}
\label{weightings}
\end{array}
\ee The value of $\tau_0$ was determined empirically at 0.6. For this number
we get the best photon-conservation for general clumpy media, when there are
no strong density gradients present. In all tests and applications
presented in this paper we use the weightings in Eq.~(\ref{weightings}),
unless otherwise noted.

In the special case of density field with strong density gradients (e.g.\ 
$n\propto r^{-2}$, as in one of the tests we discuss in \S~\ref{tests_sect})
combined with low spatial resolution, the weightings in
Eq.~(\ref{weightings}) with $\tau_0=0.6$ do not produce good photon
conservation. This is easy to understand since steep density gradients
combined with very coarse resolution lead to incorrect representation of the
density field on the grid and of significant grid-induced anisotropies, which
are only worsened by the weightings in Eq.~(\ref{weightings}). In such
cases it is better to use the weightings in
Eq.~(\ref{weightings}) with a small value for $\tau_0$
($\tau_0=\epsilon>0$), i.e.\ essentially the weightings in this case are
simply inversely proportional to the corresponding optical depths (or,
equivalently, column densities), since ${\rm max}(\tau_0,
\tau_{e,i})=\tau_{e,i}$ for $i=1,4$.  When the spatial resolution is
sufficiently high (e.g.\ $128^3$ or $256^3$ mesh in the tests discussed in
\S~\ref{tests_sect}) both types of weightings work well and produce almost
identical results, with very high level of photon conservation, generally
better than 1\%, regardless of the time resolution employed.

For $|\Delta j| > |\Delta k|$ and $|\Delta j| > |\Delta i|$ we are dealing
with a $y$-plane crossing and for $|\Delta i| > |\Delta k|$ and $|\Delta i| >
|\Delta j|$ with an $x$-plane crossing. For these two cases the calculations
are done in a similar way as above, but with the $y$ and $x$ coordinates
taking the role of the $z$ coordinate, respectively.

Both the short characteristic method (which uses the column densities
from the neighbors, as described above), and the time-averaged optical
depth calculation, require the ray-tracing to be causal. To achieve
this the mesh has to be traversed in a particular order, as described
in \citet{1999RMxAA..35..123R}: we first trace away from the source
along the $x$-axis ($y=y_{\rm s}$, $z=z_{\rm s}$), next we do strips of
constant $y$ parallel to the $x$-axis. After the source plane
($z=z_{\rm s}$) has been traced, we move to $z$-planes away from the
source, and trace each plane in the same way as the source
plane. Since each cell has its own ray segment, the short
characteristics method scales with the number of cells in the
computational mesh.

\subsubsection{Shadowing by a Dense Gas Cloud and I-front trapping}
\label{shadows_sect}
\begin{figure}
  \includegraphics[scale=0.5]{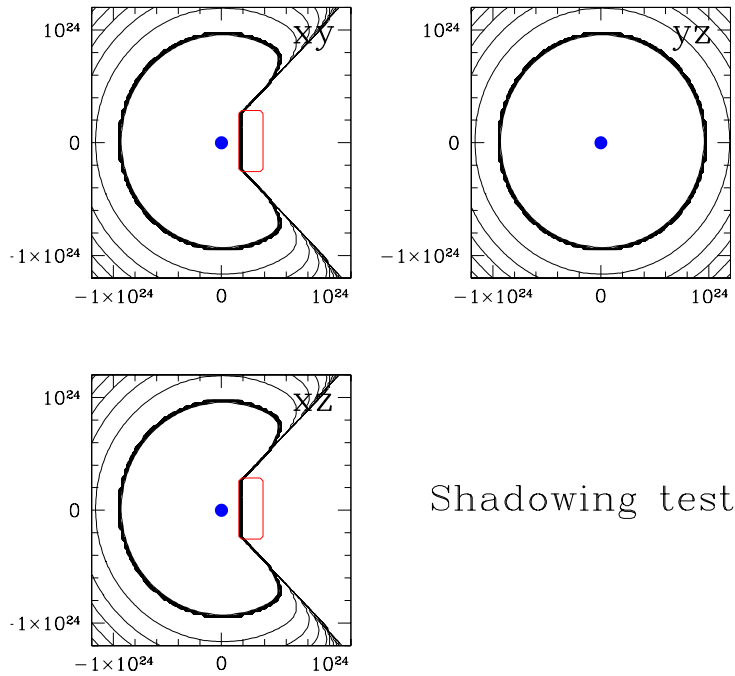}
\caption{Shadows test. Same parameters as in Test 1 above, but
  introducing a dense ($n=10^3\rm cm^{-3}$) uniform slab (red) for
  resolution of $128^3$ cells. Source is in the middle of the
  computational box (blue circle).  Contours are of the 50\% ionized
  fraction in a time-sequence from $t=20$ Myr to 100~Myr, with
  intervals of 20 Myrs. The thicker line is for $t=20$ Myrs and its
  thickness indicates the width of the ionization front.  The three
  panels for each case are cuts through the center along $x$-$y$, $y$-$z$ and
  $x$-$z$ plane, as labeled.
\label{shadows_fig}}
\end{figure}

Some of the main effects due to radiative transfer in realistic simulations
are slowing and possible trapping of the I-front in denser regions and the
corresponding casting of shadows behind these dense regions. In this section
we are testing our numerical scheme in such a situation. We use the initial
conditions from Test~1 in Table~\ref{tests}, except that the computational box
is 1 Mpc in size. We position a dense (gas number density $n=10^3\,\rm
cm^{-3}$), uniform, rectangular slab with size $2\times 20^{23}$ by $5\times
10^{23}$ by $5\times 10^{23}$~cm at a distance 0.083 Mpc from the source
(sizes are arbitrary and have no particular physical significance).
The temperature is fixed at $t=10^4$ K everywhere.
Results for a $128^3$ mesh are shown in Figure~\ref{shadows_fig}. As expected, 
the I-front is trapped inside the dense region within one cell. The
shadows cast are clean and sharp, with only modest spurious diffusion in the shadow 
behind the clump. The I-front outside the shadow is not affected by its existence 
and keeps its original, perfectly circular shape.

\end{document}